\documentclass[a4j, 11pt,dvipdfmx,autodetect-engine]{article}
\usepackage{jheppub}

\usepackage{wrapfig}
\usepackage[utf8]{inputenc}
\usepackage{braket}

\usepackage{natbib,hyperref}

\hypersetup{
    colorlinks=true,
    linkcolor=blue,
    citecolor=cyan,
    urlcolor=blue,
    pdfpagemode=FullScreen,
    }
    
\usepackage{amsmath,amsthm,amsfonts,amssymb,amscd}
\usepackage{bm}
\usepackage{bbold}
\usepackage{multirow,booktabs}
\usepackage[table,svgnames,dvipsnames]{xcolor}
\usepackage{enumerate}
\usepackage{enumitem}
\usepackage{mathrsfs}
\usepackage{wrapfig}
\usepackage{slashed}

\usepackage{empheq}
\usepackage{framed}
\usepackage[most]{tcolorbox}
\colorlet{shadecolor}{orange!15}

\edef\restoreparindent{\parindent=\the\parindent\relax}
\usepackage{parskip}
\restoreparindent

\usepackage{pgfplots}
\usepgfplotslibrary{fillbetween}

\theoremstyle{definition}

\newtheorem{protoexer}{Problem}[section]

\def\d{{\rm d}}
\def\tr{{\rm tr}}
\def\CC{\mathcal{C}}

\def\CH{\mathcal{H}}

\def\CO{\mathcal{O}}

\def\bx{\bm{x}}

\def\ben{\begin{equation}}
\def\een{\end{equation}}
\def\bena{\begin{eqnarray}}
\def\eena{\end{eqnarray}}
\newcommand{\non}{\nonumber}

\usepackage{comment}  
\usepackage{ascmac} % itembox

\vspace*{0cm}

\title{Energy Conditions and Quantum Information}

\medskip 

\author[a, b]{Norihiro Iizuka,}
\author[c, d]{Akihiro Ishibashi,} 
\author[e]{Kengo Maeda,}
\author[f]{Haruki Nakayama}
\author[f]{\\and Tatsuma Nishioka}

\affiliation[a]{\it Department of Physics, National Tsing Hua University, Hsinchu 300044, Taiwan} 

\affiliation[b]{\it Yukawa Institute for Theoretical Physics, Kyoto University, Kyoto 606-8502, Japan}

\affiliation[c]{\it Department of Physics, Nagoya University, Nagoya 464-6802, Japan}

\affiliation[d]{\it Kobayashi-Maskawa Institute, Nagoya University, Nagoya 464-8602, Japan}

\affiliation[e]{\it Faculty of Engineering, Shibaura Institute of Technology, Saitama 330-8570, JAPAN} 

\affiliation[f]{\it Department of Physics, The University of Osaka, Machikaneyama-Cho 1-1, Toyonaka 560-0043, Japan} 

\emailAdd{iizuka@phys.nthu.edu.tw}\emailAdd{ishibashi.akihiro.r7@f.mail.nagoya-u.ac.jp}\emailAdd{maeda302@sic.shibaura-it.ac.jp}
\emailAdd{hnakayama@het.phys.sci.osaka-u.ac.jp}\emailAdd{nishioka@het.phys.sci.osaka-u.ac.jp}

\abstract{The concept of energy lies at the foundation of physical science. In general relativity and quantum field theory, the positivity and conservation of energy are encapsulated by the so-called energy-momentum tensor and the energy conditions. In recent efforts to unify fundamental physics with quantum information, the energy conditions have come to play a crucial role in establishing numerous theorems. 
In this article, we review the basics of energy conditions in general relativity and their applications in gravitational physics, quantum field theory, and the holographic principle. Through these applications, we explore the profound connection between the energy conditions and quantum information. 
}

\preprint{NU-QG-10, OU-HET-1283}

\date{}

\begin{document}

\maketitle

\section{Introduction}\label{sec:1}
The notion of energy is fundamental to every scientific discipline and all technological advancements, ranging from the development of smart devices to our understanding of the universe. It is not just actual material systems that require energy; even a single piece of information needs energy to be stored or to be transferred. Quantum information has recently emerged as another pivotal concept for the next generation of science and technology. As exemplified by progress in quantum computing and quantum communication, recent efforts to integrate physical science and technology with this field are rapidly advancing. 
Even at the most fundamental level, such an attempt to unify physics and information has begun to open up a new research area where the very fabric of spacetime itself is envisioned to be described in terms of quantum information. The basic underlying idea is that energy must be assigned to a qubit---the unit of quantum information---and as a source of gravitation, energy produces spacetime curvature, according to the general theory of relativity. 
The involvement of spacetime curvature, or gravitation, as an impetus for this attempt appears to be rather inevitable, 
as the discovery of the Hawking radiation~\cite{Hawking:1974rv,Hawking:1975vcx} and the associated black hole information paradox~\cite{Hawking:1976ra} can be viewed as having initiated the crucial link between gravity, quantum, and information. According to Hawking, a black hole emits thermal blackbody radiation due to quantum field effects, loses its mass, and eventually evaporates away. However, the entire process of black hole formation and complete evaporation appears to contradict the unitarity of quantum mechanics, posing the paradox: where does the quantum information stored inside a black hole go?  

\medskip 

Any physically sensible system is expected to have positive definite energy, or at least a lower bound, to ensure a stable ground state. Also important in any physical theory is the law of the conservation of energy. 
In general relativity, the effects of matter fields on spacetime and conservation laws\footnote{%%%
The energy-momentum tensor allows us to express a local conservation of energy and momentum but it does not immediately lead to global properties of energy and momentum of a matter field. In order to discuss the positivity and conservation laws of ``total energy" of a system, one needs to introduce the notion of asymptotic flatness in order to define an isolated system (see Theorem 2.2.4 and Sec. 2.6 below).
} %%%  
are captured by the ``stress-energy-momentum tensor." 
The positivity of energy density is then prescribed by the ``energy conditions," which are inequalities to be imposed on the energy-momentum tensor. 
As will be seen, depending on the choice of observers who measure energy density and also on the specific circumstances of their imposition, one can consider various different types of energy conditions, among which the most widely used are the weak (WEC), the dominant (DEC), the strong (SEC), and the null energy conditions (NEC)~\cite{HE73,Wald84}. In other words, the choice of energy conditions is the selection of what one thinks physically reasonable matter fields are.     
In any case, roughly speaking, positive energy density generates the attractive nature of gravitation via the Einstein equations, leading to the focusing of families of world-lines of test particles or photons due to gravitational pull. This focusing effect---i.e., the convergence of causal geodesic curves---is key to understanding the physical consequences of gravitation and to establishing a number of important theorems in general relativity. For example, the singularity theorems of Hawking and Penrose~\cite{Penrose:1964wq,Hawking:1970zqf}, which predict the existence of spacetime singularities inside a black hole or at the beginning of our universe, exploit certain energy conditions and geodesic focusing in an essential way in their proof. The laws of black hole thermodynamics are shown by using a combination of various energy conditions~\cite{Hawking:1976de,Wald:1999vt}. Furthermore, by using the energy conditions, one can prohibit some exotic phenomena and structures from occurring, such as time-machines, wormholes, and faster-than-light travel~\cite{Morris:1988tu}; in other words, such an exotic or pathological spacetime structure necessitates some sort of negative energy density. 

\medskip 

The positivity of energy of a physical material system is, however, not a trivial condition, especially when quantum effects are involved; there is no shortage of examples of negative energy in quantum mechanical systems. One of the most well-known examples is the Casimir effect, which is experimentally well-established (see, e.g.,~\cite{Klimchitskaya:2009cw} and references therein). 
The Hawking radiation mentioned above also violates the positivity of energy. In fact, any locally (pointwisely) defined energy condition can, in principle, be violated by quantum field effects\footnote{%%%
It should also be noticed that even though local energy conditions can fail to hold under some quantum effects, they often admit lower bounds, hence satisfying what is called ``quantum energy inequality." See, e.g.,~\cite{Yurtsever:1994wc,Fewster:2012yh,Kontou:2020bta} for quantum energy inequalities and their relations to the (averaged) energy conditions.
}. %%%
For this reason, various types of non-local modifications of energy conditions have been proposed, such as those by taking the average of a local energy condition along an observer's world-line (called the averaged energy conditions), or by including non-local quantum field theoretic quantities, e.g., entanglement entropy of quantum fields (called the quantum null energy conditions). Such non-locally modified energy conditions turn out to play a key role in connecting quantum information with spacetime geometry. 

\medskip 

Our view of the connection between spacetime geometry and quantum matter has recently undergone revolutionary changes due to the rise of the holographic principle~\cite{tHooft:1993dmi,Susskind:1994vu}, which also emerged from the study of thermodynamic aspects of black holes. 
The basic assertion of the holographic principle is that the physics in a spacetime region can be completely described in terms of the degrees of freedom on the boundary of that region. It is fair to say that the holographic idea is still a conjecture, but a number of models exemplifying this idea have been proposed. So far the most successful and concrete model is the AdS/CFT correspondence~\cite{Maldacena:1997re,Gubser:1998bc,Witten:1998qj} (or gauge/gravity duality), which asserts that gravitational theory in a spacetime with negative curvature (called AdS) is equivalent to a certain type of quantum field theory (called CFT) residing on the boundary of the AdS spacetime. In particular, this correspondence makes it possible to analyze strongly coupled quantum fields in terms of weakly coupled classical gravity. Therefore the AdS/CFT correspondence can be a powerful tool to prove, for example, some of the energy conditions, as performed in~\cite{Kelly:2014mra}.  

\medskip 

The purpose of this article is to review the basics of energy conditions in general relativity, their uses in gravitational physics, their recent applications in quantum field theory, and relations to quantum information. This article is not intended to cover the entire range of relevant subjects, to be comprehensive, or provide a complete list of existing literature. Instead, we will introduce key ideas and methods of the uses of energy conditions and quantum information, by focusing on some specific topics.  In the first part, we mainly focus on the four locally defined energy conditions (WEC, DEC, SEC, NEC), with a particular emphasis on their role in several theorems in general relativity. In the second part, we first discuss how quantum effects lead to the violation of these locally defined energy conditions. Then we review some non-local modifications of the above mentioned energy conditions (ANEC and QNEC). We also discuss the role of (A)NEC in quantum field theory and in the context of the AdS/CFT correspondence. 

\medskip 

This article is organized as follows. In the next section, we discuss various types of the pointwise energy conditions and their uses in the context of classical theory of general relativity. In particular, we describe how the energy conditions are used in the singularity theorems, black hole mechanics and topology theorems. Most of the discussion in Sec.~\ref{sec:2} is based on Hawking-Ellis~\cite{HE73} and Wald~\cite{Wald84}. We also discuss the violation of the pointwise energy conditions and introduce the averaged energy conditions, such as ANEC. A number of examples of exotic spacetime structures, such as wormholes, superluminal travel, and their connection to the violation of the energy conditions can be found in literature, e.g.,~\cite{Lobo:2017cay}. More complete list of existing literature on the violation of locally defined, as well as averaged energy conditions can be found in, e.g., \cite{Fewster:2012yh,Kontou:2020bta,Martin-Moruno:2017exc}. 
In Sec.~\ref{sec:3}, we review the energy conditions and their derivations in quantum field theory in Minkowski spacetime.
We begin with the basic notions of quantum information such as the reduced density matrix and modular Hamiltonian for a subsystem, and introduce entanglement entropy and relative entropy as measures of quantum information.
We derive the Bekenstein bound and ANEC from the positivity and monotonicity of relative entropy by following~\cite{Casini:2008cr} and~\cite{Faulkner:2016mzt}, respectively.
We also outline the derivation of quantum null energy conditions (QNEC) by using relative entropy in~\cite{Ceyhan:2018zfg}. 
In Sec.~\ref{sec:4}, we review the connection between the energy conditions and the holographic principle. In particular, we explain the black hole information paradox in more detail in the context of the AdS/CFT correspondence, introducing the recent advancements such as the holographic entanglement entropy or the Ryu-Takayanagi formula~\cite{Ryu:2006bv,Ryu:2006ef}, quantum extremal surface (QES)~\cite{Faulkner:2013ana,Engelhardt:2014gca}, and Island formula~\cite{Almheiri:2019hni,Penington:2019npb,Almheiri:2019psf}. Then, we discuss the holographic proof of ANEC and its generalization by focusing on the role of causality in the AdS/CFT correspondence. Section~\ref{sec:5} is devoted to summary and discussions. 

\section{Energy conditions in general relativity}\label{sec:2} 

In this section, we will review basic properties of the energy conditions and their applications in general relativity and gravitational physics, including the singularity theorems, the laws of black hole mechanics, and the topological censorship. Our main focus is on the uses of four main, locally (pointwisely) defined energy conditions, but we also discuss their violations and non-local generalizations at the end of this section. 
Unless specifically stated, the assertions, theorems, and formulas discussed in this section assume a four-dimensional spacetime. Most of these formulas and theorems remain valid in general dimensions, either without modification or with obvious dimensional extensions.

\subsection{The Einstein equations} 

Spacetime geometry can be described in terms of the pair of a metric tensor $g_{\mu \nu}$ with Lorentzian signature and a connected manifold $M$. 
The Einstein equations are a set of second-order differential equations for the metric tensor, expressed as
\begin{eqnarray}
 R_{\mu \nu}- \dfrac{1}{2}R g_{\mu \nu} = 8\pi G T_{\mu \nu} \,,
\label{eq:Einstein}
\end{eqnarray}
where $R_{\mu \nu}$ and $R$ are the Ricci tensor and scalar curvature, respectively, constructed from the Riemann tensor\footnote{
The Riemann tensor is defined as follows: 
\bena 
 R^{\lambda}{}_{\sigma \mu \nu} = \partial_\mu \Gamma^\lambda_{\sigma \nu} - \partial_\nu \Gamma^\lambda_{\sigma \mu} 
 + \Gamma^\lambda_{\alpha \mu}\Gamma^\alpha_{\sigma \nu} -  \Gamma^\lambda_{\alpha \nu}\Gamma^\alpha_{\sigma \mu}  
\,, 
\eena
where $\Gamma^\lambda_{\mu \nu}= \dfrac{1}{2}g^{\lambda \sigma}(\partial_\mu g_{\sigma \nu} + \partial_\nu g_{\sigma \mu}-\partial_\sigma g_{\mu \nu})$ is the Christoffel symbol and $R_{\mu \nu} :=R^\alpha{}_{\mu \alpha \nu}$ and $R :=g^{\mu \nu}R_{\mu \nu}$. For geometric quantities, we follow the convention and notations of Wald's book~\cite{Wald84}. 
}. 
On the right-hand side are the energy-momentum (or stress-energy) tensor $T_{\mu \nu}$ for cosmic matter fields and $G$ denotes Newton's gravitational constant. Due to the Bianchi-identity the divergence of the left-hand side is automatically vanishing, and accordingly, the divergence of the right-hand side is also required to satisfy, 
\bena
 \nabla_\nu T^{\mu \nu} = 0 \,, 
\eena
where $\nabla_\mu$ denotes the covariant derivative operator\footnote{%%% 
The covariant derivative acts on a scalar field $\phi$ as $\nabla_\mu \phi = \partial \phi/\partial x^\mu =:\partial_\mu \phi$ 
and on a vector field $A^\mu$ as $\nabla_\nu A^\mu=\partial_\nu A^\mu + \Gamma^\mu_{\nu \lambda} A^\lambda$. From these formulas, one can deduce how $\nabla_\mu$ acts on more general tensor fields. 
}. %%%
This is the local law of conservation of material energy. 

\medskip 

For example, for a perfect fluid (no heat conduction and viscosity), the energy-momentum tensor is given by
\bena
 T_{\mu \nu} = (\rho + P) u_\mu u_\nu + Pg_{\mu \nu} \,, 
\label{emt:pefectfluid}
\eena
where $u^\mu$ is a unit timelike vector field, $g_{\mu \nu} u^\mu u^\nu=-1$, of the $4$-velocity of the fluid, and where $\rho$ and $P$, respectively, are the energy density and the pressure of the fluid as measured in its rest-frame. Then, from the local law of conservation above, one obtains general relativistic versions of the conservation of mass and the Euler equation. 

\medskip 

In general, when an action $S_M$ for a matter field is given, the energy-momentum tensor $T_{\mu \nu}$ is obtained by taking a variation of $S_M$ with respect to the metric tensor: 
\bena
T_{\mu \nu} := \dfrac{-2}{\sqrt{-g}}\dfrac{\delta S_M}{\delta g^{\mu \nu}} \,,  
\eena 
where $g:= \det (g_{\mu \nu})$. For example, the action for a single classical scalar field $\phi$ is given by 
\bena
S_\phi = \int \d^4 x \sqrt{-g} \left[ - \dfrac{1}{2}g^{\mu \nu}\nabla_\mu \phi \nabla_\nu \phi-V(\phi)\right] \,, 
\eena
where $V(\phi)$ denotes the potential. 
Then, the energy-momentum tensor is 
\bena
 T^{(\phi)}_{\mu \nu} &=& \dfrac{-2}{\sqrt{-g}}\dfrac{\delta S_\phi}{\delta g^{\mu \nu}} 
\non \\ 
  &=& \nabla_\mu \phi \nabla_\nu \phi - \dfrac{1}{2}g_{\mu \nu}\left[g^{\alpha \beta}\nabla_\alpha \phi \nabla_\beta \phi + 2V(\phi)\right] \,. 
\label{emt:scalar}
\eena
 
\medskip 

\subsection{Pointwise energy conditions}

Given the energy-momentum tensor for some ordinary matter field, one may expect that the energy density measured by an observer with $4$-velocity $v^\mu$ should be positive, namely, 
\bena
T_{\mu \nu}v^\mu v^\nu \geqslant 0 \,. 
\eena
For example, this implies $\rho \geqslant 0$ for the perfect fluid in the rest-frame $v^\mu=u^\mu$. This may be viewed as the simplest  form of energy conditions. 

\medskip 

The positivity of energy can, in principle, depend on an observer who measures energy and momentum. Besides the positivity, also important is the causality; in relativistic theory, energy and momentum associated with a particle form a single four-vector (four-momentum $p^\mu$), which is required to be timelike or null, and also be future pointing.  
Depending upon the choice of an observer, one may consider different types of energy conditions on the given energy-momentum tensor, as will be seen below. However, before going into more details of possible energy conditions, we should discuss why we need to impose them.  

\medskip 

There are two main reasons for considering energy conditions, both of which are motivated to choose physically reasonable energy-momentum tensors: 

\begin{enumerate}
\item[1]. Defining physically meaningful spacetime metrics. 

The first reason is that if no restrictions are imposed on the energy-momentum tensor (other than it satisfying conservation laws), virtually any metric could be considered a solution to the Einstein equations. This is because for any given metric, one can simply view the Einstein equations as the defining equations for an energy-momentum tensor\footnote{
This procedure for defining an energy-momentum tensor is sometimes known as the ``Synge G-method'' or the ``Nariai-method'' (see, e.g., \cite{Ellis:2023css}).  
}. However, such an arbitrary definition would not represent a physically meaningful spacetime. Instead, we must first impose realistic conditions on the matter fields. Only then should the Einstein equations be used to determine a physically valid metric. 

In this context, we are particularly concerned with the Einstein tensor, $G_{\mu \nu}$, as it (divided by $8\pi G$) is viewed as the energy-momentum tensor itself. The energy conditions most relevant to this discussion are the Weak Energy Condition (WEC), which requires positivity of energy, and the Dominant Energy Condition (DEC), which requires both positivity and causality. 

\medskip 

\item[2]. Probing spacetime by geodesics.  

The second reason relates to how we probe spacetime. The behavior of causal geodesic congruences---which represent a collection of observers or probe particles---is often used to understand the structure of spacetime. Energy conditions directly influence how these geodesic congruences behave through the Einstein equations. In particular, we are interested in causal geodesic congruences. As we will see below, energy conditions determine the focusing of nearby geodesics due to gravitational attractive force. Combined with energy conditions, we can formulate focusing theorems for causal geodesics, which can be then used to prove, in an essential way, a number of important theorems in general relativity, such as singularity theorems, black hole mechanics, and topological censorship. 

In this context, we are primarily concerned with the Ricci tensor, $R_{\mu \nu}$, as it directly appears in the Raychaudhuri equation as we will explain below. The energy conditions most relevant here are the Strong Energy Condition (SEC), which relates to timelike convergence, and the Null Energy Condition (NEC), which relates to null convergence.  

\end{enumerate} 

Now we state the four standard energy conditions, WEC, DEC, SEC, and NEC, locally defined within the classical general relativity. 
The energy-momentum tensor $T_{\mu \nu}$ itself can be classified essentially into four types based on the eigenvalue problem for the components $T^{(a) (b)}$ with respect to a local Lorentz/orthonormal frame $\{e^{(a)}_\mu\}$, $\eta_{ab}e^{(a)}_\mu e^{(b)}_\nu =g_{\mu \nu}$: 
\bena
 \det (T^{(a)(b)}- \lambda \eta^{ab}) = 0 \,. 
\label{eigenvalue:HEtype}
\eena 
The Hawking-Ellis classification~\cite{HE73} consists of Type I -- Type IV. We can characterize each condition, WEC, DEC, SEC, or NEC, in terms of the eigenvalues $\lambda$ for (\ref{eigenvalue:HEtype}). The most generic case is Type I, for which 
\bena
 T^{(a)(b)}= \rho e^{(a)}_0 e^{(b)}_0 + \sum_{i=1}^3 p_i e^{(a)}_i e^{(b)}_i \,,  
\eena
so that the Lorentz invariant eigenvalues of $T^{(a)}{}_{(b)}$ are given by $\{-\rho, p_1, p_2, p_3\}$~\cite{HE73,Martin-Moruno:2017exc}. Many of classically and semiclassically interesting matter fields are of Type I, encompassing, e.g., a perfect fluid (\ref{emt:pefectfluid}) and a massive scalar field (\ref{emt:scalar}) with the potential $2V(\phi)=m^2 \phi^2$. Type II -- IV involves more complications (see~\cite{HE73,Martin-Moruno:2017exc}, and see also~\cite{Maeda:2018hqu,Maeda:2022vld} for recent elaborations). In the following, for simplicity we characterize each of WEC, DEC, SEC, NEC, only in terms of Type I energy-momentum tensor. 

\medskip 
  
\begin{itembox}[l]{{Definition 2.2.1 (Weak Energy Condition)}}  

\begin{center}
$T_{\mu \nu}\xi^\mu \xi^\nu \geqslant 0$ for arbitrary timelike vector $\xi^\mu$. 
\end{center}

\end{itembox} 

\bigskip 

\begin{itemize} 
\item {For Type I case} $\rho \geqslant 0$, $\rho + p_i \geqslant 0$.

\medskip 

\item This condition simply implies that the energy density measured by any timelike observer should be non-negative and is generally believed to hold for all physically reasonable classical matter fields. 

\end{itemize} 

\medskip 

\begin{itembox}[l]{{Definition 2.2.2 (Dominant Energy Condition)}} 
For arbitrary future-directed timelike vector $\xi^\mu$, $-T^\mu{}_{\nu} \xi^\nu $ becomes a future-directed timelike or null vector.  

\end{itembox} 

\medskip 

\begin{itemize}
\item{For Type I case} $\rho \geqslant |p_i| \geqslant 0$. 

\medskip
\item{} Equivalently, for any future- (or past-) directed timelike vectors $\xi_1^\mu$ and $\xi_2^\mu$, $T_{\mu \nu}\xi_1^\mu \xi_2^\nu \geqslant 0$. 

\item{} DEC implies WEC. 

\medskip
\item{} DEC implies that the local energy density is non-negative, and the local energy flow vector is non-spaclike. 
This ensures the following conservation law:  

\smallskip 
\begin{itembox}[l]{Proposition 2.2.3 (Conservation law)}
Suppose the energy-momentum tensor for matter fields $T_{\mu \nu}$ satisfies the local conservation $\nabla_\nu T^{\mu \nu}=0$ and the DEC. 
If $T_{\mu \nu}=0$ on a closed, achronal hypersurface $\Sigma$, then $T_{\mu \nu}=0$ everywhere in the domain of dependence $D(\Sigma)$.  
\end{itembox}
\begin{itemize}
\item Lemma. 4.3.1~\cite{HE73}. 
\item For achronal surface $\Sigma$, and the domain of dependence $D(\Sigma)$, see Sec. \ref{subsec:sing} below.
\end{itemize}

\medskip 

\item{}Another important consequence of the DEC is the positivity of the {\em total energy} of an isolated system. In general relativity, an isolated system is defined in terms of {\em asymptotic flatness}, and the total energy defined at spatial infinity is called the ``Arnowitt-Deser-Misner (ADM)" energy~\cite{Arnowitt:1962hi}, whose positivity was shown by Schoen-Yau~\cite{Schon:1981vd}, Witten~\cite{Witten:1981mf}:  
%, 
\smallskip 
\begin{itembox}[l]{Theorem 2.2.4 (Positive energy theorem)}
The total energy is positive for initial data on an asymptotically flat Cauchy surface, provided the DEC holds.  
\end{itembox}
\begin{itemize}
\item As the total energy of a given isolated system, the ADM energy is a constant. One can also consider the so-called ``Bondi" energy~\cite{Bondi:1962px}, which is defined at null infinity (see Sec. 2.6) and describes the remaining energy at an instant of retarded time after the emission of gravitational radiation. For precise definitions of asymptotic flatness, as well as the ADM and the Bondi energy, see Sec. 11.2 of Wald's book~\cite{Wald84}. 

\item The positivity of the Bondi energy was shown by Horowitz-Perry~\cite{Horowitz:1981uw}, Schoen-Yau~\cite{Schon:1982re}, Ludvigsen-Vickers~\cite{Ludvigsen:1982}, again under the DEC. The positivity of these two, in particular, the Bondi energy, ensures the stability of an isolated system in general relativity.  

\item This theorem was generalized to include black holes and also for the asymptotically AdS case by Gibbons et al~\cite{Gibbons:1982jg}. 

\end{itemize}

\end{itemize}

\medskip 

\begin{itembox}[l]{{Definition 2.2.5 (Strong Energy Condition)}} 
\begin{center}
$
\left( T_{\mu \nu} - \dfrac{1}{2}T^\lambda{}_\lambda g_{\mu \nu} \right) \xi^\mu \xi^\nu \geqslant 0 
$ for arbitrary timelike vector $\xi^\mu$.
\end{center}

\end{itembox} 

\bigskip 
\begin{itemize}

\item {For Type I case} $\rho + p_i \geqslant 0$, $\rho + \sum_{i=1}^3p_i \geqslant 0$. 

\medskip 

\item Similar to WEC, SEC is introduced to study the focusing behavior of causal geodesics. 
Through the Einstein equations, SEC is equivalent to the condition $R_{\mu \nu} \xi^\mu \xi^\nu \geqslant 0$, hence leading to the focusing of timelike geodesic congruences (Theorem 2.4.4 below). 

\item For a cosmological constant, the energy-momentum tensor can be written as $8 \pi G T^{(\Lambda)}_{\mu \nu}= - \Lambda g_{\mu \nu}$ and hence SEC implies 
\bena
{\Lambda} g_{\mu \nu} \xi^\mu \xi^\nu \geqslant 0 \,.
\eena
Therefore, if a positive cosmological constant is dominant as a cosmic ingredient, the SEC is not satisfied.  
Instead of focusing, the timelike geodesic congruence has a positive expansion, expressing that a repulsive force is acting on world-lines of timelike test particles. This repulsive nature, if applying to cosmology, leads to an accelerating expansion of the universe, as necessary to describe an inflationary universe or dark energy dominated universe. For this reason, the interest in SEC lies more in its violation rather than its satisfaction. To be more concrete, let us consider the energy-momentum tensor for a scalar field (\ref{emt:scalar}). In the vacuum state (i.e., no excitation of $\phi$, and the derivative terms can be ignored), we virtually obtain $T^{(\phi)}_{\mu \nu} \approx - V(\phi) g_{\mu \nu}$. This is equivalent to the perfect fluid with $\rho=-P= V(\phi) $. Since $T_{\mu \nu}-(1/2)T^\lambda{}_\lambda g_{\mu \nu} \approx V(\phi)g_{\mu \nu}$, as far as $V(\phi)$ is positive, the SEC is violated.  
 
\medskip

\item
SEC does not imply WEC. 
%There is no containment relationship between SEC and WEC. 
For example, for a negative cosmological constant, $\Lambda<0$, SEC is satisfied, but WEC fails to hold. 

%\item For $d$-dimensions, SEC is expressed as $(T_{\mu \nu}- \dfrac{1}{d-2}T^\lambda{}_\lambda g_{\mu \nu}\right)%

\end{itemize}

\medskip 

\begin{itembox}[l]{Definition 2.2.6 (Null Energy Condition)}

\begin{center}

$T_{\mu \nu}k^\mu k^\nu \geqslant 0$ for arbitrary null vector $k^\mu $. 

\end{center}

\end{itembox}

\medskip

\begin{itemize}

\item{For Type I case} $\rho + p_i \geqslant 0$. 

\item Since NEC is equivalent to the condition $R_{\mu \nu} k^\mu k^\nu \geqslant 0$ via the Einstein equations, by applying it to the Raychaudhuri equation (see (\ref{Raych:null}) below), one can derive the focusing theorem for a null geodesic congruence (Theorem 2.4.2 below). In this sense, this is also called the {\em Null Convergence Condition} (NCC). 

\medskip 

\item NEC is analogous to WEC, with a timelike vector $\xi^\mu$ replaced by a null vector $k^\mu$. 
If WEC holds, then by continuity, NEC also holds. Similarly, if SEC conditions holds, then for any $\xi^\mu$, $T_{\mu \nu} \xi^\mu \xi^\nu \geqslant (1/2)T^\lambda{}_\lambda \xi_\mu \xi^\nu$, and hence, by continuity, NEC also holds. 

\medskip 

\item 
The violation of NEC (NCC) implies the violation of WEC. Suppose NEC is violated. Then, for some future-directed null vector $k^\mu$, there exists a constant $\delta >0$ such that 
$$
 T_{\mu \nu} k^\mu k^\nu = - \delta < 0 \,. 
$$
Let $u^\mu$ be an appropriate future-directed timelike vector. Then, $\xi^\mu := k^\mu + \epsilon u^\mu , \, (\epsilon >0)$ is also a timelike vector. For this timelike vector $\xi^\mu$, 
$$
T_{\mu \nu}\xi^\mu \xi^\nu =  -\delta + 2 \epsilon T_{\mu \nu}k^\mu u^\nu + O(\epsilon^2) \,. 
$$
Therefore for sufficiently small $\epsilon$, $T_{\mu \nu}\xi^\mu \xi^\nu <0$. 
\end{itemize}

We summarize the above four locally defined energy conditions in Table~\ref{table:1}. 
%%% Table 1
\begin{table}[h]
\centering
\setlength{\tabcolsep}{0.3pt}
\setlength{\extrarowheight}{8pt} 
\caption{\small Four locally defined energy conditions. Here $\xi^\mu$ and $k^\mu$ denote any timelike and null vector, respectively, and ``causality" means that superluminal flow of energy is not allowed. The timelike vectors $\xi_1^\mu,\: \xi_2^\nu$ appearing in DEC are assumed to share the same time orientation. 
}

{ % For twocolumn --> \tiny
\label{table:1}
\begin{tabular}{| l|c|c|c|c|c}\hline
  
  {  } & & Definition & Effects  &  Type I case    \\[8pt] \hline \hline
\,\, WEC \,\, &  & \,\,\,$T_{\mu \nu}\xi^\mu \xi^\nu \geqslant 0$ \,\, \,& positivity & \,\, $\rho \geqslant 0$, $\rho + p_i \geqslant 0$  \,\,  \\[8pt]\hline
  
\,\,  DEC & & $T_{\mu \nu}\xi_1^\mu \xi_2^\nu \geqslant 0$ & \,\, positivity and causality \,\, & \,\,
  $\rho \geqslant |p_i| \geqslant 0$ \\[8pt]\hline
   
\,\, SEC && $R_{\mu \nu}\xi^\mu \xi^\nu \geqslant 0$ & timelike focusing & \,\,$\rho + \sum_i p_i \geqslant 0$, $\rho + p_i \geqslant 0$ \,\,  \\[8pt]\hline 
 
\,\, NEC & & $R_{\mu \nu} k^\mu k^\nu \geqslant 0$ & null focusing & \,\,$\rho + p_i \geqslant 0$   \\[8pt]\hline
  
\end{tabular}
} 
\end{table}

\subsection{Geodesic congruences on curved spacetime} 

Given the metric which solves the Einstein equations, we can learn a lot about the properties of spacetime not only by directly looking into the metric tensor itself, but also---sometimes more effectively---by inspecting the behavior of causal geodesic curves, which are world-lines of probe particles such as massive test particles or photons. In this section, we first review timelike and null geodesics, and derive the formulas governing the rate of change of expansion for a family (congruence) of causal geodesics. Then, we review the (classical) focusing theorem. 

\medskip 

\begin{figure}[h]
\begin{center}
%\scalebox{0.10}{%
\includegraphics[width=80mm]{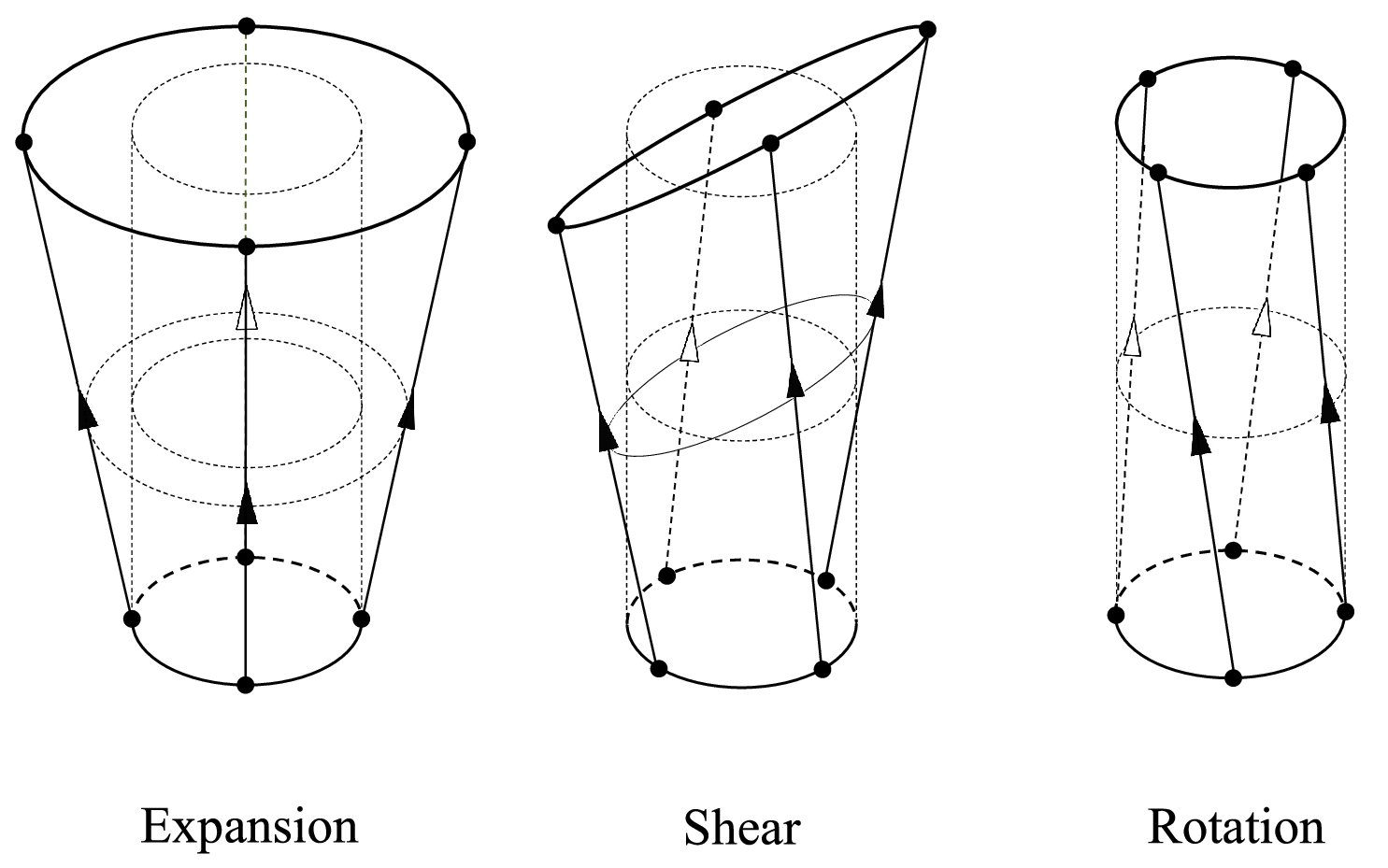} 
%              } 
\end{center}

\vspace{-5mm}
\caption{\small 
The geometric interpretation of the expansion $\theta$, the shear-tensor $\sigma_{\mu \nu}$, and the rotation $\omega_{\mu \nu}$.   
} 
\label{fig:congruence}
\end{figure}

Let us consider a timelike geodesic curve $\gamma$, whose unit tangent vector is expressed by $u^\mu=\d x^\mu/\d\tau $ with $\tau$ being the proper time. The tangent vector satisfies $g_{\mu \nu} u^\mu u^\nu=-1$ and the geodesic equation, 
\bena
 u^\nu \nabla_\nu u^\mu =0 \,. \quad
\eena 
We consider a geodesic congruence of $\gamma$, which is a bundle of mutually non-intersecting timelike geodesics nearby $\gamma$. 
Let us define the tensor field, 
\bena
h_{\mu \nu} := g_{\mu \nu} + u_\mu u_\nu \,.
\eena 
It immediately follows that $h^\mu{}_\sigma h^\sigma{}_\nu=h^\mu{}_\nu$ and $h^\mu{}_\nu u^\nu=0$, and thus the tensor field $h^\mu{}_\nu$ projects any vectors to the $3$-dimensional subspace orthogonal to $u^\mu$. It also satisfies $u^\sigma \nabla_\sigma h^\mu{}_\nu =0$. Then, we can decompose the tensor field, $\nabla_\nu u_\mu$, into its trace-part $\theta$, trace-free symmetric part $\sigma_{\mu \nu}$, and anti-symmetric part $\omega_{\mu \nu}$, as follows:  
\bena
&{}& 
 \nabla_\nu u_\mu = \dfrac{1}{3}\theta h_{\mu \nu} + \sigma_{\mu \nu}+ \omega_{\mu \nu} \,, 
\non \\ 
&{}& \theta:= h^{\mu \nu}\nabla_\nu u_\mu = \nabla_\mu u^\mu \,, \quad 
                 \sigma_{\mu \nu}:= \nabla_{(\nu}u_{\mu)}- \dfrac{1}{3}\theta h_{\mu \nu} \,, 
\non \\ 
&{}& \omega_{\mu \nu}:= \nabla_{[\nu}u_{\mu]} \,, 
\eena
where $\theta$ is called the {\em expansion} of the congruence of $\gamma$, $\sigma_{\mu \nu}$ the {\em shear}, and $\omega_{\mu  \nu}$ the {\em rotation}. Note that by definition $u^\mu \sigma_{\mu \nu}=0= u^\mu \omega_{\mu \nu}$. The geometric meaning of these quantities can be read off from Figure~\ref{fig:congruence}. 

\medskip 

Most important in the context below is the behavior of the expansion $\theta$, since it expresses the focusing of geodesics due to gravitational attraction. We are often concerned with hypersurface orthogonal geodesic congruences, for which $\omega_{\mu \nu}$ must vanish according to the Frobenius theorem\footnote{%%% 
If a vector field $n^\mu (\neq 0)$ satisfies 
\bena 
n_{[\alpha} \nabla_{\beta} n_{\gamma]}=0 \,, 
\non 
\eena 
then the orbits of $n^\mu$ are orthogonal to some hypersurface. Conversely, if $n^\mu$ is hypersurface orthogonal, then it satisfies the above equation. 
}. %%% 

\medskip 

Let us consider a congruence of timelike geodesics orthogonal to $3$-dimensional hypersurface $\Sigma$ with vanishing rotation $\omega_{\mu \nu}=0$. If we denote by $A$ an infinitesimal cross-sectional ($3$-dimensional) area element of the congruence, then 
\bena
 \theta = \dfrac{1}{A}\dfrac{\d A}{\d\tau} \,, 
\eena
and thus describes the local rate of change of cross-sectional area along the geodesic $\gamma$. 
Now, by taking the derivative of $\nabla_\nu u_\mu$ along $\gamma$, i.e., $u^\alpha \nabla_\alpha (\nabla_\nu u_\mu)$, and taking its trace, we obtain the Raychaudhuri equation
\bena
 \dfrac{\d \theta}{\d \tau} = - \dfrac{1}{3}\theta^2 - \sigma_{\mu \nu}\sigma^{\mu \nu} - R_{\mu \nu}u^\mu u^\nu \,, 
\eena
where $R_{\mu \nu}$ is the Ricci tensor. We can also derive the equations for $\sigma_{\mu \nu}$ by taking the symmetric trace-free part of $u^\alpha \nabla_{\alpha}(\nabla_\nu u_\mu)$, 
\bena
u^\alpha \nabla_\alpha \sigma_{\mu \nu}
 &=& - \frac{2}{3}\theta \sigma_{\mu \nu} 
       -\left(\sigma_{\mu \sigma}\sigma_\nu{}^\sigma - \frac{1}{3}\sigma_{\alpha \beta}\sigma^{\alpha \beta} h_{\mu \nu} \right)        
% For two column --> \non \\ &{}& 
- C_{\mu \alpha \nu \beta}u^\alpha u^\beta 
% For two columun -->\non \\ &{}& 
+ \frac{1}{2}\left( h_\mu{}^\alpha h_\nu{}^\beta - \frac{1}{3}h^{\alpha \beta}h_{\mu \nu} \right) R_{\alpha \beta} \,, 
\eena
where 
\bena
C_{\mu \alpha \nu \beta} &:=& R_{\mu \alpha \nu \beta} +g_{\mu [\beta}R_{\nu] \alpha}+ g_{\alpha [\nu}R_{\beta] \mu}
% fot twocolumn --> \non \\  &{}& 
+ \dfrac{1}{3}Rg_{\mu [\nu}g_{\beta]\alpha}
\eena
is the Weyl tensor. Note that we omit terms related to $\omega_{\mu \nu}$ since $u^\mu$ is assumed to be hypersurface orthogonal.    

\bigskip 

We can derive similar formulas for null geodesic congruences. Again, we focus on hypersurface orthogonal null geodesics, since a number of interesting spacetime structures can be prescribed by a null hypersurface, generated by null geodesics. A subtle point is that null geodesic generators of a null hypersurface ${\cal N}$ are tangent to, and simultaneously, normal to ${\cal N}$. For this reason, even though ${\cal N}$ itself is a $3$-dimensional hypersurface, a spacelike cross-section $S$ of ${\cal N}$, to which the null congruence is orthogonal, is $2$-dimensional. Let $k^\mu=\d x^\mu/\d \lambda$ be an affinely parametrized null geodesic vector field, so that 
\bena
 k^\nu \nabla_\nu k^\mu=0 \,, \quad k^\mu k_\mu =0 \,.
\eena 
We also assume that $k^\mu$ is null-hypersurface orthogonal. Let $A$ be an infinitesimal cross-sectional ($2$-dimensional) area element of the congruence. Then, the expansion of the congruence is defined by  
\bena
\theta = \nabla_\mu k^\mu = \dfrac{1}{A}\dfrac{\d A}{\d\lambda} \,.
\eena
We can also define the shear $\sigma_{\mu \nu}$ in a manner similar to that  or the timelike geodesic case. 
For this purpose, we introduce an auxiliary null vector field $l^\mu$, parallelly transported by $k^\mu$, so that it satisfies, $l^\mu l_\mu=0$, $l^\mu k_\mu =-1$, and $k^\nu \nabla_\nu l^\mu=0$. Then, we define the projection tensor $p^\mu{}_\nu$ by 
\bena
 p_{\mu \nu} :=g_{\mu \nu} + k_\mu l_\nu + l_\mu k_\nu \,. 
\eena
It satisfies $p^\mu{}_\sigma p^\sigma{}_\nu = p^\mu{}_\nu$, $p^\mu{}_\nu k^\nu =0=p^\mu{}_\nu l^\nu$, $k^\sigma \nabla_\sigma p^\mu{}_\nu =0$. 
Note that the auxiliary null vector field $l^\mu$ cannot be uniquely chosen, and accordingly $p^\mu{}_\nu$ is not uniquely defined. Then, the shear and the rotation, are defined by 
\bena
 \sigma_{\mu \nu} = \nabla_{(\mu} k_{\nu)} - \dfrac{1}{2} \theta p_{\mu \nu} \,, \quad 
 \omega_{\mu \nu}=\nabla_{[\mu}k_{\nu]}\,. 
\eena 
Again, since we are concerned with surface orthogonal null geodesics, in what follows we set $\omega_{\mu \nu}=0$. 
Then, we can derive the corresponding Raychaudhuri equation 
\bena
\dfrac{\d \theta}{\d\lambda} = - \dfrac{1}{2}\theta^2 - \sigma_{\mu \nu}\sigma^{\mu \nu} - R_{\mu \nu}k^\mu k^\nu \,.
\label{Raych:null}
\eena
We can also derive the equation for the shear tensor as
\bena
k^\alpha \nabla_\alpha \sigma_{\mu \nu}
 &=& - \theta \sigma_{\mu \nu} 
       -\left(\sigma_{\mu \sigma}\sigma_\nu{}^\sigma - \frac{1}{2}\sigma_{\alpha \beta}\sigma^{\alpha \beta} p_{\mu \nu} \right) 
% fot twocolumn --> \non \\  &{}&  
- p_\mu{}^\alpha p_\nu{}^\beta C_{\sigma \alpha \rho \beta} k^\sigma k^\rho \,, 
\label{eom:shear:null}
\eena
where $C^\mu{}_{\alpha \nu \beta}$ is the Weyl tensor. 

\subsection{Focusing theorems and energy conditions}\label{subsec:focus}

Now that we have formulas of the expansion and shear for both timelike and null geodesic congruence, we discuss the focusing effects. Below we consider first the null geodesic case in detail, and then derive essentially the same conclusion for the timelike case. 
We immediately see that the first and second terms in the right-hand side of (\ref{Raych:null}) are non-positive definite, hence always tend to decrease the expansion $\theta$ and cause the convergence of the null geodesic congruence. The third term, $-R_{\mu \nu}k^\mu k^\nu$, is non-positive if the NEC holds, thus also tends to decrease $\theta$ under the NEC. 
Consider the case that both the expansion and shear tensors vanish initially. If $R_{\mu \nu}k^\mu k^\nu \neq 0$ at some point of $\gamma$, then via the Raychaudhuri equation, $\theta$ becomes non-vanishing, which, again via the Raychaudhuri equation, makes $\theta$ start to decrease. Suppose now $R_{\mu \nu}k^\mu k^\nu=0$ (vacuum spacetime), and initially $\theta=0=\sigma_{\mu \nu}$. Even in this case, if the third term of equation~(\ref{eom:shear:null}) is non-vanishing, $p_\mu{}^\alpha p_\nu{}^\beta C_{\sigma \alpha \rho \beta} k^\sigma k^\rho \neq 0$, it makes the shear tensor non-vanishing by (\ref{eom:shear:null}) and then produces, via (\ref{Raych:null}), non-vanishing expansion. Then, again from (\ref{Raych:null}), $\theta$ starts to decrease. Now we state the following condition.   
\begin{itembox}[l]{{Definition 2.4.1 (Null generic condition)}}
A spacetime is said to satisfy the null generic condition if every null geodesic (whose tangent vector field is denoted by $k^\mu$) possesses at least one point where   
\begin{center}
$k_{[\alpha} R_{\mu]\beta \nu[\gamma} k_{\sigma ]} k^\beta k^\nu \neq 0$. 
\end{center}
\end{itembox}
This condition is equivalent to either (i) or (ii) below holds \cite{Wald84}, 
\bena
\mbox{(i) }\, R_{\mu \nu}k^\mu k^\nu \neq 0 \,, \,\, \mbox{(ii) }\, p_\mu{}^\alpha p_\nu{}^\beta C_{\sigma \alpha \rho \beta} k^\sigma k^\rho \neq 0 \,, 
\eena
at at least a single point on any null geodesic $\gamma$ and is interpreted as a non-vanishing tidal force acting on the photon described by $\gamma$. 

\medskip 

\begin{figure}[h]
\begin{center}
\includegraphics[width=50mm]{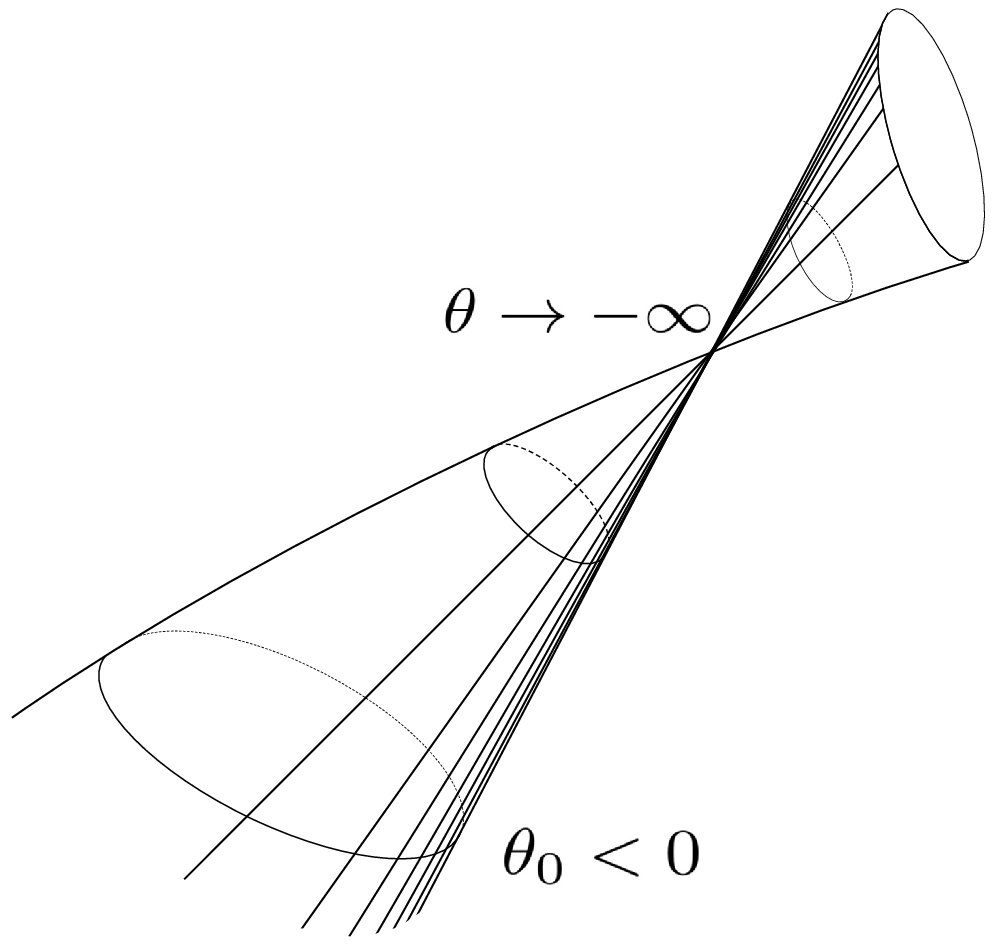} 
%\scalebox{0.10}{%
%\includegraphics{focusing.eps} 
%             } 
\end{center}
\vspace{-5mm}
\caption{\small 
A null geodesic congruence whose expansion takes initially negative value $\theta_0<0$ converges, and within the affine length $\lambda \leqslant 2/|\theta_0|$, the cross-section area $A$ of the congruence vanishes, where by definition $\theta \rightarrow -\infty$. 
} 
\label{fig:focusing}
\end{figure}

Suppose that our null geodesic curve $\gamma$ with the tangent vector $k^\mu $ is {\em complete} with respect to the affine parameter $\lambda$; namely $\gamma$ can be extended to take arbitrary value of $\lambda$ on the spacetime considered. 
Then, from the above observation, it is inevitable that under the generic conditions and the NEC, the expansion eventually negatively diverges $\theta \rightarrow - \infty$. In fact, under the NEC, we have from the Raychaudhuri equation~(\ref{Raych:null}),   
\bena
\frac{\d \theta}{\d \lambda}  \leqslant - \frac{1}{2}\theta^2 < 0 \,. 
\eena 
Assuming the initial value $\theta_0$ at $\lambda=0$, integrating the above inequality, we obtain 
\bena
    \frac{1}{\theta} \geqslant \frac{1}{\theta_0} + \frac{\lambda}{2} \,. 
\eena
In particular, if the initial value of the expansion takes some negative value $\theta_0<0$, then 
\bena
  \theta \leqslant - \frac{2}{2/|\theta_0| - \lambda} \,. 
\eena
Thus, we arrive at the following result (see also Figure~\ref{fig:focusing}): 

\begin{itembox}[l]{{Theorem 2.4.2 (Null focusing)}}
Under the NEC, any hypersurface-orthogonal null geodesic congruence is focusing:
\begin{center}
$\dfrac{\d \theta}{\d\lambda} \leqslant 0$.
\end{center}
Furthermore, if the congruence has initially a negative expansion (say, $\theta_0<0$ at $\lambda=0$), then it continues to converge and reaches a point where $\theta \rightarrow - \infty$ within a finite value of the affine length  
\begin{center}
$ \lambda \leqslant  \dfrac{2}{|\theta_0|}$. 
\end{center}
\end{itembox}

\vspace{2mm}
%%%%%%%

Suppose the above null geodesic congruence containing $\gamma$ is emanating from a point $p\in \gamma$ at some $ \lambda <0$, attains the negative value $\theta_0<0$ at $\lambda=0$, and as claimed above, reaches a point $q \in \gamma$ where $\theta \rightarrow -\infty$. Then, in general, $q$ is said to be {\em conjugate to} $p$. Then, one can claim that under the null generic condition and NEC, $\gamma$ must possess a pair of conjugate points (see, proposition 9.3.7 Wald~\cite{Wald84}). We can also consider a congruence of null geodesics with tangent vector field $k^\mu$ which are orthogonal to a $2$-dimensional spacelike surface $S$. There exist, in general, two families of such surface orthogonal null geodesics (often referred to as ``ingoing" and ``outgoing" null geodesics). In this case, a point $q$ away from $S$ along $\gamma$ where $\theta \rightarrow -\infty$ is said to be {\em conjugate to} $S$. The occurrence of a conjugate point $q \in \gamma$ to $p \in \gamma$ (or to $S$) is a natural consequence under the attractive nature of gravitation. An important and useful aspect of the occurrence of a conjugate point $q$ is that {\em a null geodesic $\gamma$ fails to remain on the boundary of causal future (or past) of $p$ (or $S$)}. 
We shall discuss this aspect in more detail in the next subsection. 

\medskip 

For a complete timelike geodesic congruence, whose tangent vector field is $u^\mu= \d x^\mu/\d\tau$ with $\tau$ being the proper time, we can obtain (in more straightforward manner) results similar to the null geodesic focusing as summarized below. Since for a timelike geodesics, the Raychaudhuri equation contains the term $R_{\mu \nu}u^\mu u^\nu$, we use, instead of NEC, the SEC as it implies, $R_{\mu \nu}u^\mu u^\nu = 8\pi G (T_{\mu \nu}-(1/2)Tg_{\mu \nu})u^\mu u^\nu \geqslant 0$ via the Einstein equations. 

\begin{itembox}[l]{{Definition 2.4.3 (Timelike generic condition)}}
A spacetime is said to satisfy the timelike generic condition if every timelike geodesic (whose tangent vector field is denoted by $u^\mu$) possesses at least one point where   
\begin{center}
$R_{\mu \alpha \nu \beta} u^\alpha u^\beta \neq 0$. 
\end{center}
\end{itembox}

\medskip 

Consider two nearby freely falling particles with the unit $4$-velocity vector fields $u^\mu = (\partial/\partial \tau)^\mu$, and let $\eta^\mu$ be the deviation vector (Jacobi field) between the two. Then the geodesic deviation equations are 
\bena
u^\alpha \nabla_\alpha (u^\beta \nabla_\beta \eta^\mu) = - (R^\mu{}_{\alpha \nu \beta}u^\alpha u^\beta) \eta^\nu \,.
\eena
Thus, the above generic condition implies that a non-vanishing tidal force acts on every freely falling test particle. 

\medskip 
\begin{itembox}[l]{{Theorem 2.4.4 (Timelike focusing)}}
Under the SEC, any hypersurface-orthogonal timelike geodesic congruence is focusing: 
\begin{center}
$\dfrac{\d \theta}{\d\tau} \leqslant 0$.
\end{center}
Furthermore, if the congruence has initially a negative expansion (say, $\theta_0<0$ at $\tau=0$), then it continues to converge and reaches a point where $\theta \rightarrow - \infty$ within a finite value of the proper time   
\begin{center}
$ \tau \leqslant  \dfrac{3}{|\theta_0|}$. 
\end{center}
\end{itembox}

\medskip 

Similar to the null geodesic case, we can say that a point $q \in \gamma$ is {\em conjugate to} a point $p$ (or to $3$-dimensional hypersurface $\Sigma$) when the congruence is emanating from the point $p\in \gamma$ (or orthogonal to $\Sigma$).   
We can also claim that under the timelike generic condition and the SEC, $\gamma$ must possess a pair of conjugate points. (See, proposition 9.3.1 Wald~\cite{Wald84}). An important aspect of the occurrence of a conjugate point $q$ for timelike geodesics is that {\em a timelike geodesic $\gamma$ fails to be a local maximum of proper time between the initial point $p$ (or $\Sigma$) and any point beyond $q$}~\cite{HE73,Wald84}.

\subsection{Singularity theorems and energy conditions}\label{subsec:sing} 

One of the most important applications of the energy conditions in general relativity is the singularity theorems. In this section, we will discuss how the energy conditions are used in the proof of the theorems. 

\medskip 

A spacetime singularity represents some extreme situation where the general relativistic description of spacetime as a manifold breaks down due to infinite energy density or infinite spacetime curvature. This, in turn, implies that spacetime singularity should not be considered as part of the spacetime manifold. This leads to the idea that if a freely falling observer reaches a singularity, her/his world-line %suddenly ends at a finite proper time and cannot be extended any further 
cannot be continued beyond a certain proper time. 
Such a world-line (i.e., endless but never attains arbitrary large values in its proper time) can be described as an {\em inextendible, incomplete} timelike geodesic curve. 
The singularity theorems employ, as the criterion for the presence of a singularity, the existence of an inextendible incomplete causal (timelike or null) geodesic curve. For further motivation, satisfactory and unsatisfactory aspects of this definition of spacetime singularities, see, e.g., \cite{HE73,Wald84}.  

\medskip 
To explain the statements of the singularity theorems and the basic ideas of their proof, we first have to introduce some notion and definitions on causal structure of spacetime. Let $S$ be any subset of spacetime manifold $M$. The {\em chronological future} of $S$, denoted $I^+(S)$, is defined as the set of all points that can be reached by a future-directed timelike curve from $S$. Similarly, the {\em causal future} of $S$, denoted $J^+(S)$, is  to be the set of all points reached by a future-directed causal curve from $S$, including $S$ itself. The chronological past $I^-(S)$, and causal past $J^-(S)$, are defined in the same manner. In Minkowski spacetime $({\Bbb R}^4, \eta_{\mu \nu})$, for any point $p$, $I^+(p)$ is the interior of the future light-cone of $p$, and $J^+(p)$ is the interior and the boundary of the future light-cone of $p$. In particular, their boundaries, $\partial I^+(p)$, $\partial J^+(p)$, and the {\em future horismos} $E^+(p):=J^+(p)\setminus I^+(p)$ of $p$ all coincide, describing the light-cone of $p$. This is also true in general curved spacetime, at least locally, but not necessarily true in global perspective. For example, $E^+(p)$ may differ from the other two. Such a subtle issue in general curved spacetime can be seen in a simple example of the Minkowski spacetime with a single point removed (see, e.g., Figure 34 in \cite{HE73}).  
% \medskip 
A subset $S$ of $M$ is said to be {\em achronal} if no two points on $S$ can be connected by a timelike curve. Note that a null (or spacelike) geodesic curve is not necessarily achronal in general curved spacetime, even though its tangent vector is always null (or spacelike), never timelike. For example, an inextendible null geodesic in compact universe (e.g., any null geodesics in locally Minkowski spacetime with torus identifications $x \leftrightarrow x+a, \, y \leftrightarrow y+b, \,z \leftrightarrow z+c$) fails to be achronal. Another non-trivial example is a circular null geodesic orbit at the radius $r=3G{\cal M}$ of the Schwarzschild spacetime with mass ${\cal M}$.  
The notion of achronality plays a role in many areas in general relativity, including non-local generalization of the NEC as we will see later (see Sec. 2.7). In the context of the singularity theorems, a crucial result related to the achronality is that \cite{HE73,Wald84}: 

\smallskip 

\begin{itembox}[l]{Proposition 2.5.1} 
Any achronal null geodesic (hence, any achronal null hypersurface) does not admit a pair of conjugate points on it.  
\end{itembox}

\smallskip 

\begin{figure}[h]
\begin{center}
%\scalebox{0.10}{%
\includegraphics[width=70mm]{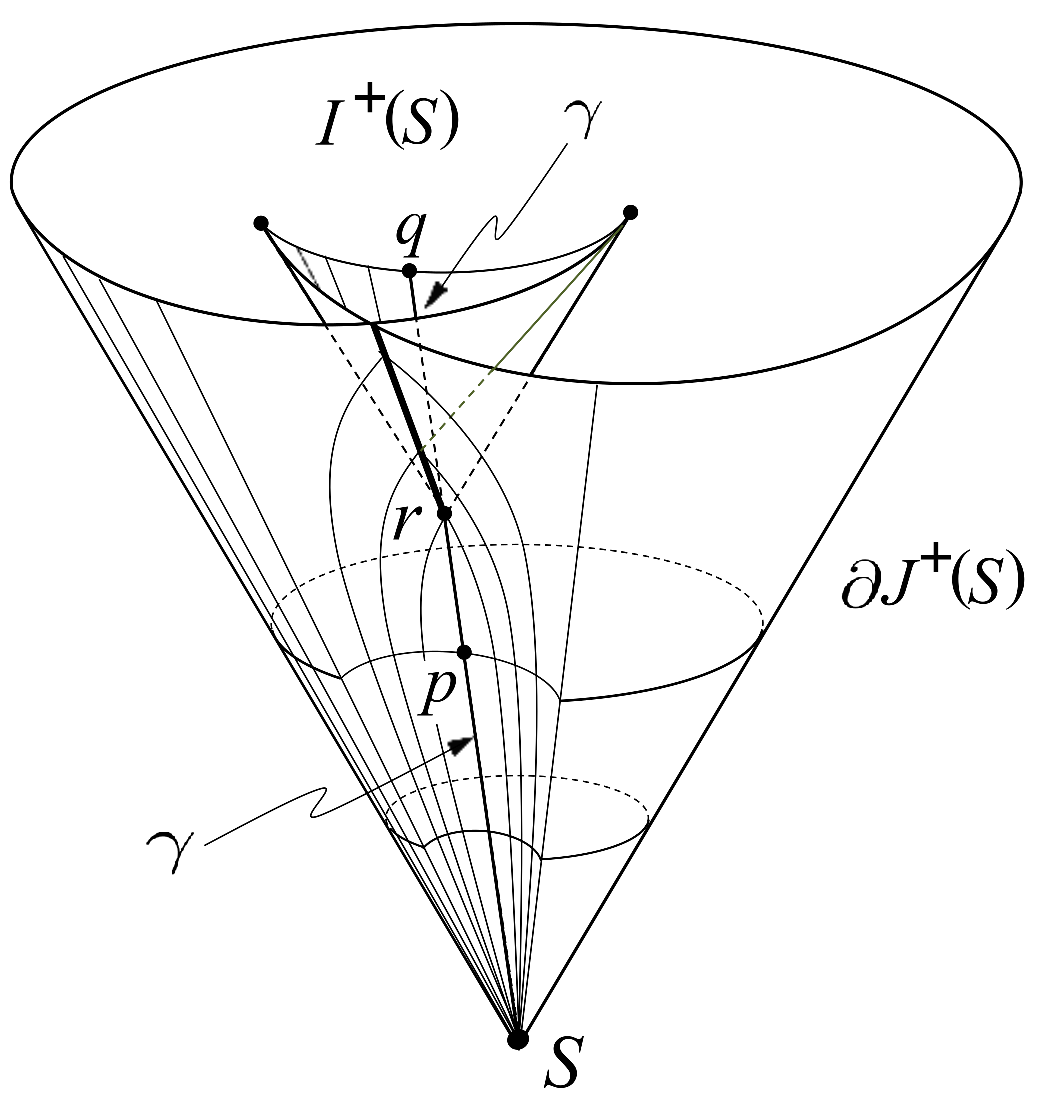}
%              } 
\end{center}
\vspace{-3mm}
\caption{\small 
Null geodesics emanating from a subset $S$ of $M$. One such null geodesic $\gamma$ is passing through $p$, where its expansion changes sign from positive to negative, and then has a conjugate point $r$ in $J^+(p)$. Beyond this conjugate point, $\gamma$ fails to remain on the boundary $\partial J^+(S)$ and enters the interior $I^+(S)$. Since any point $q$ on $\gamma$ within $I^+(S)$ can be connected to $S$ by a timelike curve, $\gamma$ no longer is achronal beyond the conjugate point $r$. See also Figure~\ref{gammaprime} below. 
 } 
\label{achronal}
\end{figure}

Figure~\ref{achronal} illustrates what happens when a conjugate point appears. 

\medskip 

\medskip 

An important example of achronal set is given by, 
\begin{itembox}[l]{Proposition 2.5.2}
The boundary $\partial J^+(S)$ of the causal future of $S$ is a closed, $3$-dimensional, achronal submanifold. 
\end{itembox}
\begin{itemize}
\item Proposition 6.3.1~\cite{HE73} or Theorem 8.1.3 in \cite{Wald84}. 
\end{itemize}

\medskip 
Let $M$ be connected and let $\Sigma$ be any closed achronal set in $M$. The {\em future domain of dependence} of $\Sigma$, denoted $D^+(\Sigma)$, is defined by the set of all points $p \in M$ such that every past-directed inextendible causal curve from $p$ intersects $\Sigma$. The past domain of dependence of $\Sigma$, denoted $D^-(\Sigma)$ is defined analogously. One often considers the union $D(\Sigma):=D^+(\Sigma) \cup D^-(\Sigma)$. In general, when $D(\Sigma)$ is a proper subset of $M$, i.e., $D(\Sigma) \varsubsetneqq M$, the boundary $\partial D^\pm(\Sigma)$ is called the {\em future (past) Cauchy horizon} of $\Sigma$, denoted by $H^\pm (\Sigma)$, respectively. A physical importance of the notion of the domain of dependence is that the physics inside $D(\Sigma)$ can be determined by the initial data on $\Sigma$, without any other information, e.g., boundary conditions outside $D(\Sigma)$. 

\medskip 

If $D(\Sigma)=M$ (i.e., $H^\pm(\Sigma)= \emptyset$), then $\Sigma$ is called a {\em Cauchy surface} of $M$. If $M$ admits a Cauchy surface, then $M$ is said to be {\em globally hyperbolic}. An important role of the global hyperbolicity in the context of the singularity theorems is concerning the length of causal curves. 
For a smooth causal curve $\lambda$ between two points $p,q  \in M$ with tangent $t^\mu=(\partial /\partial t)^\mu$, the length of $\lambda$ is defined as~\cite{HE73,Wald84},  
\bena
 L[\lambda] = \int_p^q \d t \,(-g_{\mu \nu} t^\mu t^\nu)^{1/2} \,.
\eena  
Then, one has
\begin{itembox}[l]{Proposition 2.5.3 (Maximum length curve)}
Let $N$ be globally hyperbolic (possibly as a subset of $M$). Then, for any two points $p, q \in N$ with $q \in J^+(p)$, there exists a causal geodesic curve $\mu$ from $p$ to $q$ which attains the maximum length among all causal curves from $p$ to $q$.    
\end{itembox}
\begin{itemize}
\item Proposition 6.7.1~\cite{HE73}. 
\end{itemize}
This is a consequence of the following three facts: firstly, the set $C(p,q)$ of all causal curves which connect two points $p,q$ in globally hyperbolic region $N$ is compact, secondly, the length $L[\lambda]$ is an upper semicontinuous function on $C(p,q)$, and thirdly, any upper semicontinuous function on a compact space has a maximum value. 
\hfill $\Box$

\medskip 

It is rather easy to see what happens if the assumption of global hyperbolicity is eliminated. 
For example, consider timelikely separated two points $p,q$ in Minkowski spacetime with a single point $r$ just on a straight line segment connecting the two points removed. In this case, the maximum value (in the sense of $d(p,q)$ below) is the length of the segment but there does not exist a causal curve attaining the maximum value. If we consider timelikely separated two points in anti-de Sitter (AdS) spacetime, and in particular, if these two are not lying in a globally hyperbolic subregion, then the maximum value itself does not exist~\cite{Ishibashi:2012xk}.

\medskip 

\begin{figure}[h]
\begin{center}
%\scalebox{0.10}{%
\includegraphics[width=60mm]{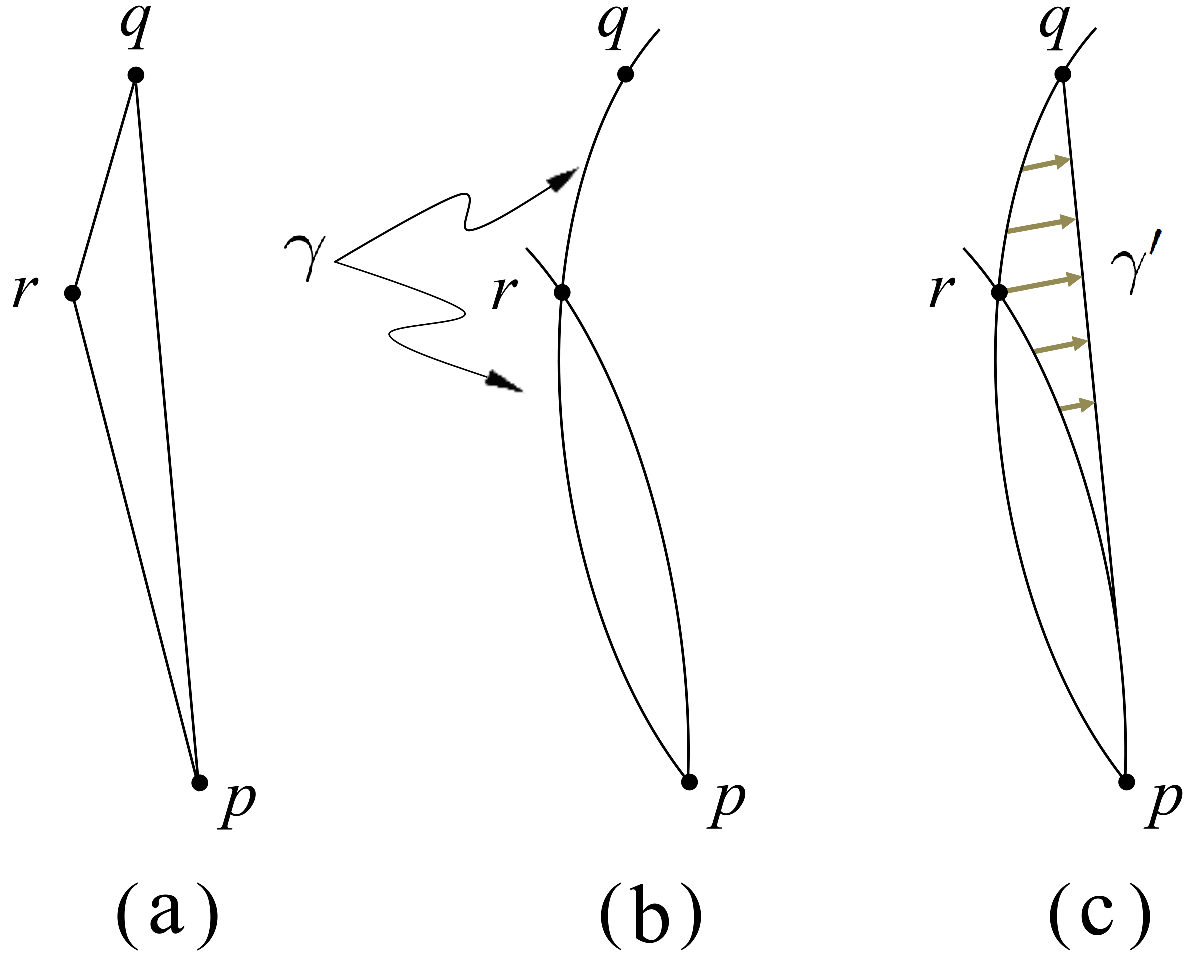} 
%          } 
\end{center}
\vspace{-3mm}
\caption{\small 
(a) Reverse triangle inequality. (b) A timelike geodesic $\gamma$ from $p$ to $q$ admits a point $r$ conjugate to $p$ between $p$ and $q$. (c) A timelike geodesic $\gamma'$ from $p$ to $q$, constructed by deforming $\gamma$ so that $\gamma'$ has a larger value of the proper time than $\gamma$. Compare Figure~\ref{achronal}. 
} 
\label{gammaprime}
\end{figure}

As mentioned briefly just below the focusing theorems in the previous subsection, if a timelike geodesic connecting two points $p,q \in M$ contains a pair of conjugate points between the two, then there exists another causal curve whose length (proper time) is larger than the original timelike geodesic. 
The Lorentzian distance function $d: M \times M \rightarrow {\Bbb R} \cup \{ \infty \}$ between two points $p,q \in M$ is defined by 
$ d(p,q):=\sup\{ L(\lambda) : \lambda \in C(p,q) \} $. If $p,q \in I^+(p), \: p \neq q$, then $d(p,q)>0$, and if $q \in E^+(p)$ or not causally related, then $d(p,q)=0$. For any $q,r \in I^+(p), \: q \in I^+(r)$, 
\bena
   d(p,q) \geqslant d(p,r) + d(r,q) \,. 
\label{rti} 
\eena  
This is the {\em reverse triangle inequality} in Lorentzian distance, which is nothing but the inequality one may see in the so-called ``twin paradox." As can be directly read off from Figure~\ref{gammaprime}, if a timelike geodesic $\gamma$ connecting points $p$ and $q$ contains a point $r$ conjugate to $p$, then there is another timelike geodesic $\gamma'$ that also connects $p$ and $q$ with a longer length (proper time), according to (\ref{rti}).  
\medskip 

Similarly, if a null geodesic connecting two points $p,q \in M$ contains a pair of conjugate points on it, then it cannot be achronal since $p,q$ can be then connected by some timelike curve. 
%In other words, it fails to be the fastest causal curve. 
Combining this observation with Prop. 2.5.3, we have the following: 

\begin{itembox}[l]{Proposition 2.5.4} 
Any causal geodesic curve which connects two points $p$ and $q$, and which attains the maximum length between them cannot have a pair of conjugate points between $p$ and $q$.   
\end{itembox}
\begin{itemize}
\item Proposition 4.5.8~\cite{HE73} and theorem 9.3.3~\cite{Wald84}. 
\end{itemize}

\medskip 

The singularity theorems predict the existence of an inextendible, incomplete causal (either timelike or null) geodesic curve, under the following three conditions concerning:
\begin{itemize}
\item[(i)] causal structure, 
\item[(ii)] strong gravity, 
\item[(iii)] geodesic focusing. 
\end{itemize} 
By strengthening or weakening these three conditions, one can formulate different versions of the singularity theorems, but the proof in all cases essentially proceeds to derive a contradiction, under the assumption that every causal geodesic be complete. 

\medskip

Let us discuss the implications of the above three conditions (i)--(iii) in more detail regarding singularity formation.

\begin{itemize}

\item[(i)] Causal structure.

The first singularity theorem by Penrose~\cite{Penrose:1964wq} assumes the global hyperbolicity, or equivalently the existence of a Cauchy surface, as a condition for global causal structure (Penrose's theorem actually requires a {\em non-compact} Cauchy surface). However, the global hyperbolicity may be considered to be too strong a causality condition. After demonstrating contradictory consequences, instead of claiming that the assumption of the geodesic completeness fails, one may rather be able to conclude that the assumption of the global hyperbolicity must fail. In this respect, it is worth noting that under the same conditions of Penrose's theorem, except the assumption of the global hyperbolicity, one can construct geodesically complete spacetimes, among which most well-known example is the Bardeen's regular black hole~\cite{2767662}.

\medskip 

For this reason, after Penrose's theorem~\cite{Penrose:1964wq}, Hawking and Penrose~\cite{Hawking:1970zqf} proposed the improved version of the singularity theorem, in which the unwanted assumption of the global hyperbolicity is dropped, and instead, the {\em chronology condition}---which demands that no closed timelike curves exist---is imposed. This is a reasonable condition as to consider a physically realistic universe. 

\medskip 

\item[(ii)] Strong gravity. 

This property is described by the requirement of the {\em trapped set} defined as follows: 
\begin{itembox}[l]{{Definition 2.5.5 (Future trapped set)}}
 A closed achronal subset $S$ is said to be a {\em future trapped set} if $E^+(S)$ is compact.  
\end{itembox}
The past trapped set is defined analogously. A simple example is a $2$-dimensional closed surface $S$ in the static Einstein universe $M={\Bbb R}\times S^3$ with the metric, 
\bena
 \d s^2 = -\d T^2 + \d R^2+\sin^2 R\, (\d\vartheta^2+\sin^2\vartheta\, \d\varphi^2) \,, 
\label{metric:SEU}
\eena 
as depicted in Figure~\ref{trpt-cmp-u}. In the cosmological context, a point $p$ whose past light cone starts converging again toward the early universe is a past trapped set (see Figure 50 in \cite{HE73}).  
   
\begin{figure}[h]
\begin{center}
%\scalebox{0.10}{%
\includegraphics[width=80mm]{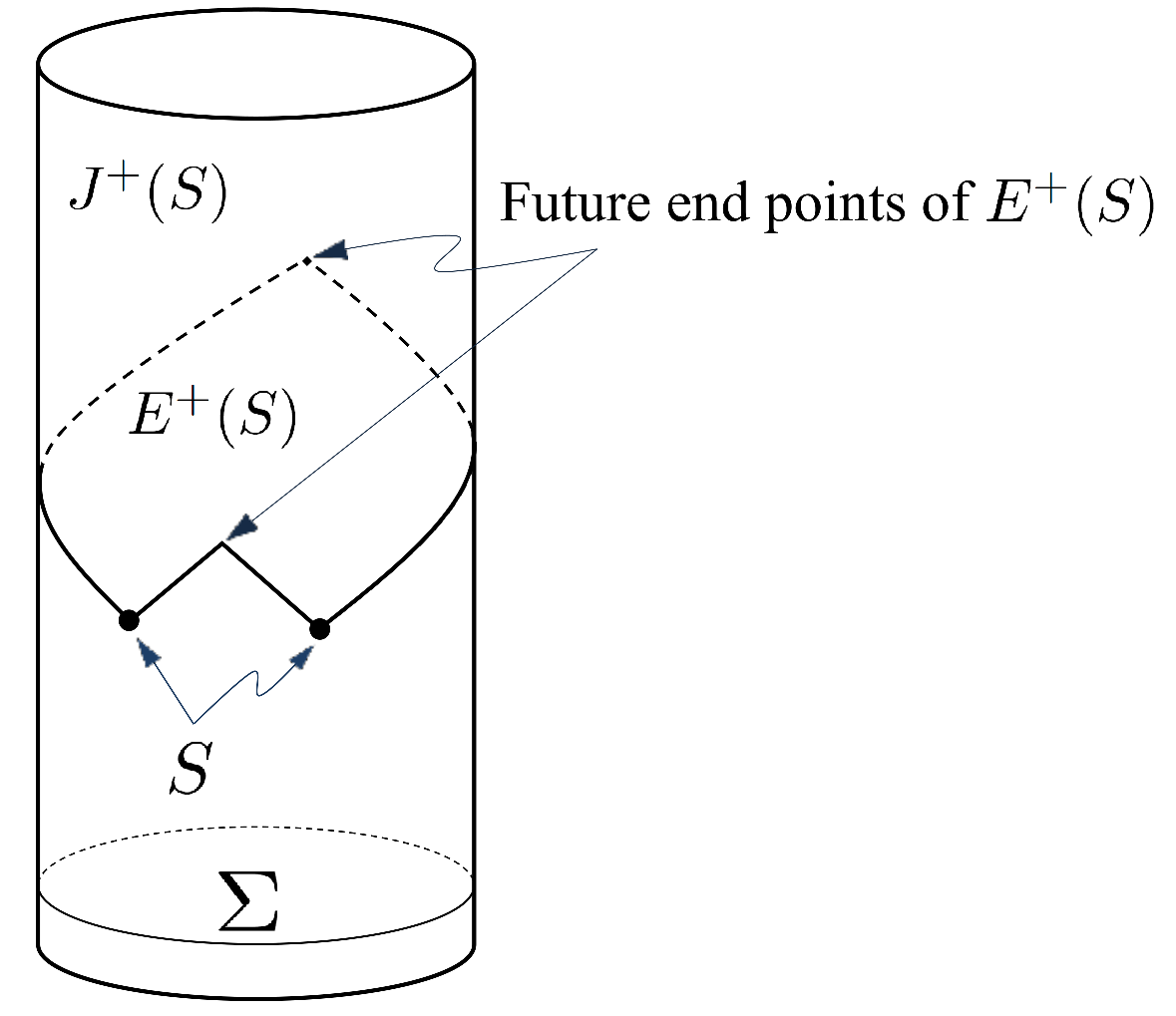}
%}
\end{center}
% \vspace{-3mm}
\caption{\small 
An achronal subset $S$ in spatially compact universe with $\Sigma$ being a compact Cauchy surface.
The null geodesics (thick curves) generating the future {\em horismos} $E^+(S)$ have the two future endpoints due to the compactness of the universe, and accordingly $E^+(S)$ also is compact, thus $S$ is a trapped set. One may think of $S$ as any $2$-sphere of constant $T$ and $R$ in the static Einstein universe. (The two points labeled as $S$ in the figure represent the single $2$-sphere $S$.) $E^+(S)$ has the two future endpoints; one for the ``ingoing" null geodesics, and the other for the ``outgoing" ones, which are orthogonal to $S$. It is important to note, however, that in a closed universe, one cannot generally distinguish between ingoing and outgoing. 
Note also that in the static Einstein universe, every Cauchy surface $\Sigma$ itself is in fact a trapped set since $E^+(\Sigma)=\Sigma$ is compact. 
} 
\label{trpt-cmp-u}
\end{figure}

\medskip 

In the context of gravitational collapse and black hole physics, the most important example of a trapped set is the {\em closed trapped surface}, which is a compact, orientable, and spacelike $2$-dimensional surface ${\cal T}$ such that the expansions $\theta_\pm$ of the two families of future (or past) null geodesic congruence orthogonal to ${\cal T}$ both are negative. An example is any $2$-sphere of constant area radius inside the Schwarzschild (horizon) radius $r_h:=2G{\cal M}$. 
To be concrete, let us express the Schwarzschild metric\footnote{ %%% 
In the standard coordinates $x^\mu=(t,r,\vartheta, \varphi)$, the Schwarzschild metric is given as 
\bena
\d s^2=-f(r)\, \d t^2+f^{-1}(r)\,\d r^2+r^2(\d\vartheta^2 +\sin^2\vartheta  \,\d\varphi^2) \,, 
\eena
where $f(r):=1-2G{\cal M}/r$, ${\cal M}=const.$. In one of the exterior region, $U<0, \:V>0$, the two coordinates are related as $U=-e^{-(t-r_*)/2r_h}, \: V=e^{(t+r_*)/2r_h}$, where $r_*= r+r_h \log (r-r_h)$.  
} %%%
in the Kruskal coordinates $\bar{x}^\mu=(U,V,\vartheta, \varphi)$, 
\bena
 \d s^2= - \dfrac{4r_h^3e^{-r/r_h}}{r}\d U \d V + r^2(\d \vartheta^2 + \sin^2 \vartheta\, \d\varphi^2) \,, 
\label{metric:Kruskal}
\eena
where the area radius $r$ is now regarded as the function of $U$ and $V$. This coordinate system covers the entire (maximally extended) Schwarzschild spacetime (see, e.g., Sec. 6.4 of Wald's book~\cite{Wald84}). In this Kruskal coordinate system, the center $r=0$ of the spherical symmetry is at $UV=1$, the horizon (i.e., $r=r_h$) is at $UV=0$, the interior region ($0<r<r_h$) consists of the black hole region $U>0, \; V>0$ and the white hole region $U<0, \; V<0$, while the exterior region ($r>r_h$) consists of two causally disjoint regions: $U<0, \: V>0$ and $U>0, \:V<0$ (see Figure~\ref{Kruskal}). 
The expansions $\theta_\pm$ of the two families (out-going and in-going) of future directed null geodesic congruences orthogonal to any $2$-sphere of constant $U$ and $V$ are calculated as 
\bena
 \theta_+ = - \dfrac{U}{r_h r} \,, \quad \theta_- = - \dfrac{V}{r_h r} \,.
\label{exp:Schwa}
\eena

\begin{figure}[h]
\begin{center}
%\scalebox{0.30}{%
\includegraphics[width=80mm]{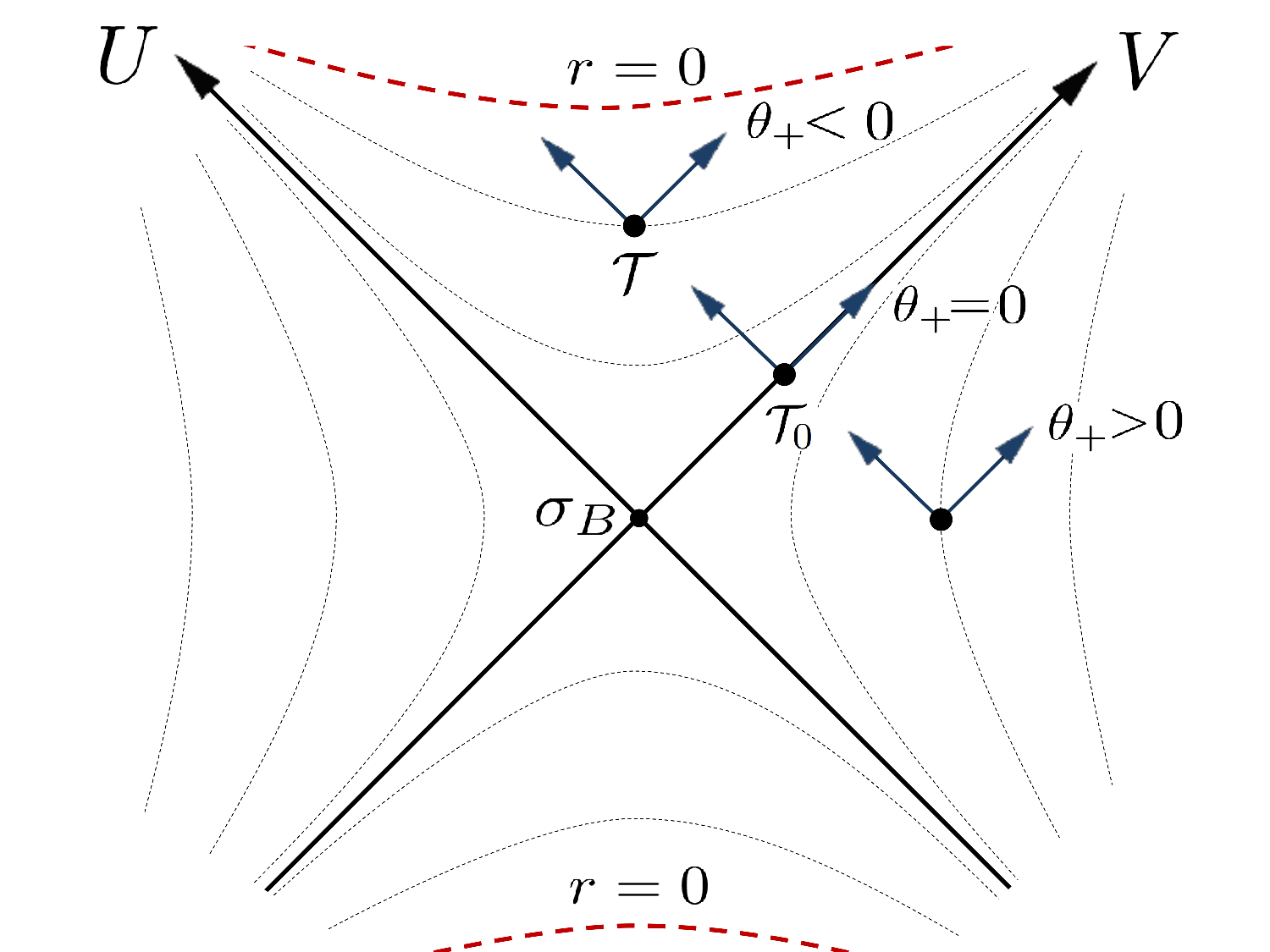} 
%              } 
\end{center}

\vspace{3mm}
\caption{\small 
The maximally extended Schwarzschild spacetime in Kruskal coordinates $(U,V)$. The future event horizon is at $U=0, V>0$ (and $V=0, U>0$), the $2$-sphere $\sigma_B$ at $U=0=V$ is the bifurcate surface. Any $2$-sphere ${\cal T}$ inside the black hole region, $U>0,V>0$, is a future trapped surface, and any $2$-sphere ${\cal T}_0$ on the future event horizon $U=0, V>0$ is a MOTS.   
} 
\label{Kruskal}
\end{figure}
Inside the black hole region $U>0, V>0$, both $\theta_\pm <0$ on ${\cal T}$, and therefore ${\cal T}$ is indeed a closed trapped surface. If every null geodesic were to be complete, then it would follow from the null focusing theorem 2.4.2 (with the NEC obviously holding in vacuum spacetime) that every future-directed null geodesic orthogonally emanating from ${\cal T}$ must have an endpoint (conjugate point to ${\cal T}$), making $E^+({\cal T})$ compact, hence ${\cal T}$ is a trapped set.  
In the present case, the actual ``endpoints" of $E^+({\cal T})$, where $\theta_\pm \rightarrow -\infty$, are located at ``$r=0$" as explicit in (\ref{exp:Schwa}), and thus not part of the spacetime manifold but correspond to the central singularity. Therefore null geodesics along $E^+({\cal T})$ are {\em incomplete}.   
Note also that for any $2$-sphere ${\cal T}_0$ on the future event horizon, say $U=0, V>0$, for the in-going null $\theta_-<0$, while for the out-going null  $\theta_+=0$ and its outward derivative $\partial_r \theta_+ \geqslant 0$. Such a closed surface ${\cal T}_0$ characterizing locally the outer boundary of trapped region is called the {\em stably marginally outer trapped surface} (MOTS)~\cite{Andersson:2005gq} (see, Sec.~\ref{subsec:top} below) or more loosely the {\em apparent horizon}.

\medskip 

\item[(iii)] Geodesic focusing. 

The null and timelike focusing theorems (Theorems 2.4.2 and 2.4.4) are a direct result of the NEC and SEC, and the Raychaudhuri equations as we have seen in the previous section. As discussed before, the null and timelike focusing theorems ensures that any complete causal geodesics inevitably contain a pair of conjugate points. 

\end{itemize} 

Let us state Penrose's version of the singularity theorem~\cite{Penrose:1964wq}.  
\begin{itembox}[l]{{Theorem 2.5.6 (Penrose 1965)}} 
Spacetime $(M,g_{\mu \nu})$ cannot be null geodesically complete if: 
 \begin{itemize}
\item[(i)] there is non-compact Cauchy surface $\Sigma$ in $M$,
\item[(ii)] there is a closed trapped surface ${\cal T}$ in $M$,   
\item[(iii)] the NEC holds.  
\end{itemize} 
\end{itembox} 

\medskip 

The proof goes roughly as follows. Suppose $M$ is null geodesically complete. From (ii) and (iii) (thus null focusing theorem), it follows that $E^+({\cal T})$ is compact. Furthermore, since $M$ is globally hyperbolic by assumption (i), $E^+({\cal T}) =\partial J^+({\cal T})$. It immediately follows from Prop. 2.5.2 that $E^+({\cal T})$ is a compact $3$-dimensional achronal, topological manifold. 
Again, from the global hyperbolicity of $M$ and the achronality of $E^+({\cal T})$, one can find a continuous one-to-one map $\psi$ from $E^+({\cal T})$ to $\Sigma$. Then, its image $\psi(E^+({\cal T}))$ must have a boundary in $\Sigma$ since $E^+({\cal T})$ is compact, whereas $\Sigma$ is non-compact. However, this contradicts the fact that $E^+({\cal T})$ is a compact $3$-dimensional achronal topological manifold without boundary. 
\hfill $\Box$

\medskip 

Next, consider Hawking and Penrose's improved version of the theorem~\cite{Hawking:1970zqf}, in which the assumption of the global hyperbolicity (i.e., the existence of a Cauchy surface) is eliminated: 
\begin{itembox}[l]{{Theorem 2.5.7 (Hawking-Penrose 1970)}} 
The following three conditions cannot all hold: 
\begin{itemize}
\item[(i)] the chronology condition holds,   
\item[(ii)] there is a trapped set ${\cal S}$, 
\item[(iii)] every inextendible causal geodesic contains a pair of conjugate points.  
\end{itemize} 
\end{itembox} 
\begin{itemize}
\item This is an alternative expression~\cite{HE73} of the Hawking-Penrose theorem~\cite{Hawking:1970zqf}. 
\item Condition (iii) results from the NEC and the SEC, the generic conditions 2.4.1 and 2.4.3, the focusing theorems 2.4.2 and 2.4.4, and the hypothetical assumption of causal geodesic completeness.   
\end{itemize}

The proof goes roughly as follows. If the above three conditions all hold, one can construct a globally hyperbolic subregion $N$ which contains inextendible timelike curves\footnote{%%%
One can always find a future inextendible timelike curve $\gamma$ in $D^+(E^+({\cal S}))$, since $E^+({\cal S})$ is compact as ${\cal S}$ is a trapped set, whereas $H^+:=\partial D^+(E^+({\cal S}))$ is non-compact under (i) and (iii) (Lemma 8.2.1~\cite{HE73}). This is essentially because if $H^+$ were to be compact, it would have to contain inextendible null geodesic generators, which however must contain a pair of conjugate points due to condition~(iii), thus contradicting the achronality of $H^+$ (Prop. 2.5.1). Consider now the compact set $F:=E^+({\cal S}) \cap \overline{J^-(\gamma)}$. 
Since $F$ turns out to be past trapped, $E^-(F)$ is compact. Hence, in a similar manner to $\gamma$, one can show that $D^-(E^-(F))$ contains a past inextendible timelike curve $\lambda$. Then, one can construct the desired globally hyperbolic subregion as $N:=D(E^-(F))$, which, by construction, contains both future and past inextendible timelike curves (constructed by suitably deforming and connecting $\gamma$ and $\lambda$ if necessary).
}. %%%
Among them, there must exist an inextendible timelike geodesic curve $\mu$ which attains the maximum length between any pair of two points on it, according to Prop. 2.5.3. Then, as such, $\mu$ cannot admit a pair of conjugate points on it according to Prop. 2.5.4. However, this contradicts (iii). 
\hfill $\Box$

\subsection{Black holes and energy conditions}\label{subsec:bh}

A black hole obeys the laws of mechanics, which correspond precisely to the ordinary laws of thermodynamics. 
This correspondence was first established within classical general relativity as a mathematical resemblance~\cite{Bardeen:1973gs}, and was soon promoted to a perfect correspondence by the discovery of Hawking radiation~\cite{Hawking:1975vcx}. 
This implies that a stationary black hole can be completely characterized by only a few parameters, just like a thermodynamic equilibrium system can be described by a few macroscopic variables. This property is established as the black hole uniqueness theorem, which states that {\em any asymptotically flat, stationary rotating black hole that is regular everywhere on and outside the event horizon, and that solves the electrovacuum Einstein equations, must be described by the Kerr-Newmann metric}. In this section, we see the key roles the energy conditions (WEC, DEC, SEC, and NEC) play in the proof of the classical black hole mechanics. 

\medskip 

In order to define a black hole as to be the region from where even light cannot escape, it is most useful to introduce the notion of {\em asymptotic flatness at null infinity} ${\mathscr I}$. 
This can be elegantly done in terms of the {\em conformal completion}\footnote{%%%
A canonical example is Minkowski spacetime itself $({\Bbb R}^4, \eta_{\mu \nu})$, which can be conformally embedded into the subregion, $-\pi < T \pm R < \pi, \: R \geqslant 0$, of the static Einstein universe ${\tilde M} \simeq {\Bbb R}\times S^3$ with the metric ${\tilde g}_{\mu \nu}$ given by (\ref{metric:SEU}). 
In fact, the two metrices are related as ${\tilde g}_{\mu \nu}=\Omega^2\eta_{\mu \nu}$ with $\Omega:=\cos T+ \sin R$. 
The null boundary $R+T=\pi $ is callled the {\em future null infinity}, denoted $\mathscr{I}^+$, which corresponds to $t, r \rightarrow +\infty$ with $t-r=const$ of Minkowski spacetime, and the null boundary $R-T=\pi$ is called the {\em past null infinity}, denoted $\mathscr{I}^-$, which does to $r \rightarrow +\infty: t \rightarrow - \infty$ with $t+r=const$. At null infinity $\mathscr{I}^\pm$, $\Omega = 0$ but $\d\Omega \neq 0$. 
}. %%%
A spacetime $(M,g_{\mu \nu})$ is said to be asymptotically flat at null infinity if there is a conformal isometry $\psi$ which maps $M$ into a larger spacetime $\tilde{M}$ with the metric ${\tilde g}_{\mu \nu} := \Omega^2 \psi_* g_{\mu \nu}$ with a smooth positive function $\Omega$ in $\psi(M)$ so that $\psi(M)$ has two (causally disjoint) null boundaries $\mathscr{I}^\pm :=\partial \psi(M) \cap J^\pm(\psi(M), {\tilde M})$, on which $\Omega =0$, $\d\Omega \neq 0$, ${\tilde g}^{\mu \nu} (\d\Omega)_\mu (\d\Omega)_\nu =0$, and the energy-momentum tensor for matter fields is assumed to decay toward null infinity $\mathscr{I}^\pm$ sufficiently rapidly (for more detail, see, e.g., \cite{Wald84}). The neighborhood of future null infinity $\mathscr{I}^+$ may be viewed as the ``region-to-escape" or ``ideal-distant-observers."
Now, having defined the future null infinity $\mathscr{I}^+$, one can define the black hole region $B$ as the complement of the causal past of the future null infinity, 
\bena
 B= M \setminus J^-(\mathscr{I}^+) \,.
\label{def:B}
\eena
The (future) {event horizon} ${\cal H}^+$ is defined as ${\cal H}^+:=\partial B=\partial J^-(\mathscr{I}^+)$. 
In what follows we further assume the {\em strong asymptotic predictability} that the closure of $M \cap J^-(\mathscr{I}^+)$ is contained in a globally hyperbolic subregion ${\tilde V}$ of ${\tilde M}$~\cite{Wald84}, so that the exterior region of $B$ does not contain causal pathology, such as a singularity, and any physics in the exterior region---including ${\cal H}^+$ and $\mathscr{I}^+$---can be determined from some Cauchy data for the subregion ${\tilde V}$.

\subsubsection{Black hole mechanics and energy conditions}

From Propositions 2.5.1 and 2.5.2, it follows that ${\cal H}^+$ is a closed, $3$-dimensional, achronal hypersurface whose null geodesic generators admit neither a pair of conjugate points nor a future endpoint in $M$. The definition of a black hole region, (\ref{def:B}), may not appear to capture the physical aspects of a black hole. However, by employing the energy conditions, we can demonstrate that $B$ certainly describes a strong gravity region, which contains a closed trapped surface.  

\smallskip 
\begin{itembox}[l]{{Proposition~2.6.1}} 
If the NEC holds, then a closed trapped surface ${\cal T}$, if it exists, cannot intersect $J^-(\mathscr{I}^+)$. 
\end{itembox} 
\begin{itemize}
\item Proposition 9.2.1~\cite{HE73}. 
\item This implies that a closed trapped surface ${\cal T}$ cannot be seen from an ideal distant observer $\mathscr{I}^+$, and therefore indicates that ${\cal T}$ should be hidden inside the black hole region $B$.  
\item Under the DEC, Schoen-Yau (1983)~\cite{Schoen:1983tiu} showed that if a sufficiently large amount of matter is condensed in a small enough region, a closed trapped surface ${\cal T}$ must occur. This, combined together with the singularity theorems and the above proposition, indicates that complete gravitational collapse results in the formation of a black hole and a singularity inside.   
\end{itemize}

\medskip 

The proof goes roughly as follows. To derive a contradiction, suppose ${\cal T} \cap J^-(\mathscr{I}^+) \neq \emptyset$. Then, a null geodesic generator $\mu$ of $\partial J^+({\cal T})$ has an intersection with $\mathscr{I}^+$ in $\tilde M$. This implies on one hand that $\mu$ is future complete, and on the other hand that $\mu$ has the past endpoint at ${\cal T}$. Then, a null geodesic congruence of $\mu$ orthogonal to ${\cal T}$ must have a negative expansion $\theta_+<0$ at ${\cal T}$, since ${\cal T}$ is a closed trapped surface. Then, by Theorem 2.4.2, $\mu$ must have a conjugate point to ${\cal T}$ in the future of ${\cal T}$. This, however, contradicts the fact that $\mu$ is the generator of the achronal hypersurface $\partial J^+({\cal T})$ (Proposition 2.5.1). 
\hfill $\Box$

\medskip 

Applying a similar type of argument to the event horizon itself, one obtains the following result: 

\smallskip 
\begin{itembox}[l]{{Theorem 2.6.2 (Area theorem: The second law)}} 
Under the NEC, the cross-sectional area of the event horizon never decreases toward future.  
\end{itembox} 
\begin{itemize}
\item Proposition 9.2.1~\cite{HE73}, Theorem 12.2.6~\cite{Wald84}. 
\item One-quarter of the horizon cross-sectional area can be interpreted as the {\em entropy} of the black hole~\cite{Bekenstein:1973ur} (see Sec~\ref{sec:4} and Formula 4.3.1), and accordingly the above property of the horizon area increase is phsyically interpreted as the {\em second law of the black hole thermodynamics}. 
\end{itemize}

\medskip 

The proof goes roughly as follows. Suppose that at some point $p \in {\cal H}^+$, the future directed null geodesic generator $\mu$ of ${\cal H}^+$ passing through $p$ has a negative expansion, $\theta_0<0$. Then, if $\mu$ is complete, it must contain a pair of conjugate points on it by the null focusing theorem 2.4.2. This contradicts Prop. 2.5.1, hence $\theta_0 \geqslant 0$. Now $\mu$ may not be complete. In that case, let us consider a cross-section $\sigma$ of ${\cal H}^+$, which includes $p$. By slightly deforming $\sigma $ outward in a neighborhood of $p$, we can construct a deformed $2$-surface $\sigma'$ in such a way that the outward null geodesics orthogonal to $\sigma'$ still has a negative expansion. However, by applying the same argument in the proof of Prop. 2.6.1, such a surface $\sigma'$ with negative expansion is not allowed to intersect $J^-(\mathscr{I}^+)$. Therefore, we can conclude that the expansion $\theta$ of any generators of ${\cal H}^+$ must be non-negative, $\theta \geqslant 0$. This immediately implies that the cross-section area of ${\cal H}^+$ must increase at least locally or be constant. By the assumption of the strong asymptotic predictability, this local increase of the horizon cross-section area toward the future direction never be terminated due to, e.g., some of the null generators hitting a (would-be) singularity on ${\cal H}^+$. 
\hfill $\Box$

\medskip

Next we consider stationary black holes, since at a sufficiently late time, any black hole formed by gravitational collapse is expected to settle down to a stationary black hole spacetime, which corresponds to an equilibrium thermodynamic system. 

\medskip 

An asymptotically flat spacetime is said to be {\em stationary}, if there is a Killing vector field, $t^\mu$ whose orbits are complete (i.e., $t^\mu$ generates an isometry group) and timelike, at least near infinity $\mathscr{I}$. 
Similarly, a spacetime is said to be {\em axisymmetric}, if there exists a Killing vector field $\varphi^\mu$ whose orbits are closed, spacelike curves. We also assume that the two Killing vector fields $t^\mu$ and $\varphi^\mu$ commute.   
An important consequence is that:  
\smallskip 
\begin{itembox}[l]{{Theorem 2.6.3 (Rigidity theorem)}} 
If an asymptotically flat, analytic, stationary spacetime solving the electrovacuum Einstein equations contains a black hole, then the event horizon of the black hole must be a Killing horizon. Furthermore, if a stationary black hole is rotating, it must also be axisymmetric. 
\end{itembox} 
\begin{itemize}
\item Proposition 9.3.6~\cite{HE73}. See also \cite{Hollands:2006rj,Hollands:2008wn} for more general analyses. 
\item A null hypersurface ${\cal N}$ is said to be a {\em Killing horizon}, if there exists a Killing vector field $\chi^\mu$ which is normal to ${\cal N}$ on ${\cal N}$. 
\item Since $t^\mu$ generates an isometry, it must be tangent to ${\cal H}^+$, hence it is either null (and normal to ${\cal H}^+$) or spacelike on ${\cal H}^+$. If $t^\mu$ is null on ${\cal H}^+$, then $\chi^\mu=t^\mu$. If $t^\mu$ is spacelike, then $t^\mu \neq \chi^\mu$ and an appropriate linear combination of these two provides the additional Killing vector field $\varphi^\mu$ whose orbits are closed, thus expressing axisymmetry. 
\item For an asymptotically flat, stationary axisymmetric black hole, we normalize $t^\mu$ such that $t^\mu t_\mu \rightarrow -1$ at infinity and 
$\varphi^\mu$ so that its closed orbits are $2\pi$-periodic.  
\item The first step of the proof is to show that for the null geodesic generators of the event horizon ${\cal H}^+$, $\theta=0$ and $\sigma_{\mu \nu}=0$, and hence all cross-sections of ${\cal H}^+$ are isometric. For this purpose, the NEC plays a role (see Proposition~9.3.1 of~\cite{HE73}).

\end{itemize} 

\medskip 

From the rigidity theorem 2.6.3, the horizon normal Killing vector field $\chi^\mu$ is related to the stationarity $t^\mu$ and axisymmetry $\varphi^\mu$ as   
\bena
 \chi^\mu = t^\mu + \Omega_H \varphi^\mu \,,
\eena
where the constant $\Omega_H$ is the angular velocity of the horizon with respect to the asymptotic stationary observers along the orbits of $t^\mu$. The parameter $v$, satisfying $\chi^\mu \nabla_\mu v=1$, is called the Killing parameter of $\chi^\mu$, which is in general different from the affine parameter $\lambda$ along the null geodesic generators of the Killing horizon.  

\medskip 

For any Killing horizon with the associated Killing vector field $\chi^\mu$, we can define a {\em surface gravity} $\kappa$ as
\bena
 \kappa^2 = - \dfrac{1}{2}(\nabla^\mu \chi^\nu) \nabla_\mu \chi_\nu\,.
\eena 
Although the surface gravity $\kappa$, defined as above, is a local notion, the following result holds: 
\smallskip 
\begin{itembox}[l]{{Theorem 2.6.4 (The zero-th law)}} 
The surface gravity $\kappa$ of a stationary black hole must be a constant over its event horizon. 
\end{itembox} 
\begin{itemize}
\item Bardeen-Carter-Hawking (1973) \cite{Bardeen:1973gs} with DEC. 
\item As will be discussed in Sec.~\ref{sec:4}, the surface gravity $\kappa$ of a stationary black hole can be interpreted as the {\em temperature} of the black hole. 
\end{itemize} 

\medskip 

If one drops a small amount of matter into a stationary, axisymmetric black hole, then its mass and angular momentum are expected to change, eventually settling down to a new stationary, axisymmetric state. Let us consider this process in a perturbative framework. Suppose a small amount of energy momentum, represented by $T_{\mu \nu}$, is ``thrown" into a stationary black hole. As a result of $T_{\mu \nu}$ crossing the event horizon ${\cal H}^+$, the mass and angular momentum of the black hole will slightly change by $\delta M$ and $\delta J$. 
These changes can be evaluated as follows:
\bena
 \delta M &=& \oint_\sigma \d S \int_{0}^{\infty} \!\! \!\! \!\! \d\lambda \; T_{\mu \nu}k^\mu t^\nu \,, 
\non \\ 
 \delta J &=& - \oint_\sigma \d S \int_{0}^{\infty} \!\! \!\! \!\! \d\lambda \; T_{\mu \nu}k^\mu \varphi^\nu \,, 
\label{delta:M:J} 
\eena
where $\sigma$ denotes a horizon cross-section with the area element $\d S$, and $k^\mu=\d x^\mu/\d\lambda$ the tangent to the horizon null geodesic generator with $\lambda$ being an affine parameter so that $k^\mu = \chi^\mu \d v/\d\lambda= \chi^\mu/(\kappa \lambda) = (t^\mu + \Omega_H \varphi^\mu)/(\kappa \lambda)$. 
At this stage, we regard $T_{\mu \nu}$ as a small perturbation, which generates non-vanishing expansion and shear via the Raychaudhuri equation. In the linear order of perturbation, neglecting the quadratic terms of the expansion and shear, we can write the Raychaudhuri equation for the horizon null geodesic generators as 
\bena
 \dfrac{\d\theta}{\d\lambda} &\simeq& - 8\pi G T_{\mu \nu} k^\mu k^\nu 
\non \\
 &=&  - \dfrac{8\pi G}{\kappa \lambda} T_{\mu \nu} (t^\mu +\Omega_H \varphi^\mu) k^\nu\,. 
\label{Ray:1st}
\eena
Then, by integrating the Raychaudhuri equation along the horizon generators\footnote{%%%
By multiplying 
$\kappa \lambda$ on both side of (\ref{Ray:1st}), and taking integration over $\sigma$ and along the null generator, 
the right-hand side immediately yields 
$- 8\pi G (\delta M - \Omega_H\delta J)$, 
while the left-hand side gives 
\bena
 \oint_\sigma \d S\int_0^\infty \!\! \!\! \!\! \d\lambda\, (\kappa \lambda)\, \dfrac{\d\theta}{\d\lambda} 
 &=&  \kappa \oint_\sigma \d S\int_0^\infty \!\! \!\! \!\!\d\lambda \left[ \dfrac{\d(\lambda \theta)}{\d\lambda}-\theta \right]
\non \\
 &=& - \kappa \oint_\sigma \d S\int_0^\infty \!\! \!\! \!\! \d\lambda\, \theta 
\non \\ 
 &=& - \kappa \delta A \,, 
\eena
where we have used the fact that before and sufficiently after one has dropped matter fields into the horizon, $\theta=0$ as a stationary state. 
}, %%% 
as well as horizon cross-section $\sigma$, we have 
\smallskip 
\begin{itembox}[l]{{Theorem 2.6.5 (The first law)}} 
\bena
 \dfrac{\kappa }{8\pi G} \delta A= \delta M - \Omega_H \delta J 
\label{bh-law:1st}
\eena
\end{itembox} 
\begin{itemize} 
\item Bardeen-Carter-Hawking (1973) \cite{Bardeen:1973gs}. 
\item In general relativity $M$ is equivalent to energy. As mentioned before, $\kappa$ and $A$ are proportional to the black hole's temperature and entropy, respectively. The formula can then be interpreted as expressing the first law of black hole thermodynamics, with $\Omega_H \delta J$ viewed as the work term. 

\end{itemize} 

\medskip 

If the DEC holds, the integrand of (\ref{delta:M:J}) satisfies $T_{\mu \nu}k^\mu t^\nu \geqslant 0$, and the change of the black hole mass $\delta M$ is non-negative ($\delta M \geqslant 0$). This makes sense since the DEC implies the positivity of energy density and at the same time, it prohibits superluminal energy flow. This implies that matter with positive energy density can be swallowed by the black hole but never be able to escape from the event horizon. On the other hand, we have seen in the area theorem 2.6.2 that under the NEC, if we throw a positive energy density into the horizon, $\delta A > 0$. Then, from the first law (\ref{bh-law:1st}), we have,  
\bena
 \delta M > \Omega_H \delta J \,.
\label{compare:delta:M:J} 
\eena

\medskip 

As a concrete example, let us consider the Kerr metric, which describes a stationary, rotating vacuum black hole with the two parameters $(M, J)$, whose surface gravity is given by,   
\bena
 \kappa = \dfrac{1}{2GM}\dfrac{\sqrt{1-(J/GM^2)^2}}{\left(1 + \sqrt{1-(J/GM^2)^2} \right)} \,.  
\eena 
According to Theorem 2.6.4, $\kappa$ corresponds to the temperature, and the absolute zero-temerature (extreme black hole) state could be achieved in the limit 
\bena
J \rightarrow GM^2 \,.
\label{lim:extremeKerr}
\eena 
Now let us attempt to attain this zero-temperature state by throwing a particle into a Kerr black hole of nearly zero temperature. Since near $\kappa=0$, the horizon angular velocity is $\Omega_H \simeq 1/2GM$, we have from (\ref{compare:delta:M:J}), 
\bena
 \delta (G M^2) > \delta J \,.
\eena
Thus, the change of the mass is always larger than that of the angular momentum, indicating that the extremal limit (\ref{lim:extremeKerr}) can never be achieved by this process. This implies the third law of black hole thermodyamics to hold under the NEC and DEC. For more elaborate analysis and discussion, see, e.g., \cite{Israel:1986gqz,Wald:1974hkz}. 

\medskip 

\subsubsection{Topology and energy conditions}\label{subsec:top}

Let us discuss the constraints that the energy conditions can place on the possible topology of spacetime. We first discuss the topology of black hole horizon and then the topology of the exterior region of an asymptotically flat black hole spacetime. 

\medskip 
To discuss horizon topology, we first recall the notion of stably marginally outer trapped surface (MOTS), which we have introduced in the previous section, as any $2$-sphere on the future event horizon of the Schwarzschild black hole. To define the notion of MOTS in more general setting, consider a closed, connected $2$-surface ${\cal T}_0$. Let $k^\mu$ be the tangent vector field for an out-going\footnote{
In general, the notion of ``out-going" or ``outward" direction is not uniquely given. For an asymptotically flat spacetime with a single asymptotic region, if ${\cal T}_0$ separates a Cauchy surface $\Sigma$ into two disconnected parts $\Sigma_{\rm in}$ and $\Sigma_{\rm out}$, then ``out-going" direction is defined as the one, say $\Sigma_{\rm out}$, that contains the asymptotic region.    
} 
null geodesic congruence orthogonally emanating from ${\cal T}_0$, and $\theta_+$ be the associated expansion.  
Let $\ell^\mu$ be a past-directed, outward null vector field on ${\cal T}_0$. If $\theta_+ =0$ and $\ell^\nu \nabla_\nu \theta_+ \geqslant 0$ on ${\cal T}_0$, then ${\cal T}_0$ is said to be a MOTS. The following theorem determines the topology of MOTS. 

\smallskip 

\begin{itembox}[l]{{Theorem 2.6.6 (Topology of horizon cross-section)}} 
Under the DEC, the horizon cross-section ${\cal T}_0$ of a stationary black hole must be topologically $2$-sphere.  
\end{itembox} 
\begin{itemize}
\item Proposition 9.3.2~\cite{HE73}. 
\end{itemize}

\medskip 

The proof consists of two parts. The first part is to show that any horizon cross-section of a stationary black hole is a MOTS, which is essentially the same as the first step of the proof of the rigidity theorem 2.6.3 (Proposition~9.3.1 of~\cite{HE73}). 
The second part is to show that MOTS must be a topologically $2$-sphere under DEC. Let us consider a horizon cross-section ${\cal T}_0$, which is assumed to be compact $2$-dimensional surface. By using the Einstein equations and the properties of the MOTS, $\theta_+ =0$ and $\ell^\nu \nabla_\nu \theta_+ \geqslant 0$, one can find the following inequality:
\bena
\oint_{{\cal T}_0} \!\! \d S \, {\cal R} \geqslant 16 \pi G \oint_{{\cal T}_0} \!\! \d S \, T_{\mu \nu} k^\mu (-\ell^\nu) \geqslant 0 \,, 
\label{ineq:GB}
\eena 
where ${\cal R}$ denotes the scalar curvature on the $2$-dimensional compact surface ${\cal T}_0$. 
Noting that $-\ell^\mu$ is future-directed, we find the inequality to hold under the DEC. 
According to the Gauss-Bonnet theorem, the left-hand side is $2\pi \chi$ with $\chi$ being the Euler characteristic number of ${\cal T}_0$. It then immediately follows that the horizon cross-section ${\cal T}_0$ must be topologically either $2$-sphere or $2$-torus. The latter case is possible only when the equality holds strictly in (\ref{ineq:GB}). Such a situation seems to rather be unstable; for example if we add some small perturbation or positive cosmological constant or if MOTS satisfies $\ell^\nu \nabla_\nu \theta_+ >0$ at a single point, then the integral (\ref{ineq:GB}) becomes strictly positive. Therefore the case of $2$-torus seems implausible. In fact, with further technical assumptions for the rigidity of MOTS~\cite{Galloway:2006ws}, one can rule out the $2$-torus case and conclude that ${\cal T}_0$ must be topologically a $2$-sphere. 
\hfill $\Box$

\medskip 

Next, let us consider the topology of the black hole exterior region. We first describe two theorems concerning topological constraints on globally hyperbolic, asymptotically flat spacetimes. 

\medskip 

\begin{itembox}[l]{{Theorem 2.6.7 (Topology and singularity)}} 
Let $(M,g_{\mu \nu})$ be a globally hyperbolic, asymptotically flat spacetime, on which the NEC holds. Suppose that $M$ contains a non-simply connected Cauchy surface $\Sigma$. Then, $M$ must be geodesically incomplete. 

\end{itembox} 
\begin{itemize}
\item Gannon (1975)~\cite{Gannon:1975}. 
\item The asymptotic flatness implies that the topologically non-trivial structure is isolated. 

\end{itemize}

\medskip 

\begin{figure}[h]
\begin{center}
%\scalebox{0.10}{%
\includegraphics[width=80mm]{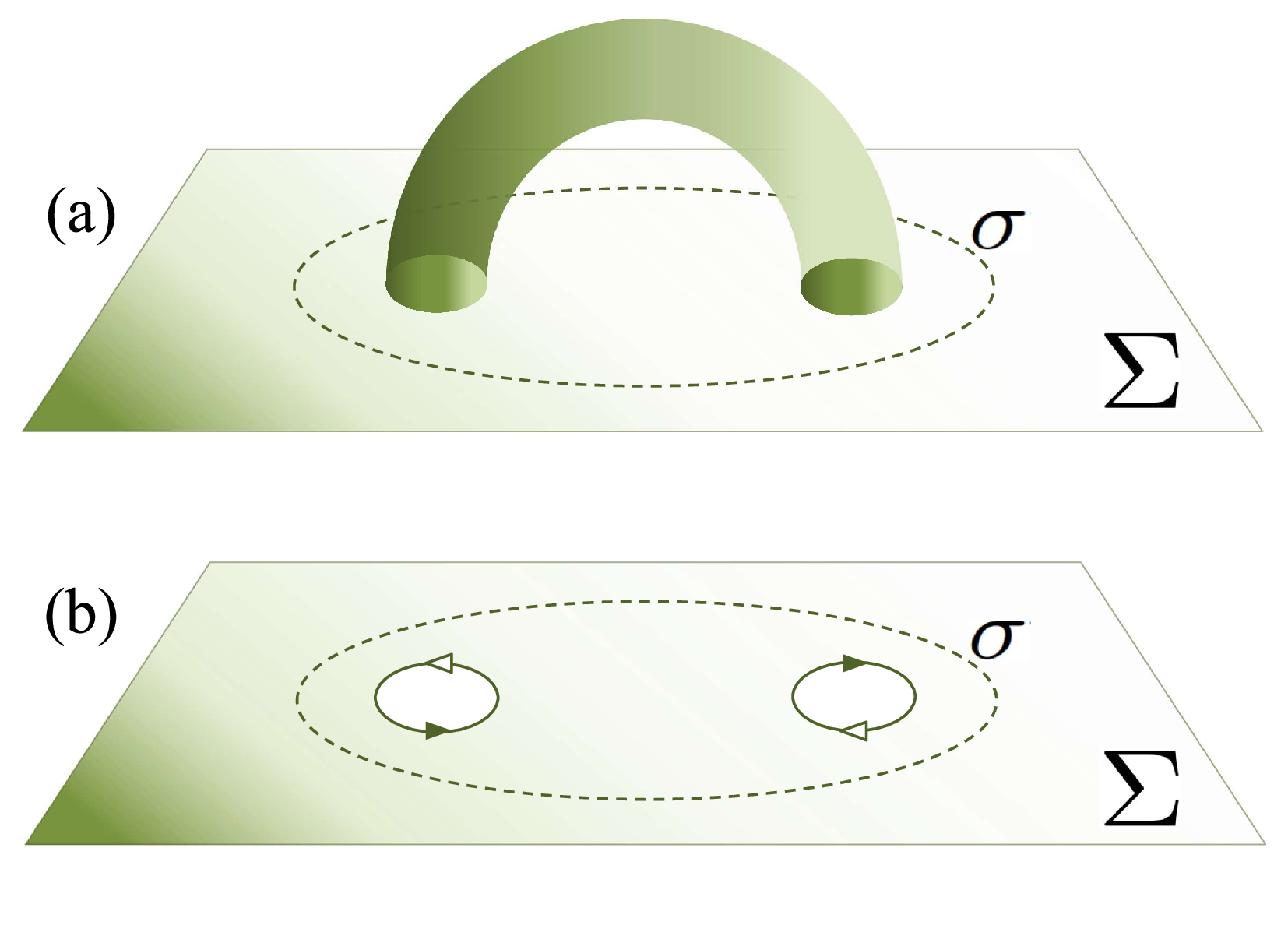}
%          } 
\end{center}
% \vspace{-3mm}
\caption{\small 
(a) A ``wormhole (handle-structure)" exists in the interior of a sphere $\sigma$ within a $3$-dimensional space $\Sigma$. 
(b) This situation can be constructed by excising two $3$-dimensional balls from the interior of $\sigma$ and identifying their $2$-sphere boundaries.
 } 
\label{top}
\end{figure}

If a non-trivial topological structure is isolated, a natural question is whether it can be detected. To make this more concrete, let us imagine a situation where a bounded region enclosed by a sphere $\sigma$ in an asymptotically flat $3$-dimensional Cauchy surface $\Sigma$ contains a ``handle" or a ``wormhole" structure (see Figure~\ref{top}). 

\smallskip 

Can a distant observer actually detect this wormhole (handle) structure within that bounded region inside $\sigma$? 
For example, we would attempt to send a probe rocket into this bounded region so that the rocket would orbit the handle, become snagged by its non-trivial topological structure, and then return to a distant location to report its findings. Again under the NEC, we have the following theorem. 
\smallskip 
\begin{itembox}[l]{{Theorem 2.6.8 (Topological censorship)}} 
Let $(M,g_{\mu \nu})$ be a globally hyperbolic, asymptotically flat spacetime. If the NEC holds in $M$, then any causal curve $\gamma$ from a point on $\mathscr{I}^-$ to a point on $\mathscr{I}^+$ can be continuously deformed to a timelike curve $\gamma_0$ which is contained in a simply connected neighborhood of $\mathscr{I}^+\cup \mathscr{I}^-$.  
\end{itembox} 
\begin{itemize}
\item Friedman-Schleich-Witt~(1993)~\cite{Friedman:1993ty}. 
\item It is equivalent to say that $M$ must have as simple a topological structure as the neighborhood of $\mathscr{I}$. Since $M$ is globally hyperbolic, 
there exists a Cauchy surface $\Sigma$ so that $M \simeq {\Bbb R}\times \Sigma$. Then, it implies $\Sigma$ must be simply connected. 

\item In fact, this theorem was shown in \cite{Friedman:1993ty} under a weaker condition, the averaged null energy condition (ANEC), which we will discuss in the next subsection.  

\item Here, we can consider $\gamma_0$ as the world-line of an observer who stays in the asymptotic region near infinity, and $\gamma$ as the world-line of a probe rocket that departs from the asymptotic region, explores the topology of spacetime $M$, and returns to the asymptotic region again. Two curves, $\gamma_0$ and $\gamma$, are said to be homotopic if one can be continuously deformed into the other. This means that the probe $\gamma$ does not get caught on a non-trivial topological structure like a ``handle."

\item ``Censorship" is used metaphorically here to describe how physical laws prohibit the observation of topology. 
\end{itemize}

\begin{figure}[h]
\begin{center}
%\scalebox{0.10}{%
\includegraphics[width=80mm]{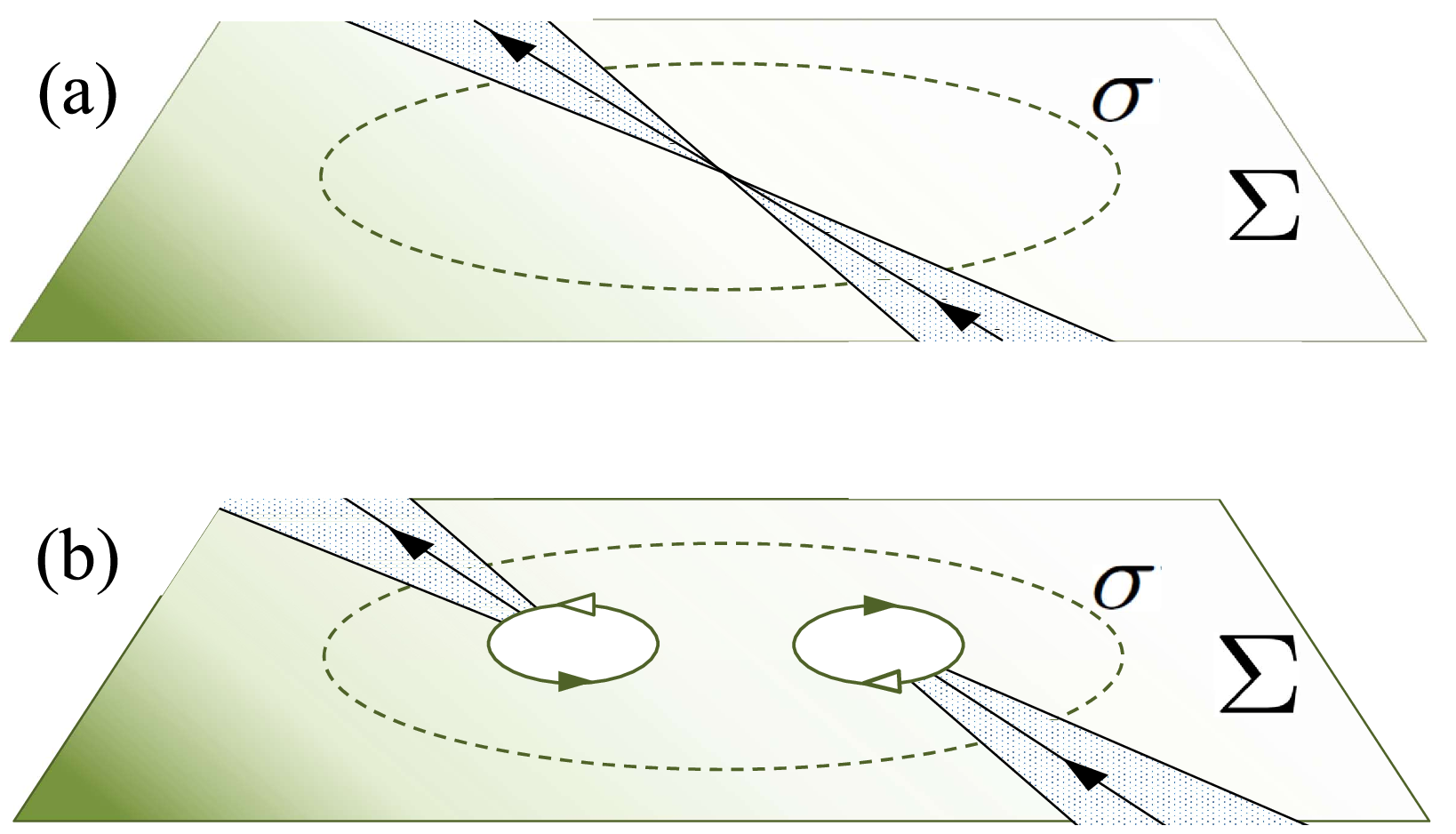}
%            } 
\end{center}
% \vspace{-3mm}
\caption{\small 
(a) Consider a bundle of light rays (i.e., null geodesic congruence) traveling in a flat spacetime, which serves as a topologically trivial space. 
An ingoing light ray bundle perpendicular to $2$-sphere $\sigma$ has a negative expansion $\theta<0$, and converges at the origin (which corresponds to the conjugate point to $\sigma$ and the expansion $\theta$ of the ingoing light rays changes from $-\infty$ to $+\infty$ there) and then emerges from $\sigma$ as an outgoing bundle with a positive expansion $\theta>0$. 
(b) Now, consider a bundle of light rays entering the interior of $\sigma$ containing a wormhole (handle-structure), depicted in Figure~\ref{top}, and subsequently exiting from $\sigma$ as an outgoing light ray. The cross-sectional area of the light rays entering one mouth of the wormhole decreases, having a negative expansion. However, after passing through the wormhole and exiting from the other mouth, the light ray bundle travels outward from $\sigma$ with its cross-sectional area increasing as it propagates towards infinity. This behavior of the light ray bundle leads to a contradiction. 
 } 
\label{top_cov}
\end{figure}

\medskip 

The idea of the proof of Theorem 2.6.8 is as follows. Suppose an ingoing null geodesic congruence is emitted orthogonally from a $2$-sphere in the vicinity of past null infinity $\mathscr{I}^-$ toward inner region enclosed by $\sigma$ in which wormhole (handle) structure is confined. As depicted in Figure~\ref{top_cov} (b), an ingoing null geodesic congruence perpendicular to $\sigma$ enters one mouth of the handle structure, and then exits from the other mouth as an outgoing bundle, propagating outward from $\sigma$, toward future null infinity $\mathscr{I}^+$. As the null congruence travels toward the entrance of the handle structure, the cross-section area $A$ of the congruence decreases (i.e., its expansion is negative $\theta<0$), but as it exits from the other mouth, $A$ increases (i.e., $\theta>0$). However, the congruence does not pass through a zero point, $A=0$ (i.e., the conjugate point where $\theta=\pm \infty$), while traversing the handle structure. Such a behavior of the expansion $\theta$ (i.e, changing its sign from negative to positive, without passing a conjugate point) contradicts the null focusing theorem 2.4.2, under the NEC. 
\hfill $\Box$ 

\medskip  

In conclusion, under the NEC, a distant observer cannot see an isolated non-trivial topology. This does not mean that asymptotically flat spacetime as a whole can only have a simple topology. Instead, it implies that if a non-trivial topological structure like a handle exists, it must be hidden within a region that is not visible from null infinity---namely, inside a black hole, just like the case of a closed trapped surface ${\cal T}$, which also cannot intersect $J^-(\mathscr{I}^-)$ (Prop. 2.6.1). In this respect, it is intriguing to note that Theorem 2.6.7 is analogous to Penrose's singularity theorem (Theorem 2.5.6) with condition (ii) replaced with the existence of an isolated non-trivial topology in an asymptotically flat Cauchy surface.

\medskip 

\begin{figure}[h]
\begin{center}
%\scalebox{0.10}{%
\includegraphics[width=60mm]{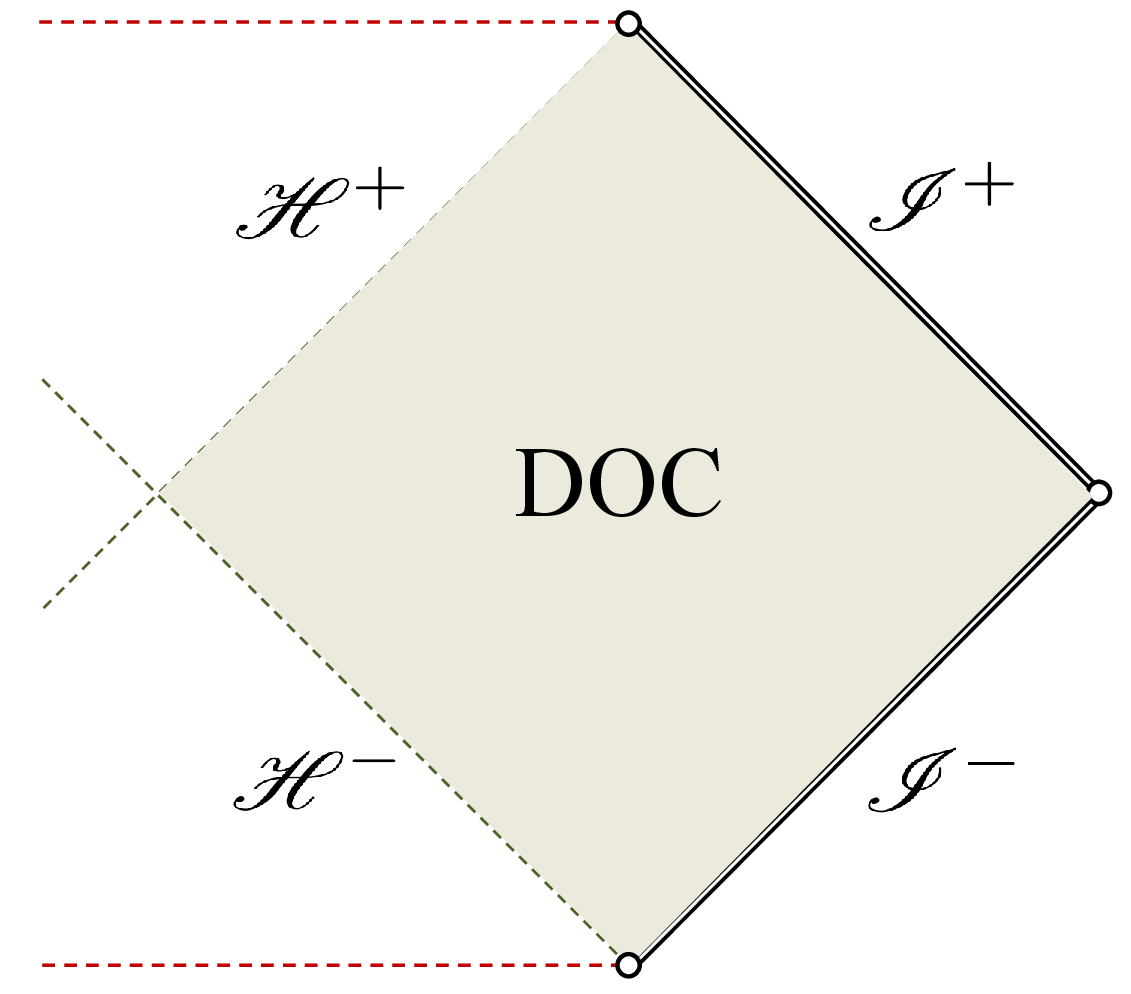}
%           } 
\end{center}
% \vspace{-3mm}
\caption{\small The domain of outer communications of an asymptotically flat black hole spacetime. 
 } 
\label{DOC}
\end{figure}

As the exterior region of an asymptotically flat black hole, we define the {\em domain of outer communications} by 
\bena
{\rm DOC}:=I^+(\mathscr{I}^-)\cap I^-(\mathscr{I}^+) \,, 
\eena
as illusrated in Figure~\ref{DOC}. 

Applying Theorem 2.6.8 to ${\rm DOC}$, one can find the following constraints on the horizon topology. 

\smallskip 
\begin{itembox}[l]{{Theorem 2.6.9}} 
Suppose $(M,g_{\mu \nu})$ is a stationary, asymptotically flat spacetime containing a single asymptotically flat region, whose ${\rm DOC}$ is globally hyperbolic. Suppose the NEC holds. Then, (i) ${\rm DOC}$ is simply connected, and (ii) each connected component of the horizon cross-section is homeomorphic to a $2$-sphere. 
\end{itembox} 
\begin{itemize} 
\item Chrusciel-Wald (1994)~\cite{Chrusciel:1994tr}. 
\item The first statement (i) results from the topological censorship (Theorem 2.6.8) and the second statement (ii) is proven by cobordism. 
\item Compare with Theorem 2.6.6, in which the DEC is used, while here NEC. 
\end{itemize}

\medskip 

The proof goes roughly as follows. Consider a partial Cauchy surface $\Sigma$ in ${\rm DOC}$ which has boundaries at $\mathscr{I}^+$ and at ${\cal H}^+$. Since $\mathscr{I}^+ \cap \Sigma$ is topologically $S^2$ and $\Sigma$ is simply connected by Theorem 2.6.8, $\sigma={\cal H}^+ \cap \Sigma$ must be cobordant to $S^2$ by a simply connected cobordism. Hence $\sigma$ must also be topologically $S^2$.    
\hfill $\Box$

\medskip 

Note that our discussion so far focuses on $4$-dimensional spacetimes. Similar techniques using NEC and cobordism in Theorem 2.6.9 can be applied to higher dimensional black holes, but they yield only weaker constraints~\cite{Helfgott:2005jn}. The generalization of the argument in Theorem 2.6.6 by using the DEC to higher dimensions yields that MOTS must admit a Riemannian metric of positive scalar curvature, accordingly the topology of MOTS must be of positive Yamabe type~\cite{Galloway:2005mf,Galloway:2006ws}, which allows for various types of topologies other than the spherical one (see~\cite{Emparan:2001wn} for a seminal example of the ``black ring" and \cite{Hollands:2010qy} for thorough analysis, and see also, e.g.,~\cite{Hollands:2012xy,Ida:2011jm} for more general discussions of black holes and their topology in $4$ and higher dimensions). 

\medskip 

\begin{figure}[h]
\begin{center}
%\scalebox{0.10}{%
\includegraphics[width=80mm]{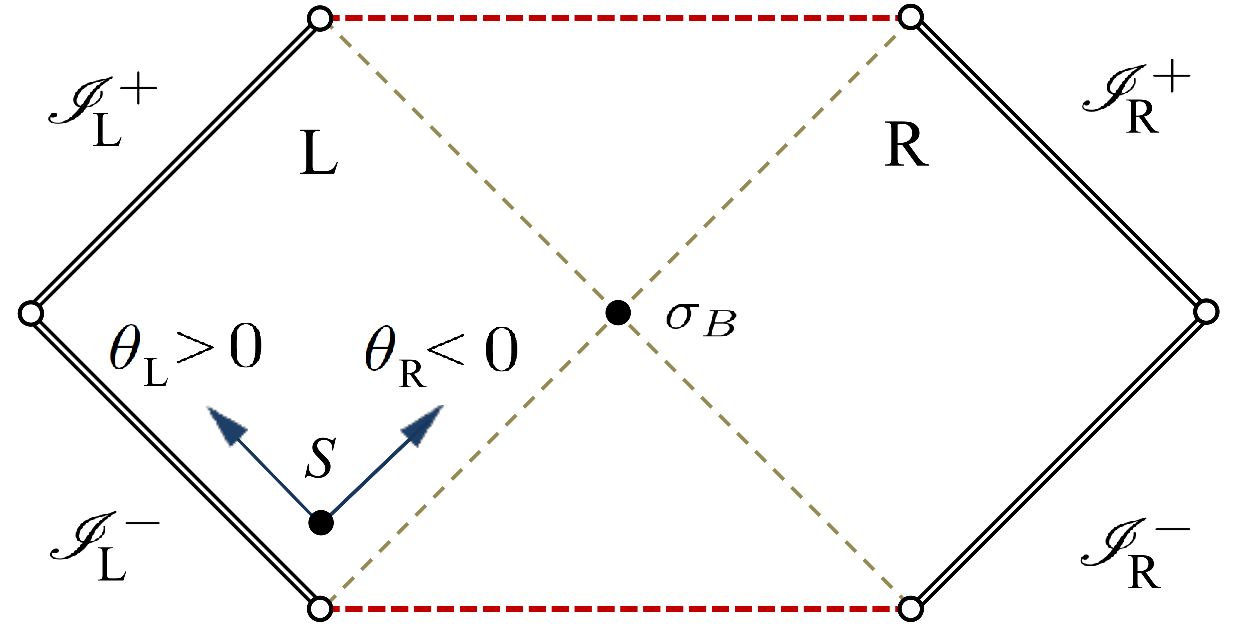}
%            } 
\end{center}
% \vspace{-15mm}
\caption{\small Penrose diagram of the maximally extended Schwarzschild spacetime. Each point describes a $2$-sphere. There are two asymptotic regions, L and R, which are causally disconnected divided by the event horizon, and $\sigma_{\rm B}$ is the bifurcate surface or Einstein-Rosen bridge (ER). 
$S$ is a $2$-sphere in ${\rm DOC}_{\rm L}$, and $\theta_{
\rm L}$ and $\theta_{\rm R}$, are the expansions of, respectively, the out-going (toward $\mathscr{I}^+_{\rm L}$) and in-going null geodesic congruences perpendicular to $S$. The in-going null geodesics with negative expansion $\theta_{\rm R}<0$ goes toward the central singularity but never enters ${\rm DOC}_{\rm R}$.  
}   
\label{Schwa-LR}
\end{figure}

Another application of topological censorship is concerning a wormhole. Even without an isolated non-simply connected (handle-like) structure, one may regard a throat-like structure as a wormhole. A typical example is the bifurcate surface of the maximally extended Schwarzschild spacetime, called the {\em Einstein-Rosen bridge} (ER) located at $U=0=V$ in the Kruskal coordinates (\ref{metric:Kruskal}). As depicted in Figure~\ref{Schwa-LR}, the maximally extended Schwarzschild spacetime has two causally disconnected asymptotic null infinities $\mathscr{I}_{\rm R}$ and $\mathscr{I}_{\rm L}$, and associated DOCs, defined respectively as ${\rm DOC}_{\rm R}:= I^+(\mathscr{I}^-_{\rm R}) \cap I^-(\mathscr{I}^+_{\rm R})$ and ${\rm DOC}_{\rm L}:= I^+(\mathscr{I}^-_{\rm L}) \cap I^-(\mathscr{I}^+_{\rm L})$. On a Cauchy surface $\Sigma \simeq S^2 \times {\Bbb R}$, e.g., the hypersurface $U=-V$, passing through ER, the area radius $r$ has the minimal value $r_h=2G{\cal M}$ at $U=0=V$, representing a ``throat" on $\Sigma$. 

\medskip 

However, since the two DOCs are spacelike separated with ER located at $\partial {\rm DOC}_{\rm R} \cap \partial {\rm DOC}_{\rm L}$, any causal curve cannot travel from inside one DOC to the other, passing through ER. From this observation, to make our discussion more precise we define a traversable wormhole structure (TW) as follows. 
Let $(M,g_{\mu \nu})$ be a connected, globally hyperbolic asymptotically flat spacetime with two (mutually disconnected) null infinities, $\mathscr{I}_{\rm R}$ and $\mathscr{I}_{\rm L}$. (The global hyperbolicity is not strictly necessary, but is assumed to simplify the argument.) Consider ${\rm DOC}_{\rm R}$ and ${\rm DOC}_{\rm L}$, associated with $\mathscr{I}_{\rm R}$ and $\mathscr{I}_{\rm L}$, respectively, as defined above.   
If there exists a non-empty set  
\bena
{\rm TW}:= {\rm int}{\rm DOC}_{\rm R} \cap {\rm int}{\rm DOC}_{\rm L} \neq \emptyset \,,
\eena
then, $M$ is said to contain a {\em traversable wormhole}. 

\medskip 

Then, applying a similar argument of the proof of Theorem 2.6.8, we immediately obtain the following:  

\smallskip 
\begin{itembox}[l]{{Proposition 2.6.10 (No traversable wormhole)}} 
The traversable wormhole structure (TW) must violate the NEC. 
\end{itembox}

\begin{figure}[h]
\begin{center}
%\scalebox{0.25}{%
\includegraphics[width=80mm]{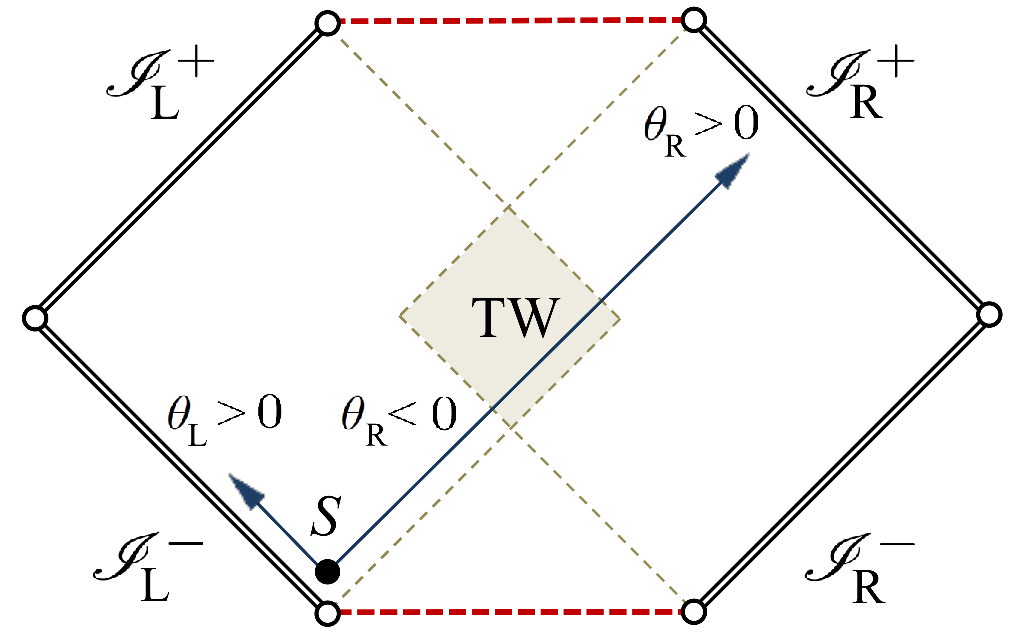}
%            } 
\end{center}
% \vspace{-15mm}
\caption{\small Penrose diagram of an asymptotically flat, traversable wormhole spacetime. There are two asymptotic regions, L and R. 
$S$ is a $2$-sphere in ${\rm DOC}_{\rm L} \cap I^-({\rm TW})$, but not in ${\rm DOC}_{\rm R}$. 
If $S$ is sufficiently close to $\mathscr{I}^-_{\rm L}$, the ``in-going" null geodesics orthogonally emanating from $S$ have a negative expansion $\theta_{\rm R}<0$ in the region ${\rm DOC}_{\rm L} \setminus {\rm TW}$, then passing through the ${\rm TW}$, eventually approach $\mathscr{I}^+_{\rm R}$ as ``out-going" null geodesics with a positive expansion $\theta_{\rm R}>0$. From $\mathscr{I}^+_{\rm R}$, $S$ may look like a ``closed trapped surface" which, however, should not be visible from $\mathscr{I}^+_{\rm R}$ under the NEC, according to Prop. 2.6.1.
} 
\label{wh-LR}
\end{figure}

The idea of the proof is as follows. The essence of the proof of Theorem 2.6.8 is that any structure which allows the expansion $\theta$ of a null geodesic congruence to change its sign from negative to positive without passing a conjugate point contradicts the null focusing theorem 2.4.2, under the NEC.
As illustrated in Figure~\ref{wh-LR}. consider a $2$-sphere $S$ in ${\rm DOC}_{\rm L} \cap I^-({\rm TW})$ in a neighborhood sufficiently close to $\mathscr{I}^-_{\rm L}$ so that a future-directed ingoing (toward ${\rm TW}$) null geodesic $\gamma$ perpendicularly emanating from $S$ has a negative expansion $\theta_{\rm R}<0$ on $S$. Passing through ${\rm TW}$, the null geodesic $\gamma$ enters ${\rm DOC}_{\rm R}$ and eventually approaches $\mathscr{I}^+_{\rm R}$. In the neighborhood of $\mathscr{I}^+_{\rm R}$, the expansion of $\gamma$ must turn to be positive $\theta_{\rm R}>0$. This leads to the contradiction mentioned above. 
\hfill $\Box$   

\subsection{Averaged energy conditions}\label{subsec:AEC}

So far we have reviewed the pointwise (locally defined) energy conditions (WEC, DEC, SEC, NEC) and their roles in classical general relativity. 
However, these conditions can be violated even within classical frameworks. For example, consider a scalar field $\phi$ with potential $V(\phi)$ with the energy-momentum tensor (\ref{emt:scalar}) in a spatially homogeneous and isotropic (FLRW) universe. Let $t^\mu = (\partial/\partial t)^\mu$ be a unit timelike vector orthogonal to any homogeneous and isotropic hypersurface. Furthermore, let us assume that $\phi$ depends only on $t$ in accord with the FLRW symmetry. Then we have 
$ \left( T^{(\phi)}_{\mu \nu}- ({1}/{2})T^{(\phi)}{}^\lambda{}_\lambda g_{\mu \nu}\right) t^\mu t^\nu = \dot{\phi}^2 - V(\phi)$. Thus, if $\dot{\phi}^2<V$, which can occur for e.g., a massive free scalar field, or inflation/dark energy models, the SEC fails to hold.  

\medskip 

In the semiclassical approach, quantum field effects can be incorporated in the expectation value of the renormalized energy-momentum tensor $\langle T_{\mu \nu} \rangle$, which replaces the classical counterpart in the right-hand side of the Einstein equations~(\ref{eq:Einstein}). There are a number of known examples in which the expectation value $\langle T_{\mu \nu} \rangle$ violates the locally defined energy conditions, even when their classical counterparts do not violate the locally defined energy conditions. In fact, Casimir effects violate {\em all} locally defined (pointwise) energy conditions~\cite{Roman:1986tp}. For example, consider two parallel reflecting planes orthogonal to $z$-axis in Minkowski spacetime. Then the vacuum expectation value of the energy-momentum tensor for a scalar field between the two planes takes the form of $\langle 0| T_{\mu \nu}|0 \rangle = - (C^2/L^4) \mathrm{diag}(1,-1,-1,3)$ with $C^2$ being a positive constant and $L$ the separation distance between the two planes~\cite{Birrell:1982ix}, for which all the four locally defined energy conditions are violated. See also, e.g., Fewster~\cite{Fewster:2012yh} and Section 10.2 of~\cite{Lobo:2017cay} (and references therein) for a more elaborated explanation that all pointwise energy conditions can be violated in quantum field theory.    

\medskip 

While the pointwise energy conditions may fail in some restricted local regions, they may still hold in the rest of the universe. This observation suggests that non-local forms of energy conditions, which are defined by taking a suitable average over spacetime, may be more robust under many circumstances of interest. 
For example, in the case of Casimir effect above, although between the two planes, the NEC is violated along $z$-axis, since the two planes supply positive energy, if we integrate $\langle 0|T_{\mu \nu}k^\mu k^\nu|0\rangle$ along $z$-axis long enough passing through the two planes, we can have a positive energy value. 
In this respect, see also \cite{Ford:1978qya} for a discussion concerning the second law of thermodynamics and averaged energy conditions.    
In what follows, we consider generalizations of the energy conditions to the averaged forms. As we have seen in Sec.~\ref{subsec:focus}, \ref{subsec:sing}, and \ref{subsec:bh}, the energy conditions play a key role mainly through the focusing theorems, we will focus on the averaged  SEC and NEC. As a weaker condition than SEC, the {\em averaged strong-energy condition} (ASEC) is simply formulated as follows. 
\smallskip 

\begin{itembox}[l]{{Definition 2.7.1 (Averaged SEC)}} 
For every inextendible timelike geodesic curve with proper time $\tau$ and the tangent $\xi^\mu$, the integral along the timelike geodesic, 
\bena
 \int \d\tau \left( T_{\mu \nu} - \dfrac{1}{2}T g_{\mu \nu} \right) \xi^\mu \xi^\nu \geqslant 0 \,. 
\eena 
\end{itembox} 
\begin{itemize}
\item Integrating the local timelike focusing $\d\theta/\d\tau \leqslant - R_{\mu \nu}\xi^\mu \xi^\nu$ in a finite interval $[0, \tau]$, and using the Einstein equations, we have
\bena
 \theta(\tau) \leqslant \theta(0) -  8\pi G \int_0^\tau \!\! \d\tau \left( T_{\mu \nu} - \dfrac{1}{2}T g_{\mu \nu} \right) \xi^\mu \xi^\nu \,. 
\eena
Thus, the ASEC expresses the timelike geodesic focusing $\theta(\tau) \leqslant \theta(0)$ in averaged sense.  
   
\end{itemize}

As for a weaker condition than NEC, we have the {\em averaged null energy condition} (ANEC) 

\smallskip 

\begin{itembox}[l]{{Definition 2.7.2 (Averaged NEC)}} 
For every inextendible null geodesic curve with affine parameter $\lambda$ and the tangent $k^\mu$, the integral along the null geodesic, 
\bena
 \int \d\lambda\, T_{\mu \nu} k^\mu k^\nu \geqslant 0 \,. 
\eena 
\end{itembox} 
\begin{itemize}
\item This is the condition used in the original statement of the topological censorship~\cite{Friedman:1993ty} (Theorem 2.6.8). 

\item Hawking-Penrose singularity theorem (c.f., Theorem 2.5.7) is generalized by using ASEC and ANEC~\cite{Tipler:1977zza,Tipler:1978zz,Fewster:2010gm}. 
See also \cite{Borde:1987qr}, for the focusing theorems under ASEC and ANEC, and further generalizations of the singularity theorems with averaged energy conditions.   
\end{itemize}

\medskip 

ANEC was shown to hold in various situations (see, e.g., \cite{Yurtsever:1994wc,Klinkhammer:1991ki,Yurtsever:1990gx,Wald:1991xn,Ford:1995gb,Kontou:2012ve}). 
In Minkowski background, ANEC was proven by a completely field theoretic argument~\cite{Faulkner:2016mzt,Hartman:2016lgu}. 
The ANEC on curved spacetime which solves the semiclassical Einstein equations is called {\em self-consistent ANEC}, and has been shown to hold for pure and mixed states under certain conditions on the curvature scale~\cite{Flanagan:1996gw}. 

\medskip 

There have also been discovered a number of cases in which the ANEC can be violated. For example, the ANEC for free scalar field can be violated if the background is locally flat but topologically non-trivial (e.g., torus compactification)~\cite{Klinkhammer:1991ki}. 
In curved spacetime, there are many known cases of the ANEC violation. For example, ANEC can be violated for a conformally coupled free scalar field in conformally flat spacetime~\cite{Urban:2009yt} (see also~\cite{Visser:1994jb,Ishibashi:2019nby}). If one considers a null geodesic of circular orbit in a Schwarzschild spacetime, (e.g. a null geodesic orbiting at the constant area radius $r=3G{\cal M}$), ANEC is not satisfied for the Hartle-Hawking state~\cite{Visser:1996iw}. Such a null geodesic of circular orbit is not achronal, while in many cases of interest in applications, e.g., the singularity theorems, the achronality of null geodesics plays a role. For this reason, there has been proposed the notion of achronal ANEC (aANEC), which asserts that the ANEC should hold for every complete {\em achronal} null geodesic but not necessarily on chronal null geodesics~\cite{Wald:1991xn,Graham:2007va} (see also for the case of the aANEC violation~\cite{Visser:1994jb,Urban:2010vr,Kontou:2015yha}). 

\medskip 

The ANEC has recently attracted increasing attention, not only in general relativity, but also in quantum field theory and in the context of the holographic principle. We will discuss ANEC and its further generalizations in more detail in the subsequent sections~\ref{sec:3} and~\ref{sec:4}. 

%%% 

\section{Energy conditions and quantum field theory}\label{sec:3}
In this section, we turn to energy conditions in quantum field theory in Minkowski spacetime.
After giving a brief review of quantum information in a bipartite system, we consider the modular Hamiltonian $K_A$ associated with the subsystem $A$ in the vacuum.
When the boundary of $A$ is planar, the Bisognano-Wichmann theorem \cite{Bisognano:1976za} states that $K_A$ generates the Rindler boost, hence be written by smearing the stress-energy tensor on $A$.
This theorem generalizes to the case when $A$ has the boundary on a null hypersurface \cite{Casini:2017roe}, which allows us to derive the ANEC from the monotonicity of the vacuum modular Hamiltonian \cite{Faulkner:2016mzt}.
We also consider QNEC, which bounds the null-null component of the stress tensor by the second null derivative of entanglement entropy \cite{Bousso:2015mna,Wall:2011kb}.
We sketch the derivation of QNEC following \cite{Kudler-Flam:2023hkl} and leave the complete proof to \cite{Ceyhan:2018zfg,Hollands:2025glm}.

\subsection{Quantum information measures}

We will consider a bipartite system with the Hilbert space $\CH = \CH_A \otimes \CH_{\bar A}$.
For a density matrix $\rho$ on $\CH$, the reduced density matrix $\rho_A$ on the Hilbert space $\CH_A$ of the subsystem $A$ is defined by
\begin{align}
    \rho_A
        :=
            \tr_{\bar A} \left[\, \rho\,\right] \ ,
\end{align}
where $\tr_{\bar A}$ means the partial trace over the Hilbert space $\CH_{\bar A}$.
Conversely, for any density matrix $\rho_A$ of a system $A$, one can find an auxiliary Hilbert space $\CH_R$ and a state $\ket{\rho} \in \CH_{A}\otimes \CH_{R}$ such that
\begin{align}
    \rho_A
        =
        \tr_R\left[\, \ket{\rho} \bra{\rho}\,\right] \ .
\end{align}
The state $\ket{\rho}$ and the auxiliary system $R$ are called the purified state of $\rho_A$ and the purifying system, respectively.
The expectation value of an operator $\CO_A$ acting on $\CH_A$ can be  written by using the purified state $\ket{\rho}$ as
\begin{align}
    \tr_A\left[\, \rho_A\,\CO_A\,\right]
        =
        \tr_{AR}\left[\, \ket{\rho} \bra{\rho}\,(\CO_A\otimes {\bf 1}_{R})\,\right]
        =
        \bra{\rho}(\CO_A\otimes {\bf 1}_{R})\ket{\rho}  \ ,
\end{align}
where we denote the union of the two subsystems $A$ and $R$ as $AR$.

\medskip 

The entanglement entropy $S(\rho_A)$ of the subsystem $A$ is defined as the von Neumann entropy of the reduced density matrix $\rho_A$:
\begin{equation}\label{EE_def}
    S(\rho_A)
        :=
            -\tr_A\left[\rho_A\, \log \rho_A \right] \ .
\end{equation}
Entanglement entropy is known to satisfy the following inequality known as the strong subadditivity (SSA) \cite{Lieb1973cp}:
\begin{align}\label{SSA}
    S(\rho_{AB}) + S(\rho_{BC}) \ge S(\rho_{A}) + S(\rho_{C}) \ .
\end{align}

\medskip 

For a pair of two density matrices $\rho_A$ and $\sigma_A$ on $\CH_A$, the (quantum) relative entropy is defined by
\begin{equation}\label{relative_entropy}
    S(\rho_A || \sigma_A)
        := 
        \tr_A\left[ \rho_A\, \log \rho_A \right] - \tr_A\left[ \rho_A\, \log \sigma_A\right] \ .
\end{equation}
Roughly speaking, the relative entropy measures the ``distance" between two quantum states $\rho_A$ and $\sigma_A$.
Indeed, it is non-negative
\begin{align}\label{Relative_positivity}
    S(\rho_A || \sigma_A) \ge 0\ ,  
\end{align}
and vanishes only when the two states are the same:
\begin{align}
    S(\rho_A || \sigma_A) = 0\qquad \Longleftrightarrow\qquad \rho_A = \sigma_A \ .
\end{align}
Moreover, the relative entropy decreases monotonically as one shrinks the size of the subsystem: \cite{lindblad1975completely,Araki1976zv,uhlmann1977relative,witten2018notes-8d6}:\footnote{The monotonicity of the relative entropy is equivalent to the strong subadditivity of entanglement entropy \eqref{SSA}.}
\begin{align}\label{RE_monotonicity}
    B \subset A\quad \rightarrow \quad  S(\rho_{B}|| \sigma_{B}) \leq S(\rho_{A}|| \sigma_{A}) \ .
\end{align}

\medskip 

For a density matrix $\sigma_A$, we can define the modular Hamiltonian as an analogue of energy:
\begin{align}\label{modular_hamiltonian}
    K^{(\sigma)}_{A} = - \log \sigma_A \ .
\end{align}
Then, the relative entropy can be rewritten as
\begin{equation}\label{Relative_K-S}
    S(\rho_A||\sigma_A)= \Delta K^{(\sigma)}_{A}-\Delta S_A \ ,
\end{equation}
where
\begin{align}
    \begin{aligned}
        \Delta K^{(\sigma)}_{A} 
            &=
                \tr_A\left[\rho_A\, K^{(\sigma)}_{A}\right]-\tr_A\left[\sigma_A\, K^{(\sigma)}_{A}\right]\ \ , \\ 
        \Delta S_A 
            &= 
                S(\rho_A)-S(\sigma_A) \ .
    \end{aligned}
\end{align}
In fact, this expression coincides with the free energy $F=\Delta E - T \Delta S$, if $\sigma_A$ is the thermal density matrix. 
When the system $\bar A$ is a purifying system of both $\rho_A$ and $\sigma_A$, we can simplify $\Delta K^{(\sigma)}_{A}$ to the following form:
\begin{align}
        \Delta K^{(\sigma)}_{A} 
            &=
                \bra{\rho}\,K^{(\sigma)}_A \otimes {\bf 1}_{\bar{A}}\,\ket{\rho}
                -
                \bra{\sigma}\,K^{(\sigma)}_A \otimes {\bf 1}_{\bar{A}}\,\ket{\sigma} \ .
\end{align}

Having applications to QFTs in mind, we introduce the full modular Hamiltonian $\hat K^{(\rho)}_A$ as follows:
\begin{align}\label{full_modular_hamltonian}
    \hat K^{(\sigma)}_A 
        &:= 
            K^{(\sigma)}_A \otimes {\bf 1}_{\bar{A}} - {\bf 1}_{A} \otimes K^{(\sigma)}_{\bar A} \ .
\end{align}
The monotonicity of relative entropy \eqref{RE_monotonicity} can be expressed as the operator inequality for the full modular Hamiltonian \cite{Borchers:1995zg,Blanco:2013lea}:
\begin{align}\label{full_modular_monotonicity_QM}
    B \subset A \quad \rightarrow \quad \hat K^{(\sigma)}_B \le \hat K^{(\sigma)}_A \ .
\end{align}

\subsection{Modular Hamiltonian in quantum field theory}

From now on, we will extend the discussion in the previous section to quantum field theory.
Let the Hilbert space $\CH$ be defined on a Cauchy surface $\Sigma$, and denote by $\CH_A$ the Hilbert space associated with a subregion $A\subset \Sigma$.
In a finite-dimensional system, the Hilbert space factorizes as $\CH = \CH_A \otimes \CH_{\bar A}$, which allows us to define the reduced density matrix $\rho_A$, modular Hamiltonian $K_A^{(\sigma)}$ and entanglement entropy.
In quantum field theory, however, such a factorization no longer holds, so these quantities are not well-defined in general.
Nevertheless, in an algebraic formulation of QFT, one can rigorously define the relative entropy $S(\rho_A||\sigma_A)$ and the full modular Hamiltonian $\hat K_A^{(\sigma)}$, which generates the modular flow of local operator algebra (see \cite{witten2018notes-8d6} for a review).

\medskip 

In what follows, we will take a less rigorous approach and assume that the Hilbert space factorizes under a suitable ultraviolet regularization, so that the reduced density matrix, modular Hamiltonian, and entanglement entropy can be treated for practical purposes.

\medskip 

\begin{figure}[ht]
\begin{center}
%\scalebox{0.10}{%
\includegraphics[width=120mm]{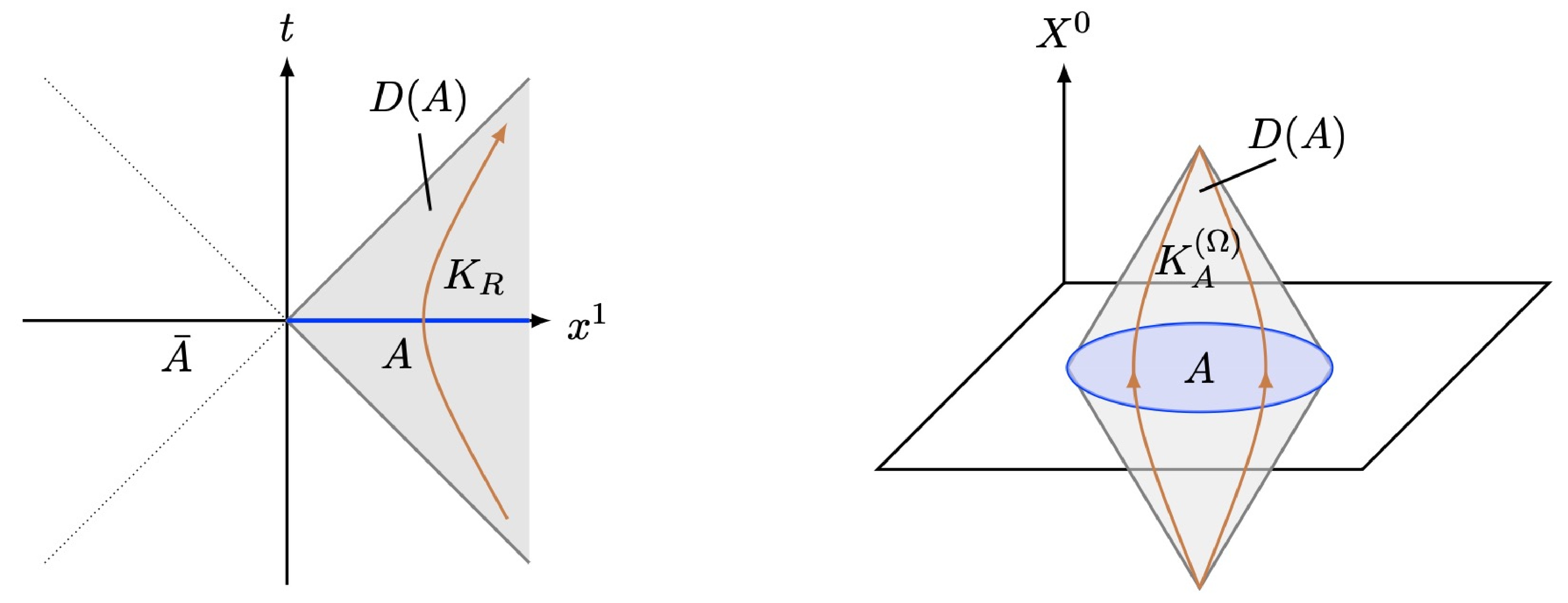} 
%            } 
\end{center}

\vspace{3mm}

    \caption{
    [Left] The right Rindler wedge is the causal domain $D(A)$ of the region $A$ as shown in gray color. The vacuum modular Hamiltonian coincides with the boost operator $K_A^{(\Omega)} = 2\pi\,K_R$ due to the Bisognano-Wichmann theorem.
    [Right] The right Rindler wedge is mapped to the causal domain of the spherical region by the conformal transformation \eqref{CHM}.}
    \label{fig:causal_domain}
\end{figure}

The {full} modular Hamiltonian defined by \eqref{full_modular_hamltonian} does not have a local expression written by an integral of a local operator in general, unlike a Hamiltonian.
However, when $\sigma$ is the vacuum density matrix $\Omega$, i.e.\, $\sigma = \Omega = \ket{\Omega}\bra{\Omega}$, the {full} modular Hamiltonian \eqref{full_modular_hamltonian} can be ``geometric," which is a local integral of the stress tensor, generating the time evolution.
A well-known example is the Bisognano-Wichmann theorem \cite{Bisognano:1976za}.
Let region $A$ be the right half-space $A = \{ t=0 , x^1 \ge 0 \}$.
We take the coordinate in Minkowski spacetime as
\begin{equation}\label{eq:coordtrans}
    t= r\,\sinh \tau\ ,\qquad  x^{1} =r\,\cosh \tau \ .
\end{equation} 
That is, the coordinate $(r, \tau)$ implies the Minkowski coordinate of the observer who is uniformly accelerating in the Minkowski spacetime, and covers only $ |t| \le x^1$, known as the (right) Rindler wedge (see the left panel of figure \ref{fig:causal_domain}). 
The Rindler wedge is the causal domain (diamond) $D(A)$ of the region $A$.
The causal domain $D(A)$ is the set of points $p$ such that all causal curves through $p$ intersect $A$, and the information at the point $p$ is completely determined by that of $A$. 
The Hamiltonian $K$ 
%\textcolor{red}{\st{for the observer}} 
with respect to the Rindler time $\tau$ is given by
\begin{align}\label{modular_hamiltonian_boost}
    K
        =
            \int_{t=0}
            \d^{d-1}x\, x^{1}\, T_{tt} \ .
\end{align}

The Bisognano-Wichmann theorem states that the vacuum full modular Hamiltonian coincides with the Rindler Hamiltonian, $\hat K_A^{(\Omega)} = 2\pi\, K$.
The physical implication of this theorem is as follows. 
Starting from $x^1 > 0$ at the initial time, the uniformly accelerating observer cannot obtain the information about $x^1 \le 0$ at the same time, since even the signal with the speed of light never reaches the observer (for this observer, $x^1 < 0$ is regarded as the inside of the black hole). 
Thus, the system as seen by this observer is described by the Hamiltonian $K_R$, which is the restriction of $K$ to the right Rindler wedge:\footnote{$K_R$ may not be a well-defined operator. See \cite{Witten:2021unn,AliAhmad:2023etg} for further discussion.}
\begin{align}\label{modular_hamiltonian_boost}
    K_R
        =
            \int_{A=\{ t=0,\, x^1\ge 0\}}
            \d^{d-1}x\, x^{1}\, T_{tt} \ .
\end{align}
On the other hand, assuming that quantum field theory has a tensor factorized Hilbert space under some UV regularization, the system of the observer may be characterized by the reduced density matrix $\Omega_A = e ^{-K_A^{(\Omega)}}$, where $K_A^{(\Omega)}$ is the modular Hamiltonian.
The Bisognano-Wichmann theorem then implies $K_A^{(\Omega)} = 2\pi\,K_R$, so that $\Omega_A = e^{-2\pi K_R}$.
This shows that the Rindler observer perceives a thermal bath at inverse temperature $\beta = 2\pi$, even though the total system is in a pure state. 

While the Bisognano-Wichmann theorem holds in any relativistic quantum field theory (under certain condition), there is another expression for the modular Hamiltonian in conformal field theory (CFT).
Let region $A$ be the solid ball of radius $R$ centered at the origin: $A=\{ X^0=0,\, 0\le r \le R\}$.
Here, $r:= \sqrt{ \sum_{i=1}^{d-1} (X^i)^2}$ is the radius of the spherical coordinates in the constant-time slice.
The Rindler wedge is mapped to the causal domain $D(A)$ of the solid ball $A$, by conformal transformation, called the CHM map \cite{Casini:2011kv} (figure \ref{fig:causal_domain}):
\begin{align}\label{CHM}
    X^{\mu} 
        = 
            \frac{x^{\mu}+(x \cdot x)\,c^{\mu}}{1+2\,(x\cdot c)+(x\cdot x)\,(c\cdot c)} - 2\,R^2\, c^{\mu} \ ,
\end{align}
where $x^0=t$ and $c^{\mu}=(0,1/(2R),\vec{0}\,)$.
This means that, in CFT, the vacuum reduced density matrix of the right half-space is related to $\Omega_A$ by a unitary transformation.
Then, the modular Hamiltonian of $\Omega_A$ is given by \cite{Casini:2011kv,Hislop:1981uh}
\begin{equation}\label{spherical_modular_hamiltonian}
    K^{(\Omega)}_A = 2\pi \int_{X^0=0,\,0\le r\le R}\d^{d-1}X\,  \frac{R^{2}-r^{2}}{2R}\, T_{X^0X^0} \ .
\end{equation}
It follows that the full modular Hamiltonian $\hat K^{(\Omega)}_A$ takes the same form as \eqref{spherical_modular_hamiltonian} with the integration over the entire time slice $X^0=0$.
While the derivation of $\hat K^{(\Omega)}_A$ was not fully rigorous, the resulting expression can be shown to hold exactly in the case of a free massless scalar field \cite{Longo:2020amm}.

\medskip 

\begin{figure}[ht]
\begin{center}
%\scalebox{0.10}{%
\includegraphics[width=100mm]{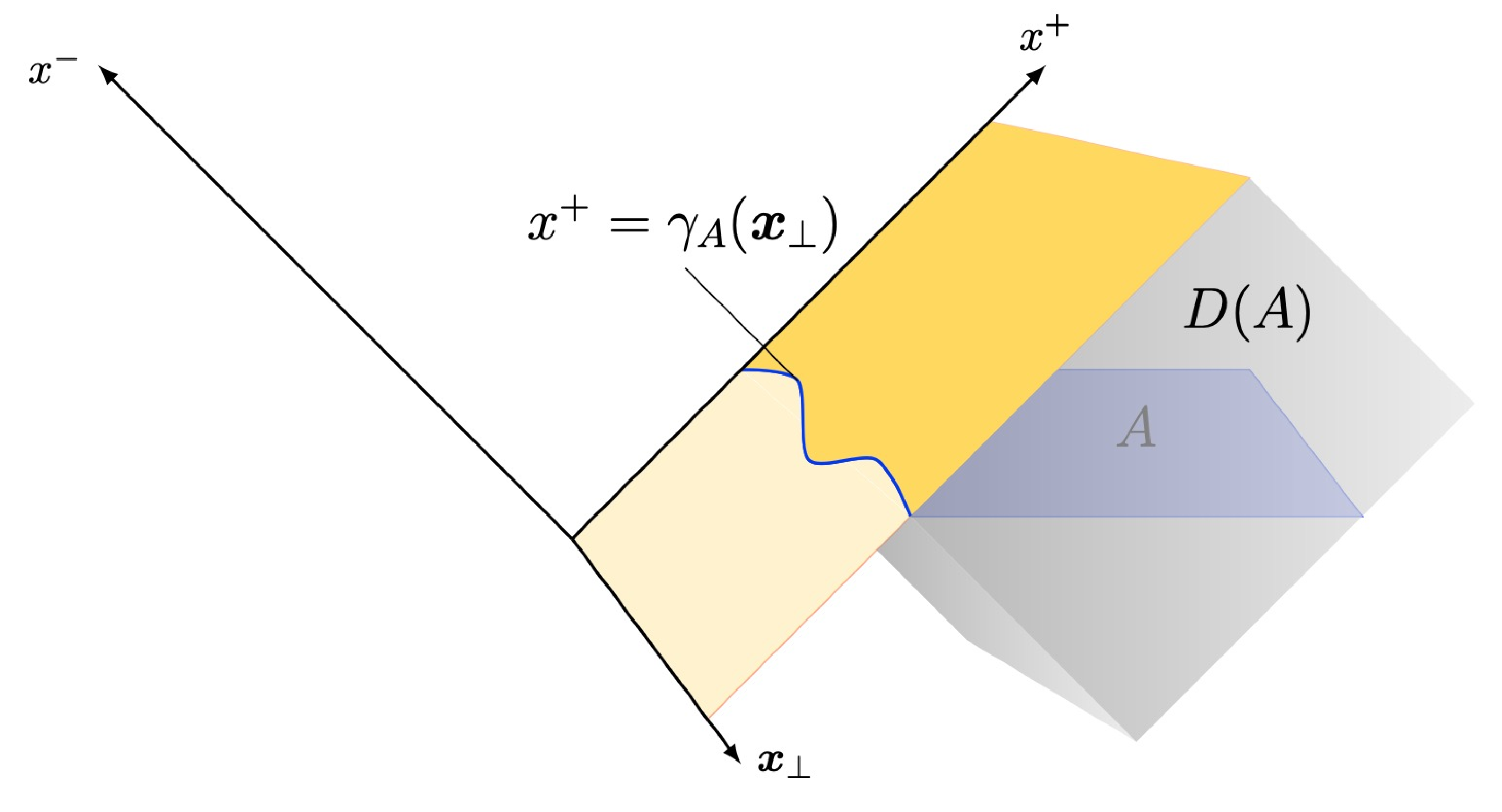} 
%         } 
\end{center}

\vspace{3mm}
\caption{The null deformed Rindler wedge as the causal domain $D(A)$ of a spacelike hypersurface $A$ with $x^-\le 0$ bounded by $\partial A = \{ x^- =0,\, x^+=\gamma_A(\bm x_\perp)\}$ .}
    \label{fig:null_plane}
\end{figure}

Let us define the light-cone coordinates $x^\mu = (x^+, x^-, \bx_\perp) \ ,~ x^\pm := t \pm x^1$ in Minkowski spacetime.
We can consider the null deformed Rindler wedge as the causal domain $D(A)$ of a spacelike hypersurface $A$ with $x^-\le 0$ bounded by $\partial A = \{ x^- =0,\, x^+=\gamma_A(\bm x_\perp)\}$ as shown in figure \ref{fig:null_plane}.
The full modular Hamiltonian associated with such a region $A$ in quantum field theory is also shown, at a physicist's level of rigor in \cite{Casini:2017roe}, to have a local expression similar to \eqref{modular_hamiltonian_boost}:
\begin{align}\label{modular_hamiltonian_null_full}
    \hat{K}^{(\Omega)}_A = 2\pi \int \d^{d-2}\bx_\perp\int_{-\infty}^\infty \d x^+ (x^+ - \gamma_A (\bx_\perp))\, T_{++}\left(
    x^+, x^-=0, \bx_\perp\right) \ .
\end{align}
We will use this expression in deriving averaged null energy condition on flat space in section \ref{sec:ANEC}.

\subsection{Bekenstein bound}

Bekenstein proposed the following conjecture (Bekenstein bound) \cite{Bekenstein:1980jp}:
\begin{align}\label{Bekenstein_bound}
    S \lesssim R\,E \ ,
\end{align}
where $S$ and $E$ are the entropy of the matter and energy in the region $A$ with typycal size $R$.
This bound arose through the following thought experiment.
Consider an object with energy $E$ and entropy $S$, and drop into the black hole of radius $R$ in the three-dimensional space for simplicity (a similar argument holds in an arbitrary dimension).
It is known that the black hole entropy is proportional to its area.
As the radius of the black hole increases by $\Delta R \propto E$, the entropy also increases by $\Delta S_\text{BH} \propto R \Delta R$.
The second law of thermodynamics implies that the entropy of the total system including an object and the black hole is bounded: $S \le \Delta S_\text{BH} \propto R\,E$.

\medskip 

However, the Bekenstein bound becomes ambiguous in QFT, because the energy $E$ and the entropy $S$ have ultraviolet divergence.
That is, the bound \eqref{Bekenstein_bound} is ill-defined.

\medskip 

To give a well-defined formulation of the Bekenstein bound, we apply the Bisognano-Wichmann theorem \cite{Casini:2008cr}.
Suppose that the region $A$ is the solid ball of the radius $R$ in CFT.
Consider the reduced density matrices of an excited state and the vacuum state: $\rho_A, \Omega_A$.
The positivity of the relative entropy \eqref{Relative_positivity} and the relation to {the} modular Hamiltonian \eqref{Relative_K-S} show
\begin{align}\label{S_less_K}
    \Delta S_A \le \Delta K^{(\Omega)}_A \ ,
\end{align}
where the $K^{(\Omega)}_A$ in the right hand side is the vacuum modular Hamiltonian in CFT \eqref{spherical_modular_hamiltonian}.
Let $E$ be the energy of the excited state in the region $A$, then we can estimate $T_{X^0X^0}\sim E/R^{d-1}$, since $T_{X^0X^0}$ is the energy density.
Integrating over the region $A$, we obtain $\Delta K_A^{(\Omega)} \sim  R\, E$.
Therefore, we can think of \eqref{S_less_K} as the quantum version of the Bekenstein bound \eqref{Bekenstein_bound}.
In addition, a different form of the Bekenstein bound was discovered using the monotonicity of the relative entropy in \cite{Blanco:2013lea}.

\medskip 

In the next two sections, we show the applications of the monotonicity of the relative entropy \eqref{RE_monotonicity} (or the modular Hamiltonian \eqref{full_modular_monotonicity}) to deriving energy conditions in quantum field theory.

\subsection{Averaged null energy condition}
\label{sec:ANEC}

Energy is an important and universal notion in physics. The system that has the nonnegative energy density should be classically stable, if the vacuum energy is set to zero. 
However, QFTs do not satisfy the local energy condition: 
the inequality $\bra{\rho} \,T_{tt}\, \ket{\rho} \ge 0$ for any state $\ket{\rho}$ is violated by the quantum effect \cite{Epstein:1965zza}.
On the other hand, those negative energy density is localized at some regions in spacetime and the averaged energy-momentum tensor over spacetime can be nonnegative. One example of such energy conditions is the averaged null energy condition (ANEC):
\begin{align}\label{ANEC}
    \int_{-\infty}^\infty \d x^+\,\bra{\rho}\,T_{++}(x^+,\, x^-=0,\, \bx_\perp)\,\ket{\rho} \ge 0 \ .
\end{align}
The ANEC holds for any QFT in Minkowski spacetime, which was proved in special cases (free theory, two dimensions, holography) \cite{Klinkhammer:1991ki,Wald:1991xn,Folacci:1992xg,Verch:1999nt,Kelly:2014mra}, and in interacting theories \cite{Faulkner:2016mzt,Hartman:2016lgu}. 
Here, we give a proof using the monotonicity of the relative entropy along the line of \cite{Faulkner:2016mzt}. 

\medskip 

\begin{figure}[ht]
\begin{center}
%\scalebox{0.20}{%
\includegraphics[width=100mm]{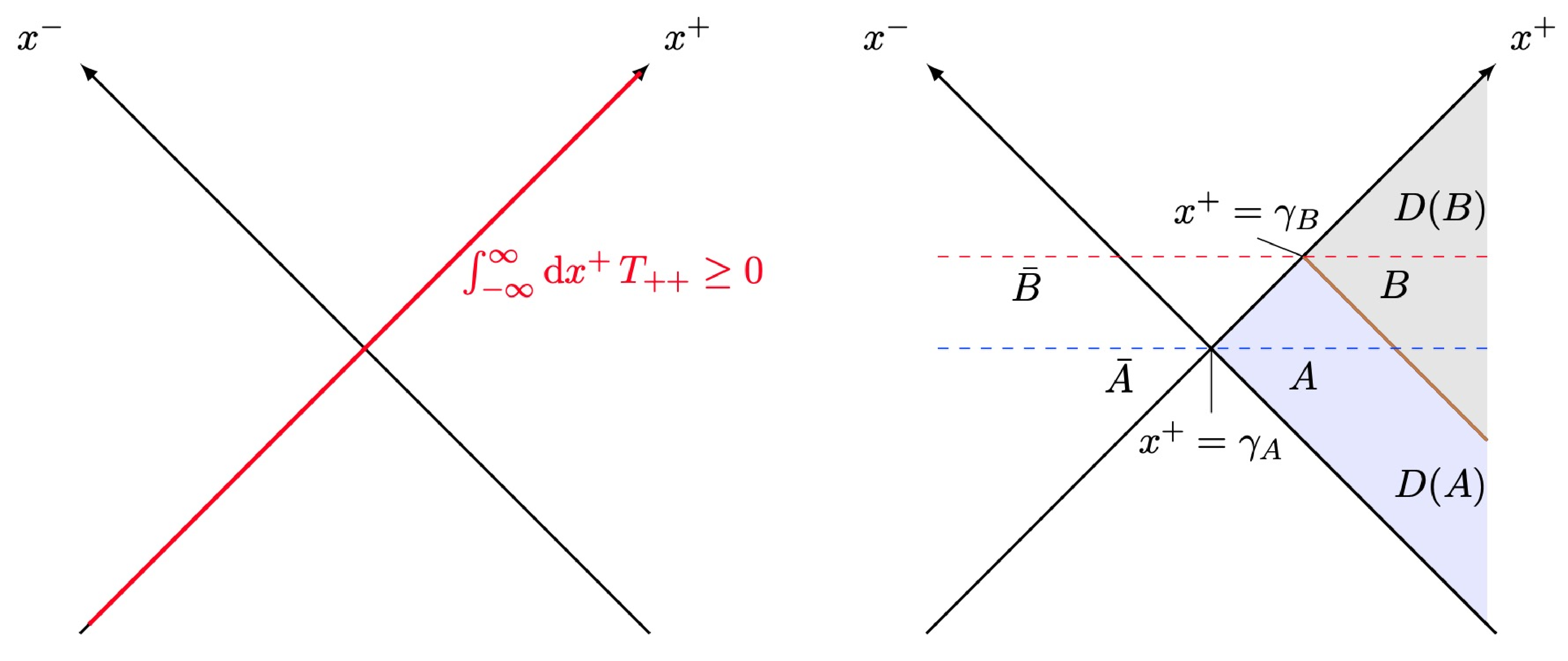} 
%              } 
\end{center}

\vspace{3mm}
\caption{[Left] The averaged null energy condition (ANEC). [Right] The regions $A$ and $B$ used to prove the ANEC.}
    \label{fig:ANEC}
\end{figure}

We begin with the full modular Hamiltonian \eqref{modular_hamiltonian_null_full} for the null deformed Rindler wedge $D(A)$. 
We also consider another region $B$, whose boundary satisfies $\partial B = \{x^-=0, x^+=\gamma_B(\bx_\perp)\}$ and $\Delta\gamma(\bx_\perp) := \gamma_B(\bx_\perp) - \gamma_A(\bx_\perp) >0$ (figure \ref{fig:ANEC}). 
In quantum field theory, \eqref{full_modular_monotonicity_QM} implies
\begin{align}\label{full_modular_monotonicity}
    D(B) \subset D(A) \quad \rightarrow \quad \hat K^{(\rho)}_B \le \hat K^{(\rho)}_A \ ,
\end{align}
where $D(A)$ and $D(B)$ are the causal domains of spacelike hypersurfaces $A$ and $B$, respectively.
Thus, in the present setup, 
\eqref{modular_hamiltonian_null_full} and \eqref{full_modular_monotonicity} yield the inequality
\begin{align}
    \begin{aligned}
        0 &\le \hat K^{(\Omega)}_A - \hat K^{(\Omega)}_B \\
            &= 2\pi \int \d^{d-2}\bx_\perp\int_{-\infty}^\infty \d x^+\,\Delta\gamma (\bx_\perp)\,T_{++}(x^+, x^-=0,\bx_\perp) \ . 
    \end{aligned}
\end{align}
Since this (operator) inequality holds for any positive function $\Delta \gamma(\bx_\perp)$, the ANEC \eqref{ANEC} must be satisfied for any state $\ket{\rho}$.

ANEC has some applications. 
For example, $(3+1)$-dimensional CFTs have two central charges characterizing conformal anomaly (which is the violation of the conformal symmetry in a curved background), and the ratio of those is bounded from ANEC \cite{Hofman:2008ar,Hofman:2016awc}. 
In addition, the irreversibility of renormalization groups, called the $\CC$-theorem, has been rederived in two- and four-dimensional QFTs by means of ANEC \cite{Hartman:2023qdn,Hartman:2023ccw}. 
In CFTs, there are some generalizations of ANEC, such as higher-spin ANEC \cite{Hartman:2016lgu,Kravchuk:2018htv,Meltzer:2018tnm} and ANEC in de Sitter and anti-de Sitter backgrounds \cite{Rosso:2019txh}.

\subsection{Quantum null energy condition}

While ANEC is a kind of global energy conditions, there is a local bound on the null energy density. 
This is called the quantum null energy condition (QNEC) \cite{Bousso:2015mna}, whose statement is as follows:
\begin{itembox}[l]{{Definition 3.5.1 (QNEC)}}
\vspace{-5mm} 
\begin{align}\label{QNEC}
    \bra{\rho}\,T_{++}(x^+=\gamma_A(\bx_\perp),\, x^-=0,\, \bx_\perp)\,\ket{\rho} \ge \frac{1}{2\pi} \frac{\delta^2 S(\rho_A)}{\delta \gamma_A(\bx_\perp)^2} \ .
\end{align}
\end{itembox} 
Here $\rho_A = \tr_{\bar{A}}\left[\, \ket{\rho} \bra{\rho}\,\right]$ and $S(\rho_A)$ is the entanglement entropy of a spacelike hypersurface $A$ in $x^- \le 0$ whose boundary is anchored on $x^- = 0, x^+=\gamma_A(\bm{x}_\perp)$.
Note that the second-order null shape derivative of the entanglement entropy in the right-hand side is not necessarily positive.
ANEC follows from QNEC by integrating both sides of \eqref{QNEC} along, assuming that the surface term $\delta S / \delta \gamma$ vanishes.
QNEC arose in the context of quantum focusing conjecture (QFC) \cite{Bousso:2015mna}, and was proven in free theory \cite{Bousso:2015wca,Malik:2019dpg}, in holographic theories \cite{Koeller:2015qmn}, and in interacting theories \cite{Balakrishnan:2017bjg, Ceyhan:2018zfg, Hollands:2025glm}.\footnote{For interacting CFT, QNEC is conjectured to be saturated \cite{Leichenauer:2018obf,Balakrishnan:2019gxl}.}
QNEC also has applications to quenches \cite{Mezei:2019sla,Kibe:2024icu,Kibe:2025cqc}, and renormalization group flow \cite{Casini:2023kyj}.

\medskip 

Let us give a sketch of a physicist's derivation of QNEC (we closely follows that of \cite{Kudler-Flam:2023hkl}).
We will use the expression of the relative entropy \eqref{Relative_K-S} and the vacuum modular Hamiltonian corresponding to \eqref{modular_hamiltonian_null_full}
\begin{align}\label{modular_hamiltonian_null}
    K^{(\Omega)}_A = 2\pi \int \d^{d-2}\bx_\perp\int_{\gamma_A (\bx_\perp)}^\infty \d x^+ (x^+ - \gamma_A (\bx_\perp))\, T_{++}\left(
    x^+, x^-=0, \bx_\perp\right) \ .
\end{align} 
By taking $\sigma$ in \eqref{Relative_K-S} to be the vacuum reduced density matrix, the second-order null shape derivative of the relative entropy becomes
\begin{align}\label{relative_second_variation}
    \frac{\delta^2}{\delta \gamma_A (\bx_\perp)^2} S(\rho_A||\Omega_A)
    &= \frac{\delta^2}{\delta \gamma_A (\bx_\perp)^2} 
    ( \Delta K^{(\Omega)}_{A}-\Delta S_A ) \notag \\  
    &= 2\pi\bra{\rho} \, T_{++}\left(
    x^+=\gamma_A (\bx_\perp),\, x^-=0,\, \bx_\perp\right) \, \ket{\rho}
    - \frac{\delta^2 S(\rho_A)}{\delta \gamma_A (\bx_\perp)^2}
\end{align}
where we used $\bra{\Omega} \, T_{++} \, \ket{\Omega} = 0$ and the fact that the vacuum entanglement entropy $S(\Omega_A)$ is independent of $\gamma_A (\bx_\perp)$ \cite{Casini:2018kzx}. 
Thus, a proof of QNEC amounts to showing the following inequality:
\begin{align}\label{QNEC_rephrased}
    \frac{\delta^2 S(\rho_A||\Omega_A)}{\delta \gamma_A (\bx_\perp)^2} \ge 0 \ .
\end{align}
This inequality is mathematically well-defined while the original QNEC \eqref{QNEC} is not as the latter involves entanglement entropy explicitly.
    \eqref{QNEC_rephrased} is guaranteed by the ``ant-formula":
\begin{align}\label{ant-formula}
    -\frac{\delta S(\rho_A||\Omega_A)}{\delta \gamma_A (\bx_\perp)}
    = 2\pi\, \inf_{\tilde{\rho}} \int_{-\infty}^\infty \d x^+\,\bra{ \tilde{\rho}} \,T_{++}(x^+,\, x^-=0,\, \bx_\perp)\,\ket{\tilde{\rho}} \ ,
\end{align}
where $\ket{\tilde{\rho}}$ reduces to $\rho_A$ when restricted to the region $x^+ \ge \gamma_A (\bx_\perp)$ $(\rho_A = \tr_{\bar{A}} \ket{\tilde{\rho}} \bra{\tilde{\rho}})$.
The ant-formula was conjectured in \cite{Wall:2017blw}, and then proven rigorously in \cite{Ceyhan:2018zfg,Hollands:2025glm}.
That is, the right-hand side is monotonically decreasing functional with respect to $\gamma_A (\bx_\perp)$, which leads to \eqref{QNEC_rephrased}. 
While a proof of this formula is technically complicated, we can easily bound the right-hand side from below by the left-hand side \cite{Bousso:2019dxk}, as we see below.  

\medskip 

By differentiating \eqref{modular_hamiltonian_null_full} with respect to $\gamma_A$ and taking the expectation value, the expectation value of the ANEC operator can be written as 
\begin{align}\label{ANEC_modular}
   \int_{-\infty}^\infty \d x^+\,\bra{\tilde{\rho}}\,T_{++}(x^+,\, x^-=0,\, \bx_\perp)\,\ket{\tilde{\rho}}
   = -\frac{1}{2\pi} \frac{\delta}{\delta \gamma_A (\bx_\perp)}
   \bra{\tilde{\rho}} \, \hat{K}^{(\Omega)}_{A} \, \ket{\tilde{\rho}} \ .
\end{align}
On the other hand, the monotonicity of the relative entropy \eqref{RE_monotonicity} implies that the relative entropy of the complementary region $\bar A$ becomes larger as we increase $\gamma_A$:
\begin{align}
    \frac{\delta S(\tilde\rho_{\bar{A}}||\Omega_{\bar{A}})}{\delta \gamma_A (\bx_\perp)} \ge 0
    \ .
\end{align}
Using \eqref{Relative_K-S} and repeating a similar argument as in \eqref{relative_second_variation}, this inequality can be written as
\begin{align}\label{relative_first_variation}
    \frac{\delta S(\tilde\rho_{\bar{A}}||\Omega_{\bar{A}})}{\delta \gamma_A (\bx_\perp)} 
            =
                \frac{\delta \langle K^{(\Omega)}_{\bar{A}} \rangle_{\tilde\rho_{\bar{A}}}}{\delta \gamma_A (\bx_\perp)}
            -
                \frac{\delta S(\tilde\rho_{\bar{A}})}{\delta \gamma_A (\bx_\perp)} 
            \ge 
                0 \ .         
\end{align}
Moreover, the strong subadditivity $S(\tilde\rho_{\bar A B}) + S(\tilde\rho_{BC}) \ge S(\tilde\rho_{\bar A}) + S(\tilde\rho_{C})$ for the three regions
\begin{align}
    \bar A &= \{ x^- = 0, -\infty \le x^+ \le \gamma_A (\bx_\perp) \} \ , \notag \\ 
    B &= \{ x^- = 0, \gamma_A (\bx_\perp) \le x^+ \le \gamma_A (\bx_\perp)+\delta \} \ , \\
    C &= \{ x^- = 0, \gamma_A (\bx_\perp)+\delta \le x^+ \le \notag \infty \}
    \ ,
\end{align}
yields the following inequality in the limit $\delta \rightarrow 0$:
\begin{align}\label{SSA_first_variation}
    \frac{\delta S(\tilde\rho_{\bar{A}})}{\delta \gamma_A (\bx_\perp)}
    \ge
    \frac{\delta S(\rho_{A})}{\delta \gamma_A (\bx_\perp)} \ ,
\end{align}
where we used the relation $\tilde\rho_{A} = \rho_A$ which follows from the definition of the purified density matrix $\tilde\rho$ for $\rho_A$.
Combining \eqref{relative_first_variation} and \eqref{SSA_first_variation}, we obtain
\begin{align}\label{monotonicity_SSA}
    \frac{\delta \langle K^{(\Omega)}_{\bar{A}} \rangle_{\tilde\rho_{\bar{A}}}}{\delta \gamma_A (\bx_\perp)}
    \ge
    \frac{\delta S(\rho_{A})}{\delta \gamma_A (\bx_\perp)} \ .
\end{align}
By adding ${- \delta \langle K^{(\Omega)}_A \rangle_{\rho_{A}}} / {\delta \gamma_A (\bx_\perp)}$ to both sides of \eqref{monotonicity_SSA}, and using \eqref{Relative_K-S}, we find
\begin{align}
    - \frac{\delta}{\delta \gamma_A (\bx_\perp)}
   \bra{\tilde{\rho}} \, \hat{K}^{(\Omega)}_A \, \ket{\tilde{\rho}} 
   \ge
    - \frac{\delta S(\rho_A||\Omega_A)}{\delta \gamma_A (\bx_\perp)} \ .
\end{align}
By combining this inequality with \eqref{ANEC_modular} and taking the infimum with respect to the purified state $\tilde\rho$, we finally obtain 
\begin{align}
    2\pi\, \inf_{\tilde{\rho}} \int_{-\infty}^\infty \d x^+\,\bra{\tilde{\rho}}\,T_{++}(x^+, x^-=0, \bx_\perp)\,\ket{\tilde{\rho}}
    \ge
    -\frac{\delta S(\rho_A||\Omega_A)}{\delta \gamma_A (\bx_\perp)} \ .
\end{align}
The ant-formula follows if one can prove the inequality in the opposite direction.
An interested reader may refer to the papers \cite{Ceyhan:2018zfg,Hollands:2025glm} for the detail of the proof.

%%%

\section{Energy conditions and holographic principle}\label{sec:4}

A recent development of the holographic principle, more specifically the AdS/CFT correspondence, serves as another powerful catalyst for integrating fundamental physics with quantum information theory. In this section, we will provide a brief overview of the basic aspects of the holographic principle, its relation to the black hole information paradox, and its applications to the ANEC.  

\subsection{Black hole entropy and holography}\label{subsec:bh:holography}

Although our main focus is on the energy conditions, a deeper understanding of their quantum extensions is incomplete without addressing holography. This subsection introduces the holographic principle, which---though seemingly peripheral---sheds crucial light on the emergence of energy, entropy, and spacetime itself in quantum gravity. In particular, black holes provide a striking arena where these ideas converge and force us to reconsider basic physical notions.

\medskip

The path to holography begins with the thermal nature of black holes. Hawking's discovery~\cite{Hawking:1974rv,Hawking:1975vcx} that black holes emit blackbody radiation implies that they possess a well-defined temperature. Building on Bekenstein's insight that horizon area encodes entropy~\cite{Bekenstein:1973ur}, the four laws of black hole mechanics were formulated by Bardeen, Carter, and Hawking~\cite{Bardeen:1973gs}, yielding a consistent thermodynamic description:
\begin{align}
T_H & = \dfrac{\hbar \kappa}{2\pi k_{\rm B}} \quad  
\mbox{(Hawking temperature)} 
%\\  
%S_{\rm BH} &= \frac{k_{\rm B} A}{4 G \hbar} \quad \mbox{(Bekenstein-Hawking entropy)}
\end{align}
which is called the {\em Hawking temperature} given by the surface gravity $\kappa$, and 

\begin{itembox}[l]{{Formula 4.1.1 (Bekenstein-Hawking entropy)}}

\vspace{-3mm} 

\begin{equation}
\label{BHentropyformula}
S_{\text{BH}} = \frac{k_{\rm B} A}{4 G \hbar} \,.
\end{equation}
\end{itembox}
This is called the {\em Bekenstein-Hawking entropy formula}, proportional to the area $A$ of a cross-section of the event horizon as seen in Sec.~\ref{subsec:bh}.

\medskip

This thermodynamic interpretation of black holes, though initially counterintuitive, finds natural footing in quantum field theory in curved spacetime. As clarified by the work of Bogoliubov and Unruh~\cite{Unruh:1976db}, particle content depends on the observer and the causal structure of spacetime. When a causal horizon is present, such as a black hole event horizon, it limits access to information and thus redefines the vacuum state and the notion of particles.

\medskip

By dimensional analysis, since Schwarzschild black hole horizon radius scale as $r_h \sim GM$, the corresponding thermal energy satisfies
\begin{equation}
k_{\rm B} T_{\rm H} \sim \frac{\hbar}{G M},
\end{equation}
and applying the first law of thermodynamics yields
\begin{equation}
S \sim \frac{k_{\rm B} G}{\hbar} \int M \d M \sim \frac{k_{B} r_h^2}{\hbar G},
\end{equation}
thus reproducing the area law.

\medskip

What is remarkable about this result is that the entropy scales with area, not volume, suggesting that the fundamental degrees of freedom responsible for this entropy reside not in the bulk, but on the boundary. This observation gave rise to the holographic principle, originally proposed by 't Hooft~\cite{tHooft:1993dmi} and Susskind~\cite{Susskind:1994vu}, which posits that all the information contained in a volume of space can be represented by degrees of freedom on its boundary.

\medskip

Since black holes are believed to be the densest and most entropic objects in nature, they offer a natural upper bound on the information content in a given region. This inspired the conjecture that the fundamental theory underlying quantum gravity is {\it holographic}: 
\begin{itembox}[l]{{Conjecture 4.1.2 (Holographic principle)}}
The complete description of a gravitational system in a bulk spacetime is equivalently formulated as a lower-dimensional, non-gravitational quantum field theory living on its boundary.
\end{itembox}
A concrete and precise realization of this principle was proposed by Maldacena~\cite{Maldacena:1997re} in 1997, now famously known as the AdS/CFT correspondence.

\medskip 

Anti-de Sitter (AdS) spacetime $(M,g_{M N})$ is a maximally symmetric solution to the vacuum Einstein equations with a negative cosmological constant $\Lambda<0$, and its (conformal) boundary $\partial M$ is a spacetime of one dimension lower\footnote{As a maximally symmetric spacetime, a $(d+1)$-dimensional AdS spacetime is represented as the hyperboloid $M \simeq S^1 \times {\Bbb R}^d$ embedded in $(d+2)$-dimensional flat spacetime with the metric of signature $(-,-,+,+,\cdots,+)$. In the context of the AdS/CFT correspondence, by ``AdS bulk" $(M, g_{MN})$ we mean its {\em universal covering space}, i.e., $M \simeq {\Bbb R}^{d+1}$, and therefore globally $\partial M \simeq {\Bbb R}\times S^{d-1}$.  
}. 
As discussed in Sec. 2.2 (see below Definition 2.2.5), since a cosmological constant $\Lambda$ can be viewed as a type of matter with the energy-momentum tensor $T^{(\Lambda)}_{M N}=-(\Lambda/8\pi G_{d+1})g_{MN}$, where $G_{d+1}$ denotes the bulk gravitational constant, AdS spacetime itself violates the WEC and the DEC, but satisfies the SEC and the NEC. Hence the timelike and null focusing theorems both hold in AdS spacetime. 

\medskip 

\begin{figure}[h]
\begin{center}
%\scalebox{0.10}{%
\includegraphics[width=80mm]{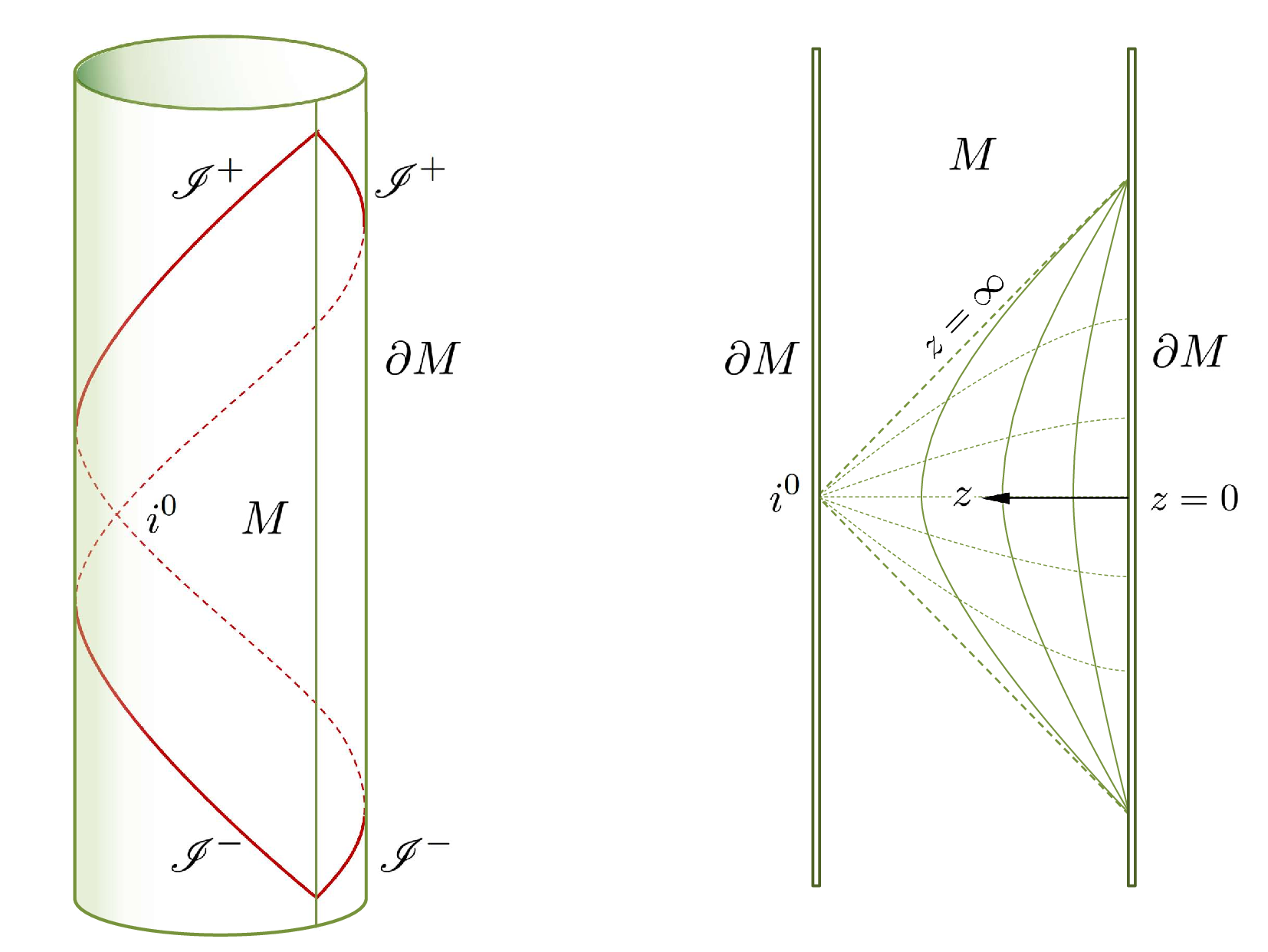} 
%            } 
\end{center}

\vspace{0mm}
\caption{\small 
[Left] The cylinder depicts a conformally compactified AdS spacetime. The cylinder's interior corresponds to the AdS bulk $M$ and its surface represents the conformal boundary $\partial M \simeq {\Bbb R}\times S^{d-1}$. The diamond-shaped subregion on $\partial M$ corresponds to $d$-dimensional Minkowski spacetime $({\Bbb R}^d,\eta_{\mu \nu})$. This subregion is enclosed by the red curves, which represent the future and past null infinity ($\mathscr{I}^\pm \subset \partial M$), and a single point $i^0 \in \partial M$, which represents spatial infinity. 
[Right] $2$-dimensional section of the conformal diagram of AdS spacetime. Two vertical double lines represent conformal infinity (i.e., the surface of the cylinder).  
The triangular subregion inside $M$ is covered by the horospherical chart (also known as the Poincar\'e chart) given by (\ref{AdS:Poincare}), where $\infty >z>0$ inside $M$. Each vertical hyperboloid represents a constant-$z$ hypersurface, which is a $d$-dimensional Minkowski spacetime with the metric $(\ell/z)^2\eta_{\mu \nu}$. The boundary is located at $z=0$.   
} 
\label{AdS_chart}
\end{figure}

The AdS metric $g_{MN}$ in $(d+1)$-dimension can be expressed in the horospherical (Poincar\'e) chart as
\bena
 \d s^2_{(d+1)}= g_{MN}\,\d x^M \d x^N= \dfrac{\ell^2}{z^2}\left( \d z^2 + \eta_{\mu \nu} \d x^\mu \d x^\nu \right) \,,  
\label{AdS:Poincare} 
\eena
where $\ell$ denotes the AdS curvature radius defined by $\ell^2 = -d(d-1)/2\Lambda$ and where $z \rightarrow 0$ corresponds to infinity. 
Multiplying $g_{MN}$ by ${z^2}/\ell^2$ and taking the limit $z \rightarrow 0$, we obtain the $d$-dimensional Minkowski spacetime $({\Bbb R}^d,\eta_{\mu \nu})$ as part of the conformal boundary $\partial M$, which is considered as the arena for a quantum field theory (see Figure~\ref{AdS_chart}). 
As (part of) the $d$-dimensional boundary spacetime $(\partial M,g_{\mu \nu})$, one can also consider different geometries, such as $d$-dimensional de Sitter, AdS, or the static Einstein universe ${\Bbb R}\times S^{d-1}$ (see the metric  (\ref{metric:SEU}) for $d=4$ case), other than the Minkowski spacetime. 

\medskip 

As for the $(d+1)$-dimensional bulk geometry $(M,g_{M N})$ in the AdS/CFT correspondence, one can, in principle, consider any asymptotically AdS spacetime, such as those including black holes inside but at large distances approaching the above AdS metric. 
In general, the AdS/CFT correspondence asserts the following: 
\begin{itembox}[l]{{Conjecture 4.1.3 (AdS/CFT correspondence)}}
A duality between a $(d+1)$-dimensional string theory defined in asymptotically Anti-de Sitter (AdS) spacetime and a $d$-dimensional conformal field theory (CFT) without gravity residing on its conformal boundary. 
\end{itembox}

In this framework, string theory or quantum gravity phenomena in the bulk AdS spacetime can be fully encoded in the quantum dynamics of a conventional QFT on the boundary. In fact, the AdS/CFT correspondence is widely interpreted as providing a precise non-perturbative definition of quantum gravity in AdS via a lower-dimensional, non-gravitational field theory. Strictly speaking, the boundary theory must be a matrix-valued quantum field theory, typically a theory of $N \times N$ matrices such as $\mathcal{N}=4$ supersymmetric Yang--Mills theory, and the correspondence becomes exact in the large-$N$ limit. In this limit, the bulk Newton constant scales as $G \propto 1/N^2$, meaning that classical gravity emerges as the leading order approximation in a $1/N$ expansion. While these details are crucial for precision, we will not delve further into them here.

\medskip  

As an application to the energy conditions, the AdS/CFT correspondence is practically a powerful tool in computing the renormalized energy-momentum tensor for quantum fields.
In general, any asymptotically AdS metric can be expressed near the conformal boundary---similar to (\ref{AdS:Poincare})---in the following form,  
\bena
 g_{MN}\,\d x^M \d x^N= \dfrac{\ell^2}{z^2}\left( \d z^2 + g_{\mu \nu} (z,x)\, \d x^\mu \d x^\nu \right) \,, 
\label{metric:gen:d+1}
\eena
from whichi the $d$-dimensional metric on the conformal boundary at $z \rightarrow 0$ is read off as $g^{(0)}_{\mu \nu} = \lim_{z \rightarrow 0}g_{\mu \nu}(z,x)$. Then, near the conformal boundary, one can expand the $d$-dimensional metric $g_{\mu \nu}(z,x)$ in $z$-coordinate as
\bena
 g_{\mu \nu} (z,x)= \sum_{n=0}^d g^{(n)}_{\mu \nu} (x) z^n + h^{(d)}_{\mu \nu} (x) \log z^2 + \cdots \,,  
\label{FGexp}
\eena
where the logarithmic term appears only when $d$ is even, and $g^{(2k+1)}_{\mu \nu} (x)=0$ for any integer $k$ satisfying $0 \leqslant 2k+1 <d$. This is called the {\em Fefferman-Graham (FG) expansion}. For given $g^{(0)}_{\mu \nu} (x)$, all the subleading coefficients $g^{(n)}_{\mu \nu} (x)$ with $n<d$ are determined solely by the boundary metric $g^{(0)}_{\mu \nu} (x)$ and the curvature tensors with respect to $g^{(0)}_{\mu \nu} (x)$, while the coefficient 
$g^{(d)}_{\mu \nu} (x)$ is related to the expectation value of the boundary field stress-energy tensor $\langle T_{\mu \nu} \rangle$\cite{deHaro:2000vlm}: 
\begin{itembox}[l]{{Formula 4.1.4 (Holographic stress-energy tensor)}}
The renormalized stress-energy tensor on the boundary quantum field is given in terms of the FG expansion coefficient as, 
\bena
 \langle T_{\mu \nu} \rangle = \dfrac{d \ell^{d-1}}{16 \pi G_{d+1}} g^{(d)}_{\mu \nu} (x) + {X}_{\mu \nu} \,. 
 \label{holo:SET}
\eena
\end{itembox}
Here $X_{\mu \nu}$ represents gravitational anomaly, which vanishes when the boundary dimension $d$ is odd. 
This formula stems from the fact that the AdS/CFT correspondence essentially equates the partition functions---and thus their effective actions---of the bulk and boundary theories. It is generally a formidable task to derive the left-hand side by using conventional quantum field theoretic methods. In the above formula, this task is considerably simplified by the use of the AdS/CFT correspondence, which replaces it with the FG expansion of the classical metric in the AdS bulk spacetime. In Sec.~\ref{subsec:causality} and \ref{subsec:canec}, we will discuss how the holographic stress-tensor derived from this formula can be used to examine the ANEC for strongly coupled quantum fields. However, before that, we will discuss more on recent progress in the holographic principle and understanding the black hole information paradox in the next two subsections. 

\subsection{AdS/CFT correspondence and the black hole information paradox} 

From the viewpoint of energy, the AdS/CFT correspondence implies that energetic processes in the bulk, such as black hole formation and evaporation, have a dual description in terms of unitary evolution within the boundary CFT. In this sense, the unitarity of the boundary theory guarantees the unitarity of the bulk gravitational dynamics. 
However, the situation is more subtle. If one takes the AdS/CFT correspondence at face value, then black hole evaporation must indeed proceed as a unitary process, in full accordance with quantum mechanical principles. Yet, it is crucial to emphasize that the AdS/CFT correspondence, despite the overwhelming body of evidence and many successful checks, remains a conjecture rather than a rigorously proven theorem.
Moreover, powerful and long-standing arguments challenge the idea of unitary black hole evaporation. These stem from Hawking's original semiclassical analysis, which predicts that Hawking radiation is purely thermal and uncorrelated with the matter that collapsed to form the black hole. This suggests an irreversible loss of information, in apparent violation of unitarity.
This conflict constitutes the core of the black hole information paradox. If unitarity and semiclassical gravity cannot be simultaneously maintained, then one of our foundational assumptions about spacetime or quantum mechanics must give way. Far from a technical inconsistency, the paradox offers a crucial window into the fundamental nature of quantum gravity.

\medskip

In the remainder of this subsection, we introduce a series of recent developments, particularly the concepts of quantum extremal surfaces and the island formula, that offer a striking resolution to the paradox within the holographic paradigm.

\medskip

To appreciate the tension between semiclassical gravity and unitarity, let us consider Hawking's original argument in a concrete setting. Suppose we have a large black hole, such as a supermassive one at the center of a galaxy, with a mass on the order of $10^5$ solar masses. The horizon radius in such a case is comparable to the size of the Earth. Consequently, the curvature near the event horizon is extremely small, and the semiclassical approximation, that is, treating spacetime classically while quantizing matter fields, should be valid to high precision.

\medskip

\begin{figure}[h]
\begin{center}
%\scalebox{0.25}{%
\includegraphics[width=70mm]{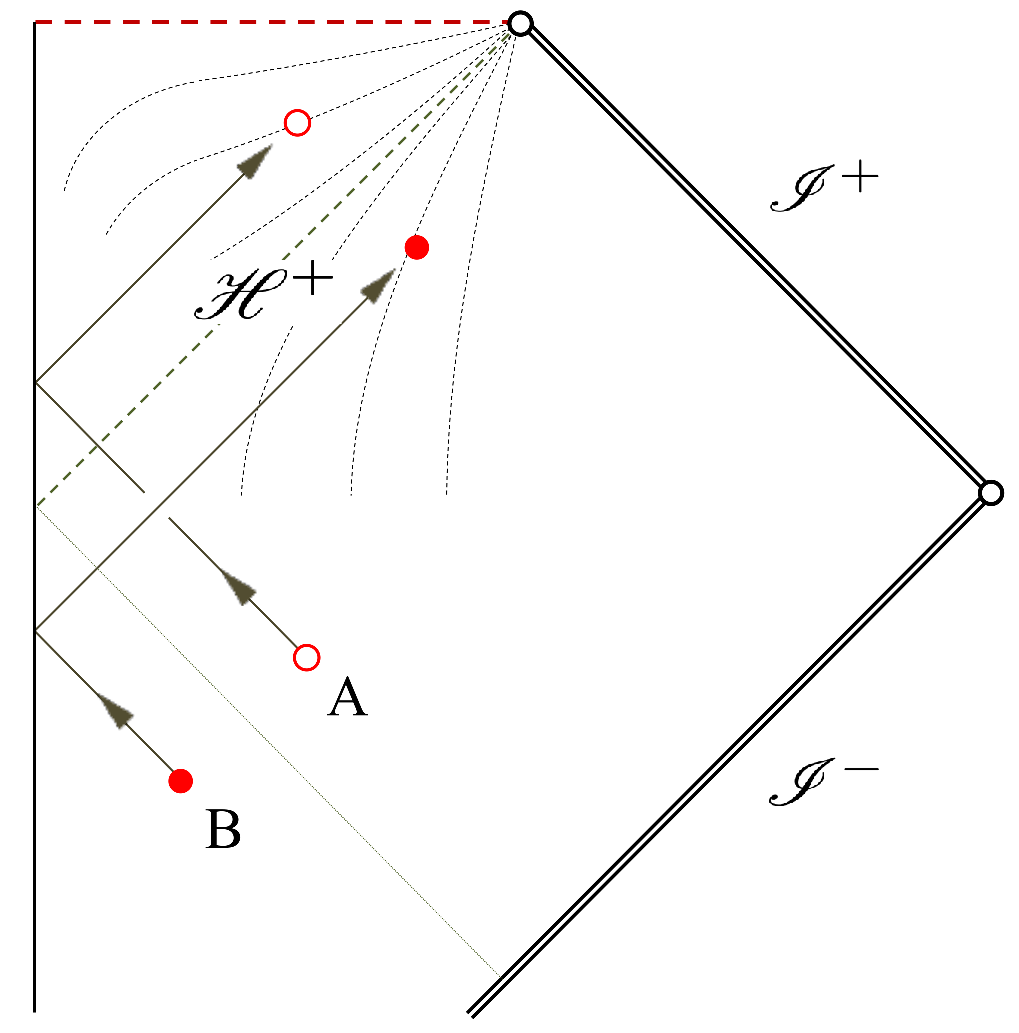}
%            } 
\end{center}
\vspace{5mm}
\caption{\small 
An entangled pair of particles A and B in a spherically symmetric spacetime of black hole formation. The thin dotted hyperboloids depict curves of constant area radius $r$ and ${\mathscr H}^+$ represents the event horizon. Particle A falls into the black hole and moves toward smaller $r$, whereas B moves toward larger $r$. As a result, the distance between the two particles increases with time. Particle B eventually escapes to $\mathscr{I}^+$ as Hawking radiation. This process generates entanglement between the black hole interior and the outgoing Hawking radiation.
} 
\label{Hawking-particles}
\end{figure}

In this approximation, the region near the horizon effectively resembles an expanding spacetime, in the sense that the proper distance between points just inside and just outside the horizon increases with time. This effective ``expansion" is not due to a global cosmological expansion, but rather arises from the diverging world-lines of near-horizon observers: the local geometry induces a relative stretching across the horizon, which facilitates the production of entangled pairs (see Figure~\ref{Hawking-particles}). Quantum fluctuations in this region give rise to entangled particle pairs, one falling into the black hole and the other escaping to null infinity $\mathscr{I}^+$ as Hawking radiation. As discussed in earlier sections, this process generates entanglement between the black hole interior and the outgoing radiation, causing the entanglement entropy of the radiation to increase steadily over time.
If the process continues in this fashion throughout the evaporation, the entropy of the Hawking radiation would grow monotonically. Since the geometry near the horizon remains nearly classical during most of the black hole's lifetime, especially for such large black holes, there appears to be no mechanism within the semiclassical framework that could halt or reverse this entropy growth. 

\medskip

However, as Page pointed out in \cite{Page:1993df, Page:1993wv}, if the evaporation process is ultimately unitary, the entanglement entropy should not increase indefinitely. Rather, it should follow a specific curve, now known as the \textit{Page curve}: it grows initially but must eventually reach a maximum and then decrease to zero as the black hole disappears, restoring a pure final state.
It is worth emphasizing here that this turning behavior---growth followed by decrease---is a natural consequence when one considers the mutual nature of entanglement entropy. The turning point is called the \textit{Page time}, which roughly corresponds to the moment when the number of degrees of freedom in the emitted radiation equals that of the remaining black hole. This typically occurs when the black hole has lost about half of its initial mass.

\medskip

This leads to a striking tension: at the Page time, the black hole is still macroscopically large, and the semiclassical approximation near the horizon should remain valid. Yet, the Page curve requires the entanglement entropy to start decreasing at this stage. This apparent contradiction---between the need for unitary evolution implied by the Page curve and the expected validity of semiclassical gravity---is at the core of the black hole information paradox.

\medskip

A breakthrough in resolving this paradox came with the application of the Ryu-Takayanagi (RT) formula and its quantum-corrected version, which incorporates so called islands in the gravitational path integral. According to the island formula, the entanglement entropy of the Hawking radiation is given not just by the von Neumann entropy of the radiation in quantum field theory, but also includes contributions from a region inside the black hole, the {\it island}, bounded by an extremal surface. This surface plays the role of a generalized RT surface, and its inclusion leads to the Page curve behavior in entropy, thus restoring consistency with unitarity.

\subsection{Entanglement entropy, the Ryu-Takayanagi formula, and the Island paradigm}

A pivotal insight in the effort to resolve the black hole information paradox is that entropy, particularly \emph{entanglement entropy}, plays a fundamental role in the structure of quantum gravity. In semiclassical gravity, the entropy of a black hole is famously captured by the Bekenstein-Hawking formula \eqref{BHentropyformula}.
%\begin{itembox}[l]{{Formula 4.3.1 (Bekenstein-Hawking entropy)}}
%\begin{equation}
%S_{\text{BH}} = \frac{k_{\rm B} A}{4 G \hbar} \,,
%\end{equation}
%\end{itembox}
%where $A$ denotes the area of the event horizon as mentioned at the beginning of this section.
This area-law scaling stands in sharp contrast to the volume-law behavior characteristic of entropy in local quantum field theory, hinting at a radical departure from conventional locality. It suggests that the underlying degrees of freedom in quantum gravity must be organized in a nonlocal fashion.

\medskip

This conceptual leap gained concrete form in the context of the AdS/CFT correspondence, especially with the introduction of the Ryu-Takayanagi (RT) formula~\cite{Ryu:2006bv,Ryu:2006ef}. For a spatial subregion $A$ of the boundary CFT, the RT formula computes its entanglement entropy as\footnote{In the Ryu--Takayanagi formula, the entanglement entropy has a UV divergence of the form $S_A \propto \mathrm{Area}(\gamma_A)/\epsilon^{2}$. Here $\epsilon$ is a short-distance cutoff in the boundary CFT: in AdS Poincaré coordinates, instead of extending the minimal surface all the way to the boundary at $z=0$, one terminates it at $z=\epsilon$, which regulates the divergence.
}
\begin{itembox}[l]{{Formula 4.3.1 (Ryu-Takayanagi formula)}}
\begin{equation}
\label{RTformula}
S_A = \min_{\gamma_A} \left[ \dfrac{k_{\rm B} \mathrm{Area}(\gamma_A)}{4 G \hbar} \right] \,.
\end{equation}
\end{itembox}
Here $\gamma_A$ is the codimension-2 minimal surface in the bulk AdS geometry, anchored to $\partial A$ and homologous to $A$. This prescription geometrizes entanglement: it relates a quantum information-theoretic quantity to a geometric property of spacetime, suggesting that entanglement is not merely a diagnostic tool but a foundational principle underlying gravitational physics.

\medskip

To further appreciate the physical meaning of the RT formula, one can ask: given a bipartition of the boundary system into regions $A$ and $\bar{A}$, what is the corresponding object in the bulk that computes the entanglement entropy $S_A$ in the boundary theory? Since entanglement entropy is a quantity defined entirely in terms of the boundary degrees of freedom, its bulk dual must also be determined solely from boundary data. The RT formula proposes that this dual is the area of a minimal surface $\gamma_A$ in the bulk, anchored to $\partial A$ and homologous to $A$.

\medskip

\begin{figure}[h]
\begin{center}
%\scalebox{0.25}{%
\includegraphics[width=100mm]{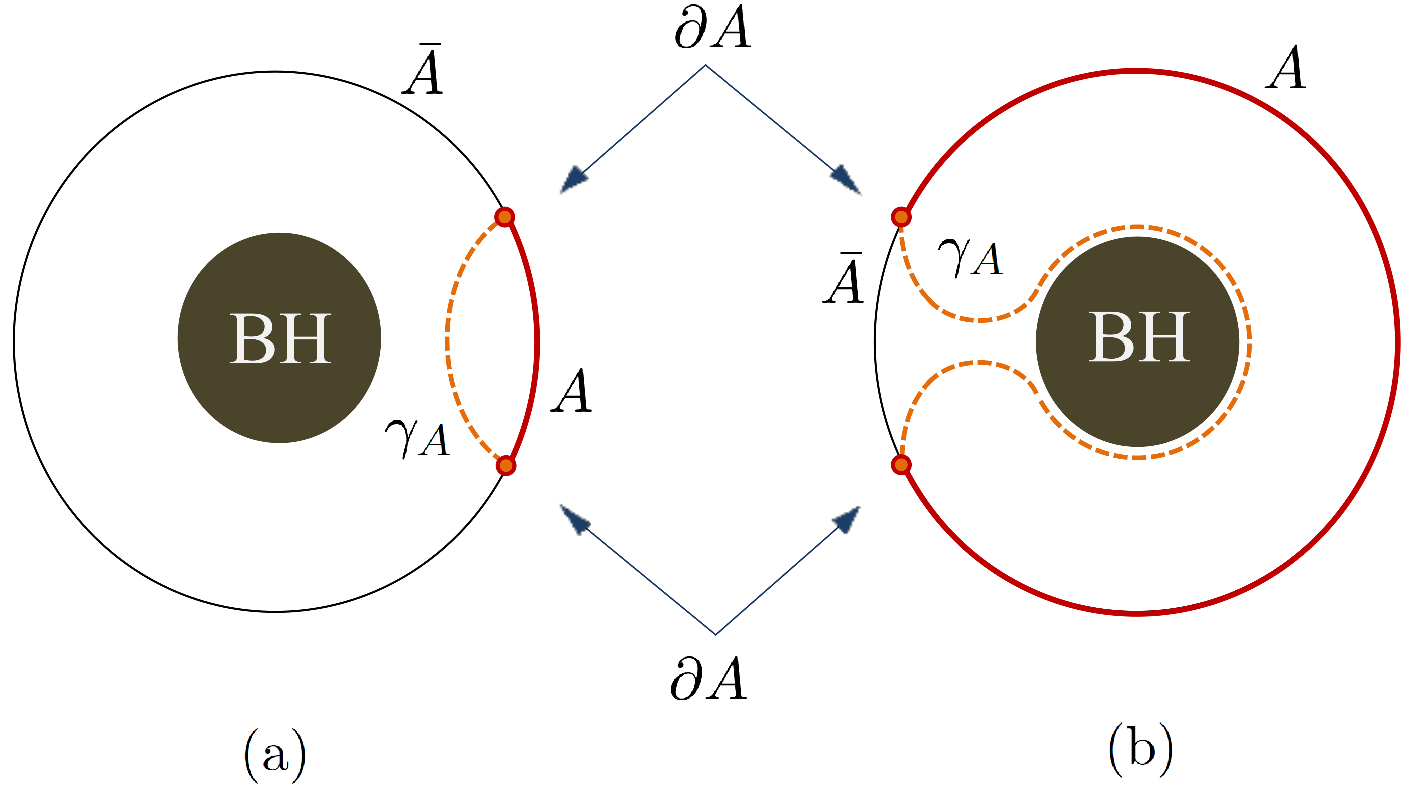}
%            } 
\end{center}
\vspace{-5mm}
\caption{\small 
A conformal diagram of a spatial section of AdS spacetime containing a black hole. The entire circle depicts the spatial section of the conformal boundary $\partial M$. (a) When $A$ (red curve) is small, the minimal surface $\gamma_{A}$ (orange dashed curve) in $M$ anchored to $\partial A$ lies close to the boundary $\partial M$. (b) When $A$ is large, occupying nearly the entire boundary $\partial M$, $\gamma_A$ nearly wraps around the black hole horizon. 
} 
\label{RT-surface}
\end{figure}

To gain intuition about how this minimal surface behaves, consider a thought experiment in which 
AdS spacetime contains a black hole in the bulk and on the boundary the size of region $A$ is gradually increased, starting from an infinitesimal region. Initially, $A$ is very small and $\bar{A}$ occupies nearly the entire boundary. In this regime, the corresponding minimal surface $\gamma_A$ lies close to the boundary and does not probe deep into the bulk. As the size of $A$ grows, and $\bar{A}$ correspondingly shrinks, the minimal surface $\gamma_A$ begins to dip further into the bulk and eventually wraps around the black hole horizon (see Figure~\ref{RT-surface}). 

\medskip

In the limiting case where region $A$ covers the entire boundary (i.e., $\bar{A}$ is empty), the entanglement entropy $S_A$ corresponds to the von Neumann entropy of the global mixed state seen by the external observer. In a holographic theory with a black hole in the bulk, this entropy must match the Bekenstein-Hawking entropy of the black hole:
\begin{equation}
S_A = S_{\text{BH}} = \dfrac{k_{\rm B}A_{\text{horizon}}}{4G\hbar} \,.
\end{equation}
This matching of entropies in the fully mixed limit strongly supports the identification of the black hole horizon itself as the minimal surface $\gamma_A$ in the RT formula \eqref{RTformula}.

\medskip

Thus, the behavior of $\gamma_A$ across varying bipartitions encodes a consistent geometric realization of quantum entanglement in holographic theories. The RT prescription not only satisfies the area law for black hole entropy in the appropriate limit, but also ensures a smooth interpolation across partial regions, reflecting the structure of entanglement entropy in quantum field theory. This consistency under varying partitions reinforces the idea that $\gamma_A$ serves as a geometric surrogate for boundary entanglement, and is a key element in the broader program of interpreting spacetime geometry as emergent from quantum entanglement.

\medskip

The RT formula thus supports the now widespread intuition that ``entanglement builds geometry.'' It suggests that spacetime geometry may emerge from the entanglement pattern in a dual quantum system, embodying the holographic principle. However, the original RT prescription applies only to static configurations and is valid only at leading order in $1/N$, where bulk quantum effects are neglected.
To describe dynamical spacetimes, such as those involving black hole formation and evaporation, the RT formula must be generalized to the covariant {Hubeny-Rangamani-Takayanagi} (HRT) prescription \cite{Hubeny:2007xt}, which replaces minimal surfaces with \emph{extremal} ones. Furthermore, to incorporate quantum corrections---crucial in scenarios like black hole evaporation---one must go beyond the classical approximation and include subleading contributions in $1/N$, leading to the quantum extremal surface (QES) prescription.

\medskip

These bring us to the notion of \emph{quantum extremal surfaces} (QES), developed by Faulkner, Lewkowycz, and Maldacena \cite{Faulkner:2013ana} and further formalized by Engelhardt and Wall \cite{Engelhardt:2014gca}. The QES prescription replaces the classical minimal-area surface with a surface that extremizes the generalized entropy: 
\begin{equation}
S_{\mathrm{gen}}(\gamma)
= \dfrac{k_{\rm B}\mathrm{Area}(\gamma)}{4G\hbar} + S_{\mathrm{bulk}}(\Sigma_\gamma) \,, 
\end{equation}
where $\Sigma_\gamma$ is the bulk region on an achronal slice bounded by $\gamma$ and the boundary region $A$, and $S_{\text{bulk}}(\Sigma_\gamma)$ is the von Neumann entropy of quantum fields in that region. 
The quantum extremal surface $\gamma_A$ is then defined as follows: 
% with $\delta_{\gamma_A} S_{\mathrm{gen}}(\gamma)=0$. If there are many extremal surfaces $\gamma_A$, we should take the one which achieve the global minimum:  
\begin{itembox}[l]{Formula 4.3.2 (Quantum extremal surface)}
\vspace{-5mm} 
\begin{equation}
S(A) = \mbox{min-ext}_\gamma S_{\mathrm{gen}}(\gamma) .
%\min_{\substack{\gamma_A:\ \partial\gamma_A=\partial A\\[2pt]\delta_{\gamma_A} S_{\mathrm{gen}}=0}}
%S_{\mathrm{gen}}(\gamma_A) .
\end{equation}
\end{itembox}
Here $\mbox{min-ext}_\gamma$ represents the following process: first search for extremal surface(s) $\gamma_A$ (homologous to $A$) that satisfy the extremality $\delta_{\gamma_A} S_{\mathrm{gen}}=0$, and then choose the one which attains the minimum among all such extremal surfaces if there are many such $\gamma_A$. The surface obtained by this \emph{min-ext} prescription (among QES homologous to $A$) provides the correct entanglement entropy $S(A)$~\cite{Engelhardt:2014gca}. 
Note that the inclusion of matter entanglement in this way is conceptually rooted in Bekenstein's generalized entropy \cite{Bekenstein:1974ax}.
There, Bekenstein proposed that the total entropy should consist of both the area term and the matter entropy outside the horizon, anticipating the structure of $S_{\text{gen}}$ long before the advent of the AdS/CFT correspondence, even before the discovery of Hawking radiation.

\medskip

This formalism lays the foundation for understanding the \emph{entanglement wedge}, a concept central to the program of holographic quantum error correction \cite{Almheiri:2014lwa, Dong:2016eik, Harlow:2016vwg}. According to the QES prescription, the region $\Sigma_\gamma$ bounded by $A$ and the QES $\gamma$ defines the entanglement wedge of $A$. Bulk operators in $\Sigma_\gamma$ can, in principle, be reconstructed from data in $A$. 

\medskip

\begin{figure}[h]
\begin{center}
% \scalebox{0.25}{%
\includegraphics[width=70mm]{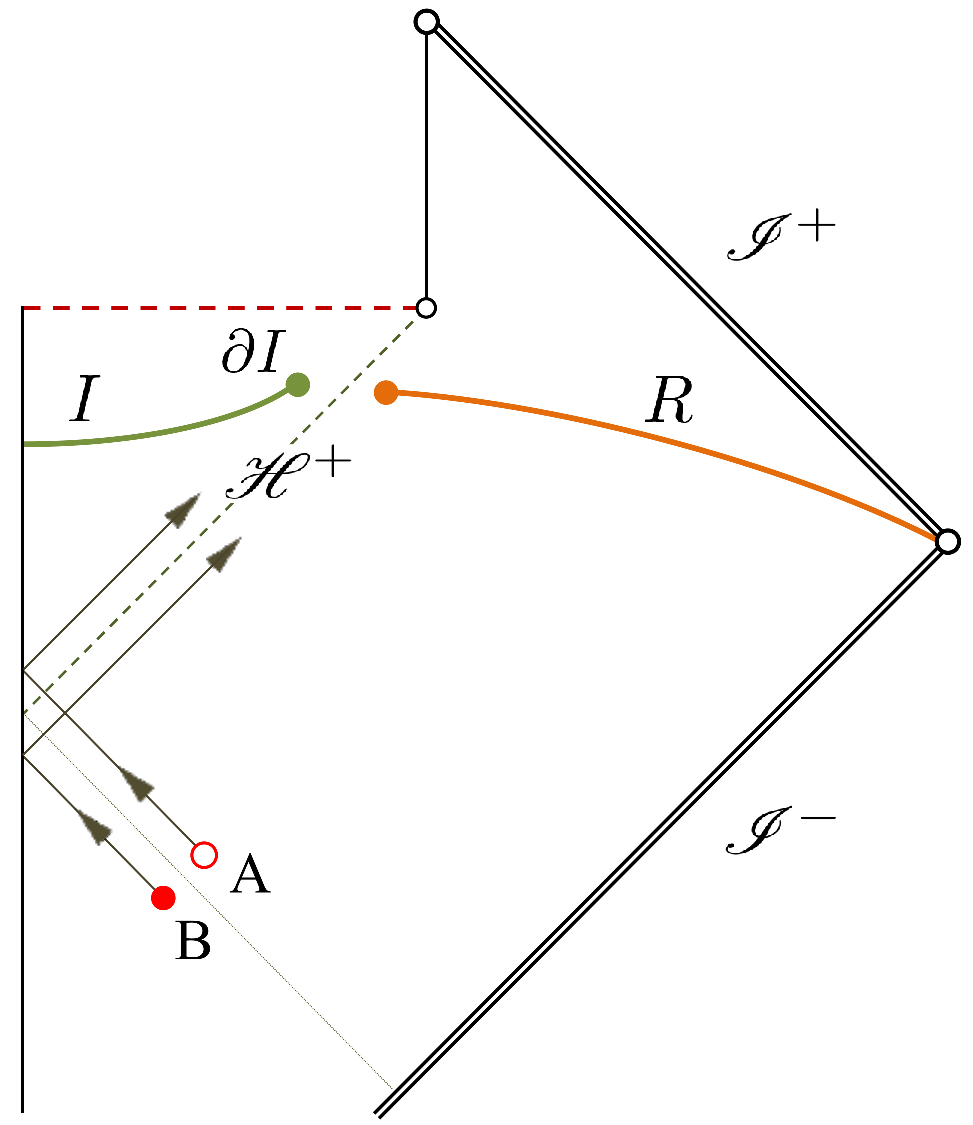} 
%              } 
\end{center}

\vspace{0mm}
\caption{\small 
An evaporating black hole due to Hawking radiation $R$ and the emergence of the island $I$. The entanglement entropy of Hawking radiation is determined by the quantum states not only on the region $R$ (orange curve) but also on the island $I$ (green curve), whose boundary $\partial I$ 
determined by the island formula~(\ref{formula:Island}) is located near the event horizon ${\cal H}^+$. A highly entangled pair of particles A and B, which may be interpreted respectively as Hawking-partner particle and late time Hawking particle. 
} 
\label{Island}
\end{figure}

A major conceptual advance came with the development of the \emph{island formula}, especially in the context of evaporating black holes coupled to an external non-gravitating bath~\cite{Almheiri:2019hni}. In practice, this means working in a regime where gravity is dynamical only in the black hole region, while the bath lives on a fixed background. This framework was pioneered by Almheiri, Engelhardt, Marolf, and Maxfield~\cite{Almheiri:2019hni} and Penington~\cite{Penington:2019npb}, and further developed by~\cite{Almheiri:2019psf}. In such a setup, the entanglement entropy $S_R$ of the Hawking radiation $R$ is given by
\begin{itembox}[l]{{Formula 4.3.3 (Island formula)}}
\begin{equation}
S_R = \mbox{min-ext}_I \left[ \frac{k_{\rm B} \mathrm{Area}(\partial I)}{4 G} + S_{\text{bulk}}(R \cup I) \right] \,. 
\label{formula:Island}
\end{equation}
\end{itembox}
Here $I$ denotes a candidate island region inside the black hole, and $\partial I$ is its boundary (see Figure~\ref{Island}). At early times, the minimal configuration has no island, reproducing Hawking's prediction of monotonically increasing radiation entropy. However, after the Page time, a new configuration appears where a nontrivial island emerges near the black hole horizon that becomes dominant, yielding a smaller generalized entropy. As a result of this island, the entropy of the radiation begins to decrease after the Page time, successfully reproducing the \emph{Page curve}\footnote{In this argument, we assume the existence of a nice foliation of regular (partial) Cauchy surfaces which cover the entire process, ranging from black hole formation to the final stage (an immediate vicinity) of black hole evaporation. For asymptotically flat spacetimes, each of these time slices can be anchored at some boundary surface in the asymptotic region, sometime referred to as cutoff surface or anchor cruve~\cite{Penington:2019npb}. The calculation of QES and the determination of the Page time can be performed based on such time slices.
}. 

\medskip

\medskip 

The reason for this transition from growth to decay in the radiation entropy lies in the emergence of the island near the horizon. Once the island is included in the entanglement wedge of the radiation, the interior degrees of freedom of the black hole---those within the island---are effectively treated as part of the same quantum system as the Hawking radiation itself. Consequently, the entanglement entropy between the radiation and the black hole begins to decrease, as the radiation now purifies the interior. 
Furthermore, the island formula contains a geometric term---the area of the island's boundary---as its first contribution. As the black hole evaporates, this area term shrinks accordingly. Since both the area term and the bulk entropy $S_{\text{bulk}}(R \cup I)$ diminish over time, the generalized entropy of the radiation decreases and eventually vanishes. This reflects the restoration of purity in the final state, as required by unitarity. 

\medskip

This transition restores consistency with unitarity and provides a semiclassically controlled resolution to the black hole information paradox. It indicates that, despite the semiclassical appearance of the geometry, the interior of the black hole becomes encoded in the radiation through the quantum entanglement. 

\medskip

The emergence of islands thus implies a striking nonlocality: information that appears to reside deep inside the black hole is accessible from the Hawking radiation outside. In holographic gravity, spacetime is not merely a stage on which entanglement plays out; it is itself a consequence of entanglement. The RT formula, its quantum extension via QES, and the island paradigm together unify geometry, entropy, and quantum information.

\medskip

Moreover, this flow of insights is not unidirectional. While boundary entanglement patterns determine the bulk geometry, the consistency and causal structure of the bulk spacetime also impose powerful constraints on the boundary theory. A notable example is the derivation of boundary energy conditions, such as the \emph{averaged null energy condition} (ANEC), from bulk causality. In the next section, we explore how this bidirectional logic further deepens our understanding of quantum gravity.

\subsection{Causal constraints and the ANEC in holography}\label{subsec:causality}

A landmark result demonstrating how bulk geometric principles constrain boundary dynamics is the holographic proof of the ANEC on the Minkowski background by Kelly and Wall~\cite{Kelly:2014mra}. Their derivation relies on the requirement of bulk causal consistency, and establishes that the ANEC holds in any boundary CFT admitting a semiclassical gravitational dual.

\medskip

As we had reviewed before, the ANEC asserts that the null-null component of the stress-energy tensor, integrated along a complete null geodesic, is non-negative:
\begin{equation}
\label{bANEC}
\int_{-\infty}^{\infty} \d\lambda \, \langle T_{\mu \nu} k^\mu k^\nu (\lambda) \rangle \geqslant 0 \,, 
\end{equation}
where $k^\mu$ is a future-directed null vector and the expectation value is given by the holographic Formula 4.1.4 \cite{deHaro:2000vlm}. The key insight in the Kelly-Wall argument is that any violation of the ANEC would imply a breakdown of boundary causality via its holographic encoding in bulk geometry. More precisely, if bulk signals could propagate in a manner that violates the causal structure of the boundary theory, then the duality itself would become inconsistent. To formalize this, they consider null deformations of boundary subregions and analyze the corresponding changes in the bulk extremal surfaces computing their entanglement entropy. A violation of the ANEC would induce deformations that allow extremal surfaces to access regions outside the causal domain of the boundary, thereby violating entanglement wedge nesting and the monotonicity of relative entropy.

\medskip 

This perspective is closely related to a foundational result by Gao and Wald~\cite{Gao:2000ga}, who showed that in asymptotically AdS spacetimes satisfying the NEC, bulk causal propagation respects boundary causal structure. More precisely, consider an asymptotically AdS spacetime $(M, g_{MN})$ and its conformal boundary $\partial M$. Suppose a pair of boundary points $p,q \in \partial M$ are connected by a null geodesic $\gamma$ which is lying entirely in $\partial M$ and is achronal with respect to the boundary metric $g^{(0)}_{\mu \nu}$. 
If there is a timelike curve through the bulk $M$ which also connects the boundary two points $p$ and $q$, then the entire spacetime $M \cup \partial M$ is said to admit a {\em bulk-shortcut}. 
\begin{itembox}[l]{{Theorem 4.4.1 (Gao-Wald theorem)}} 
Consider an asymptotically AdS spacetime $M$ which has no causal pathology, such as a naked singularity or causality violation, in the bulk. 
Then, under the (A)NEC and the null generic condition in the bulk, there exists no bulk-shortcut and the fastest possible causal curve connecting two boundary points $p, q$ is an achronal null geodesic $\gamma$ lying entirely on the boundary $\partial M$. 
\end{itembox} 
\begin{itemize}
\item If the null generic condition (Definition 2.4.1) is dropped, the exact AdS spacetime is included. For the exact AdS spacetime case, besides the boundary null geodesic $\gamma$, there exists an achronal null geodesic in the bulk which also connects $p,q \in \partial M$.  

\item The lack of this no shortcut property results in the violation of the weak cosmic censorship~\cite{Ishibashi:2019nby}.  
\end{itemize} 

\begin{figure}[h]
\begin{center}
%\scalebox{0.30}{%
\includegraphics[width=80mm]{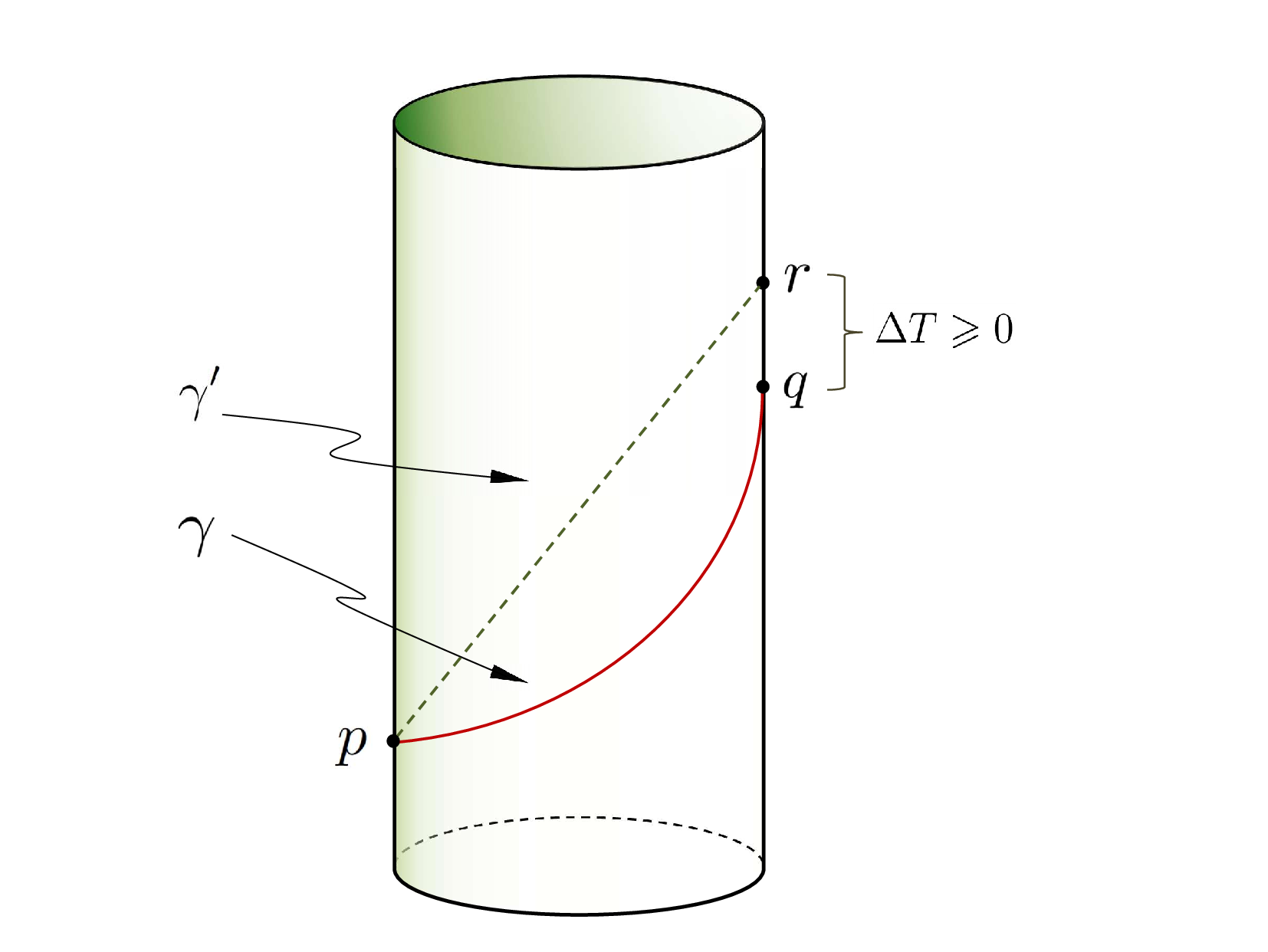} 
%  } 
\end{center}

\vspace{0mm}
\caption{\small 
A conformal diagram of an asymptotically AdS spacetime $M$ and its conformal boundary $\partial M$. Consider a boundary achronal null geodesic $\gamma$ connecting two points $p,q$ in $\partial M$ entirely lying on $\partial M$, and also a bulk null geodesic $\gamma'$ starting from $p$ to a point $r \in J^+(q,\partial M)$. Gao-Wald theorem implies that, in general, $r \neq q$ but instead lies in the chronological future of $q$, i.e., $r \in I^+(q, \partial M)$. If the null generic condition is dropped, it is possible for $r$ to coincide with $q$ (i.e., $r=q$). This happens only when $\gamma'$ is an achronal null geodesic in the bulk, which is the case of the exact AdS bulk. This implies that the propagation of light through an asymptotically AdS spacetime will always be delayed $\varDelta T \geqslant 0$ relative to propagation through the exact AdS spacetime.  
} 
\label{Gao-Wald}
\end{figure}

Gao-Wald (Theorem~2 of \cite{Gao:2000ga}) implies that the propagation of light through an asymptotically AdS spacetime will always be delayed relative to propagation through the exact AdS spacetime (see Figure~\ref{Gao-Wald}). In other words, causal curves in the bulk cannot allow information to travel outside the light-cone defined by the boundary metric. This bulk-to-boundary causality theorem implies that any violation of the ANEC in the boundary theory would necessarily reflect a breakdown of this property in the bulk.

\medskip 

This leads to the remarkable conclusion that in any holographic theory governed by semiclassical Einstein gravity with matter obeying the NEC, boundary causality enforces the ANEC \eqref{bANEC}. That is, the causal structure of the bulk spacetime, via its encoding in extremal surfaces and entanglement wedges, imposes physical energy constraints on the dual quantum field theory.

\medskip 

\subsection{Conformally invariant ANEC}\label{subsec:canec}

As briefly discussed in Sec.~\ref{subsec:AEC}, it is important to note that the ANEC can be violated in conformally coupled free scalar fields on conformally flat spacetimes \cite{Urban:2009yt} (see also~\cite{Visser:1994jb} for more general curved backgrounds), as well as in 
strongly coupled field theories with gravity duals in the holographic setting~\cite{Ishibashi:2019nby}.
A key feature of these examples of the ANEC violation is that a local violation of the NEC can be enhanced via conformal transformation, primarily because the ANEC is not conformally invariant. As a result, the ANEC may be violated in the conformally transformed spacetime.
In this subsection, as a generalization of the ANEC in the context of the AdS/CFT correspondence, we review a weighted ANEC or conformally invariant ANEC~\cite{Iizuka:2019ezn,Iizuka:2020wuj}, which applies to spatially compact spacetimes. 

\medskip 

A canonical example of spatially compact spacetimes is the static Einstein universe. 
As explained in Sec.~\ref{subsec:bh:holography}, the conformal boundary $\partial M$ of $(d+1)$-dimensional AdS spacetime is globally the static Einstein universe in $d$-dimension, $\partial M \simeq {\Bbb R}\times S^{d-1}$. 
Let us consider a pair of boundary points $p, q \in \partial M$.
Fix $p\in\partial M$ and choose $q\in J^{+}(p, {\partial M})$ so that there exists an
achronal boundary null geodesic $\gamma\subset\partial M$ with tangent
$k^\mu=\mathrm{d}x^\mu/\mathrm{d}\lambda$ connecting $p$ to $q$; take $q$ to be the
antipodal point of $p$ on $S^{d-1}$ (Fig.~\ref{Gao-Wald}).
According to Gao-Wald theorem, under the bulk NEC, there is no bulk-shortcut connecting $p,q$ and $\gamma$ must be the fastest causal curve connecting the two point. Let $\lambda_-$ and $\lambda_+$ be the affine parameter values of $\gamma$ at the point $p$ and $q$, respectively, and $\eta$ be the magnitude of the boundary Jacobi field along $\gamma$, which vanishes at $p,q$, i.e., $\eta(\lambda_\pm)=0$ but is non-vanishing at any other points between $p$ and $q$, due to the achronality of $\gamma$ (Proposition 2.5.1). On this curved boundary, we can evaluate the expectation value of the energy-momentum tensor $\langle T_{\mu \nu} \rangle$ by using the holographic formula~(\ref{holo:SET}). Note that when the boundary dimension $d$ is odd, $X_{\mu \nu}$ in (\ref{holo:SET}) identically vanishes, while when $d$ is even, $X_{\mu \nu}$ must be taken into account. Then, applying Gao-Wald theorem, we find the following inequality: 
\begin{itembox}[l]{{Definition 4.5.1 (Holographic weighted ANEC)}} 
\bena
\int_{\lambda_-}^{\lambda_+} \!\!\ \d\lambda \: \eta^d \langle  T_{\mu \nu} \rangle k^\mu k^\nu \geqslant C \,. 
\label{weightANEC}
\eena 
\end{itembox} 
where $\eta$ is the magnitude of the Jacobi field of the null geodesic congruence $\gamma$, representing the separation of points between the two adjacent null geodesics on $\partial M$. 
\begin{itemize}
\item When the boundary dimension $d$ is odd, the lower bound given by constant $C$ in the right-hand side vanishes, and this inequality is manifestly conformally invariant\cite{Iizuka:2019ezn}, and is called the {\em conformally invariant ANEC} (CANEC). In the framework of the conformal boundary field theory, all physical quantities should be described in a conformally invariant way. In this sense, the CANEC is expected to be more useful than the ANEC to put restrictions on the boundary stress-energy tensors (see also \cite{Rosso:2019txh,Rosso:2020cub} for similar analysis of the weighted ANEC). 

\item When the boundary dimension $d$ is even, the lower bound constant $C$ is expressed in terms of the boundary integration along $\gamma$ of a certain combination of boundary geometric quantities such as the boundary curvature tensor and the expansions of the boundary null congruence of $\gamma$ (see \cite{Iizuka:2020wuj} for the detailed expression). 

\end{itemize} 

The lower bound $C$ in (\ref{weightANEC}) contains a quasi-local gravitational energy~\cite{Hayward:1994bu} of the boundary spacetime, and therefore when the boundary curvatures are small enough, the holographic weighted ANEC is expected to be bounded by some local energy density. In this respect, it is interesting to note that for the Schwarzschild-AdS bulk with the mass parameter ${\cal M}$ (whose conformal boundary is still the static Einstein universe), the holographic weighted ANEC implies that \cite{Iizuka:2020wuj} 
\begin{align}
{\cal M} \geqslant 0 \,. 
\end{align}
This suggests that the Gao-Wald theorem is related to the positive energy theorem (see Theorem 2.2.4) in asymptotically AdS bulk spacetimes~\cite{Gibbons:1982jg}. This connection arises because the Schwarzschild-AdS spacetime with negative mass ${\cal M}<0$ contains a naked singularity in the bulk, but under the conditions of the Gao-Wald theorem, the occurrence of a bulk naked singularity is forbidden~\cite{Ishibashi:2019nby}. 
In fact, the non-negative time delay $\varDelta T \geqslant 0$, which is a consequence of Gao-Wald theorem, can be used to derive not only the boundary weighted ANEC (\ref{weightANEC}) but also the positivity of the bulk weighted time-averaged energy~\cite{Page:2002xn}. In this way, the causality of the bulk and boundary spacetimes relates the energy positivity of the bulk and boundary.

\medskip 

\section{Summary and discussions}\label{sec:5} 

In this article, we reviewed various applications of the energy conditions in general relativity, quantum field theory, and the holographic principle, and discussed their profound connection to quantum information.   
In Section~\ref{sec:2}, we first described the locally defined energy conditions: WEC, DEC, SEC, and NEC. The former two primarily concern the positivity aspects of matter field energy-momentum, while the latter two the geometric aspects via the geodesic focusing. We explained how these energy conditions are used---mainly through the focusing theorems---in the proofs of the singularity theorems, the black hole mechanics (or the laws of black hole thermodynamics), and theorems on spacetime topology. We also briefly discussed the violation of these local energy conditions and described their generalizations to the non-local, averaged energy conditions: ASEC and ANEC. 

\medskip 

In Section~\ref{sec:3}, we described the ANEC and QNEC in quantum field theory. The key to their derivations is the positivity and monotonicity of relative entropy. A R\'enyi generalization of QNEC is proposed and discussed in \cite{Lashkari:2018nsl,Moosa:2020jwt,Roy:2022yzm}, and it would be interesting to see if a similar derivation to QNEC would work for the R\'enyi generalization. ANEC is also assumed to derive non-trivial constraints in defect CFTs, i.e., CFTs with non-local operators (such as lines, surfaces and boundaries) in e.g.,~\cite{Jensen:2018rxu,Meineri:2019ycm,Herzog:2020bqw,Chalabi:2021jud}. It remains open whether ANEC holds in the presence of defect as the proof in subsection \ref{sec:ANEC} assumes that the state is in the vacuum and a generalization of ANEC to defect CFTs is left for future investigations. 

\medskip 

In Section~\ref{sec:4}, we first briefly described the holographic principle and the AdS/CFT correspondence. We then reviewed recent progress in understanding the black hole information paradox within the framework of holography, introducing key concepts such as the Ryu-Takayanagi formula, QES, Island formula. 
We then discussed the relationship between ANEC and holography. The AdS/CFT correspondence equates bulk and boundary field theories, which naturally leads to the question of how their respective energy conditions are related. We showed, via the Gao-Wald theorem~\cite{Gao:2000ga}, that these two energy conditions are related by bulk and boundary causalities. As a key application of this theorem, we described the holographic proof of ANEC in Minkowski background~\cite{Kelly:2014mra}, and also, as its generalization to curved backgrounds, the holographic weighted and conformally invariant ANEC~\cite{Iizuka:2019ezn,Iizuka:2020wuj}. 
For the even-boundary dimension case, the holographic weighted ANEC (\ref{weightANEC}) has the non-vanishing lower bound $C$, and thus takes a similar form to the quantum energy inequality (QEI) (see, e.g.,~\cite{Fewster:2012yh,Kontou:2020bta}). The clarification of the relationship between the holographic weighted ANEC and QEI is still ongoing. 
In the standard formulation of the AdS/CFT correspondence, the boundary metric is fixed, not considered as a dynamical variable, and so far, all holographic proofs of ANEC are performed on a fixed background.  
It remains an open question whether the self-consistent ANEC can be shown by using the holographic semiclassical Einstein equations~\cite{Compere:2008us,Ishibashi:2023luz}.   

\medskip 

As we have seen throughout this paper, locally defined energy conditions are often violated by quantum effects, such as vacuum fluctuations or the Casimir effect. It is especially noteworthy that the ANEC is not a universal feature of quantum field theories. Yet, in holographic theories with consistent bulk duals, the ANEC is restored as a consequence of deeper geometric and information-theoretic principles.
Perhaps most strikingly, this result reverses the typical logic of holography: rather than using boundary data to constrain the bulk, bulk causality is used to derive nontrivial dynamical constraints on the boundary theory. The argument elegantly ties together extremal surfaces, entanglement entropy to derive a physical energy condition from first principles.

%%%

\bigskip 

\goodbreak
\centerline{\bf Acknowledgments} We wish to thank Prof. Masahiro Hotta for proposing the idea of this review article. 
The work of N.I. was supported in part by MEXT KAKENHI Grant-in-Aid for Transformative Research Areas (A) ``Extreme Universe" No.~JP21H05182, No. 21H05184. The work of N.I. was also supported in part by NSTC of Taiwan Grant Number 114-2112-M-007-025-MY3. 
The work of A.\,I. and K.\,M. was supported in part by MEXT KAKENHI Grant-in-Aid for Transformative Research Areas (A) ``Extreme Universe" No.~JP21H05182, No.~JP21H05186. 
The work of K.\,M. was supported in part by JSPS KAKENHI Grant-in-Aid for Scientific Research (C) No. 25K07306. 
The work of T.\,N. was supported in part by the JSPS Grant-in-Aid for Scientific Research (B) No.\,24K00629, Grant-in-Aid for Scientific Research (A) No.\,21H04469, and Grant-in-Aid for Transformative Research Areas (A) ``Extreme Universe'' No.\,21H05182 and No.\,21H05190.

\end{document}